\documentclass[useAMS,usenatbib]{mn2e}
\usepackage{graphicx}
\usepackage{amssymb}

\title[Global asteroseismic properties from Kepler]
  {Global asteroseismic properties of solar-like oscillations observed by {\em Kepler}\,: A comparison of complementary analysis methods}
\author[G. A. Verner et al.]
{G. A. Verner$^{1,2}$\thanks{Email: g.verner@qmul.ac.uk}, Y. Elsworth$^{2}$, W. J. Chaplin$^{2}$, T. L. Campante$^{3,4}$, E. Corsaro$^{5}$,
\newauthor P. Gaulme$^{6}$, S. Hekker$^{7,2}$, D. Huber$^{8}$, C. Karoff$^{4}$, S. Mathur$^{9}$, B. Mosser$^{10}$, 
\newauthor T. Appourchaux$^{6}$, J. Ballot$^{11,12}$, T. R. Bedding$^{8}$, A. Bonanno$^{5}$, A-M. Broomhall$^{2}$, 
\newauthor R. A. Garc\'ia$^{13}$, R. Handberg$^{4}$, R. New$^{14}$, D. Stello$^{7}$, C. R\'egulo$^{15,16}$, 
\newauthor I. W. Roxburgh$^{1}$, D. Salabert$^{15,16}$, T. R. White$^{8}$, D. A. Caldwell$^{17}$,
\newauthor J. L. Christiansen$^{17}$, M. N. Fanelli$^{18}$ \\
$^1$Astronomy Unit, Queen Mary University of London, UK \\
$^2$School of Physics and Astronomy, University of Birmingham, UK \\
$^3$Centro de Astrof\'isica, Universidade do Porto, Rua das Estrelas, 4150-762 Porto, Portugal \\
$^4$Department of Physics and Astronomy, Aarhus University, DK-8000 Aarhus C, Denmark \\
$^5$INAF Osservatorio Astrofisico di Catania, Italy \\
$^6$Institut d'Astrophysique Spatiale, UMR 8617, Universit\'e Paris Sud, 91405 Orsay Cedex, France \\
$^7$Astronomical Institute, University of Amsterdam, Science Park 904, 1098 XH Amsterdam, The Netherlands \\
$^8$Sydney Institute for Astronomy (SIfA), School of Physics, University of Sydney, Australia \\
$^9$High Altitude Observatory, Boulder, CO, USA \\
$^{10}$LESIA, CNRS, Observatoire de Paris, France \\
$^{11}$CNRS, Institut de Recherche en Astrophysique et Plan\'etologie, 14 Avenue Edouard Belin, 31400 Toulouse, France \\
$^{12}$Universit\'e de Toulouse, UPS-OMP, IRAP, Toulouse, France \\
$^{13}$Laboratoire AIM, CEA/DSM-CNRS-U, Paris Diderot, IRFU/SAp, Gif-sur-Yvette, France \\
$^{14}$Materials Engineering Research Institute, Sheffield Hallam University, UK \\
$^{15}$Instituto de Astrof\'isica de Canarias, La Laguna, Tenerife, Spain \\
$^{16}$Universidad de La Laguna, La Laguna, Tenerife, Spain \\
$^{17}$SETI Institute/NASA Ames Research Center, Moffett Field, CA 94035, USA \\
$^{18}$Bay Area Environmental Research Inst./NASA Ames Research Center, Moffett Field, CA 94035, USA}

\date{\today}

\pagerange{\pageref{firstpage}--\pageref{lastpage}} \pubyear{2010}

\begin{document}

\def\vl#1{$\vcenter{\hbox{#1}}$\ignorespaces}

\label{firstpage}

\maketitle

\begin{abstract}
We present the asteroseismic analysis of 1948 F-, G- and K-type main-sequence and subgiant stars observed by the NASA {\em Kepler Mission}.  We detect and characterise solar-like oscillations in 642 of these stars.  This represents the largest cohort of main-sequence and subgiant solar-like oscillators observed to date.  The photometric observations are analysed using the methods developed by nine independent research teams.  The results are combined to validate the determined global asteroseismic parameters and calculate the relative precision by which the parameters can be obtained.  We correlate the relative number of detected solar-like oscillators with stellar parameters from the {\em Kepler Input Catalog} and find a deficiency for stars with effective temperatures in the range $5300 \lesssim T_\mathrm{eff} \lesssim 5700$\,K and a drop-off in detected oscillations in stars approaching the red edge of the classical instability strip.  We compare the power-law relationships between the frequency of peak power, $\nu_\mathrm{max}$, the mean large frequency separation, $\Delta\nu$, and the maximum mode amplitude, $A_\mathrm{max}$, and show that there are significant method-dependent differences in the results obtained.  This illustrates the need for multiple complementary analysis methods to be used to assess the robustness and reproducibility of results derived from global asteroseismic parameters.
\end{abstract}

\begin{keywords}
  stars: fundamental parameters -- stars: interiors -- stars: oscillations -- stars: solar-type.
\end{keywords}

\section{Introduction}

Many stars are observed to exhibit `solar-like' oscillations that are stochastically excited by near-surface convection \citep[\textit{e.g.}][]{brown1994,jcd2004}.  These global oscillations can be observed either in spectroscopic Doppler velocity measurements or by detecting small variations in brightness.  Similar global oscillations taking place on the Sun have been studied in great detail and have provided a wealth of information on the solar interior structure \citep[see][for a review]{chaplin2008}.  Solar-like oscillations have been detected on a number of main-sequence stars using multi-site ground-based spectroscopy \citep{bedding2007} and the recent high-precision photometric observations of the French-led {\em COROT} space telescope \citep{baglin2007}.  This has led to the first detailed asteroseismic studies of solar-like oscillations in main-sequence stars \citep{appourchaux2008,benomar2009,deheuvels2010,mathur2010a}.

The NASA {\em Kepler} space telescope simultaneously monitors the brightness of over 150,000 stars with the primary mission of detecting the characteristic dips in brightness caused by transiting exoplanets \citep{borucki2010,koch2010}.  The high precision available in the data enables them to also be used for asteroseismic studies \citep{gilliland2010a}.  The majority of stars observed by {\em Kepler} have photometric measurements taken at a mean `long' cadence of 29.4 minutes.  However, a subset of up to 512 stars at any given time are observed on a mean `short' cadence of 58.85 seconds\footnote{The actual cadence varies between 58.74 and 59.00 seconds once the observations are corrected for movement around the solar-system barycentre.}.  These high-precision, short-cadence observations \citep{gilliland2010b} are ideal for detecting solar-like oscillations, which have typical periods of the order of a few minutes.

In order to fully characterise a star using asteroseismology, it is desirable to have accurate estimates of individual $p$-mode parameters -- including the frequencies, amplitudes and lifetimes of a large number of modes for which the angular degree, $l$, and radial order, $n$, have been identified.  This can be achieved using a variety of classical maximum-likelihood peak-bagging techniques \citep{toutain1994,appourchaux1998} or similar alternative methods such as a Bayesian approach \citep[\textit{e.g.}][]{brewer2007}.  However, this process is often time intensive and only appropriate for data above a certain signal-to-noise level.  The average asteroseismic parameters, indicative of the global structure, are readily obtainable using automated analysis methods and can incorporate data with a lower signal-to-noise ratio for which a full peak-bagging analysis is not always possible.

In this paper, we compare the global asteroseismic parameters obtained from the analysis of 1948 F-, G- and K-type stars observed during the first seven months of {\em Kepler} operations and carried out using the methods of nine teams.  This incorporates all short-cadence observations of stars identified as potentially showing solar-like oscillations, which includes a number of giants and subgiants in addition to stars on the main sequence.  We use simulated data to test the precision and accuracy of each method and develop a process to detect spurious outliers and ensure that data from different teams are self-consistent within the stated confidence bounds.  The fraction of stars for which we can detect oscillations depends on the stellar parameters themselves and we correlate our detections with the parameters available from the {\em Kepler Input Catalog} (KIC, \citealt{brown2011}).  We then derive power-law relationships between the verified global parameters and draw conclusions from our results.

\begin{table*}
  \centering
  \begin{tabular}{|l|l|l|l|l|}
  \hline
  Method & Method for $\Delta\nu$ & Method for $\nu_\mathrm{max}$ & Method for $A_\mathrm{max}$ & Method for $\delta\nu_{02}$ \\
  \hline
  A2Z$^1$ & Peak of PSPS & Fit to SPS & Fit to SPS (K08) & -- \\
  AAU & Peak of PSPS & Peak of SPS & Peak of SPS & -- \\
  COR$^2$ & Fit to TSACF & Fit to SPS & Fit to TSACF & Fit to TSACF \\
  IAS     & Peak of TSACF & Fit to SPS & Fit to SPS (K08) & -- \\
  OCT$^3$ & 1: PSPS (full PS)  & 1: Fit to SPS & 1: Peak of SPS & Individual frequency fits \\
          & 2: PSPS (Bayesian on full PS)$^\dagger$ & 2: Peak of SPS$^\dagger$ & 2: Fit to SPS & \\
          & 3: PSPS (small $\nu$ range)$^\dagger$ & & 3: Fit to SPS (K08) & \\
          & 4: PSPS (Bayesian on small $\nu$ range)$^\dagger$ & & & \\
  ORK$^4$ & COMB function & CLEAN algorithm & -- & -- \\
  QML$^5$ & 1: Peak of TSACF & Fit to PSACF & Fit to PSACF & Fit to PSACF \\
          & 2: Fit to PSACF & & & \\
  KAB$^{6,7}$ & Fit to PSACF & 1: Fit to PSACF & Fit to SPS & Individual frequency fits \\
          & & 2: Fit to SPS & & \\
  SYD$^8$ & Fit to PSACF & 1: Peak of SPS & 1: Peak of SPS (K08) & -- \\
          & & 2: Fit to SPS & 2: Fit to SPS (K08) & \\
  \hline
  \multicolumn{5}{|l|}{$^1$\,\cite{mathur2010b}; $^2$\,\cite{mosser2009}; $^3$\,\cite{hekker2010}; $^4$\,\cite{bonanno2008};} \\
  \multicolumn{5}{|l|}{$^5$\,\cite{verner2010}; $^6$\,\cite{campante2010}; $^7$\,\cite{karoff2010a}; $^8$\,\cite{huber2009};} \\
  \multicolumn{5}{|l|}{$^\dagger$\,only used for the analysis of simulated data.} \\
\end{tabular}
\caption{Brief description of the methods used by each analysis team.  Abbreviations used are: SPS -- smoothed power spectrum, PSPS -- power spectrum of the power spectrum, TSACF -- time series autocorrelation function, PSACF -- power spectrum autocorrelation function, K08 -- method based on \protect\cite{kjeldsen2008b}.}
\label{tab:methods}
\end{table*}

\section{Data}

We have used the short-cadence data from the first seven months of observations made by {\em Kepler}.  These data consist of 2229 light curves obtained from stars of {\em Kepler}-band magnitude $Kp=6.9$ to $12.1$, each of approximately one-month duration and obtained with a mean cadence of 58.85 seconds.  These stars were predetermined to be potential solar-like oscillators based on their characteristics in the KIC.  Some of the stars were observed during multiple one-month runs, resulting in 1948 unique stars in the complete set.  This represents 74\,\% of the stars observed at short cadence during the initial survey phase of the {\em Kepler Mission}.

The light curves were first prepared in the manner described in \citet{garcia2011} starting with the raw data provided from the {\em Kepler Mission} pipeline \citep{jenkins2010}.  By using a common pre-processing method to remove spurious points and detrend each light curve, we ensured that all of the analysis teams started with the same input data for their methods.  We had found that variations in the treatment of outlying data points and different methods used to interpolate missing points or remove long-term trends introduced a significant bias into the asteroseismic parameters obtained for each star.

\subsection{Global parameters}

The prepared time series and corresponding power spectra were analysed to give estimates of the asteroseismic parameters sensitive to the global structure of the observed star.  The most prominent feature of an acoustic power spectrum is the near-regular spacing between modes of the same angular degree and consecutive orders.  The average of this so-called large frequency separation ($\Delta\nu$) gives an estimate of the total acoustic radius of the star, \textit{i.e.} the sound travel time from the centre to the surface:
\begin{equation}
  \tau_t = \int_0^R \frac{dr}{c} \approx \frac{1}{2\:\Delta\nu}.
  \label{eq:acrad}
\end{equation}
The average large frequency separation is also known to scale with the square root of the mean density of the star (e.g. \citealt{jcd1993}), giving the scaling relation
\begin{equation}
  \frac{\Delta\nu}{\Delta\nu_\odot} \approx \left( \frac{M}{M_\odot} \right)^{0.5} \left( \frac{R}{R_\odot} \right)^{-1.5}.
  \label{eq:lsep}
\end{equation}

The heights of the $p$-mode peaks in a power spectrum are modulated by an envelope that attains its maximum value at a frequency we denote as $\nu_\mathrm{max}$.  It is usual to make the assumption that $\nu_\mathrm{max}$ scales with the acoustic cutoff frequency \citep{brown1991,kjeldsen1995,mosser2010}, which when determined for an isothermal stellar atmosphere gives a scaling relation in terms of mass, radius and effective temperature of
\begin{equation}
  \frac{\nu_\mathrm{max}}{\nu_{\mathrm{max},\odot}} \approx \frac{M}{M_\odot} \left( \frac{R}{R_\odot} \right)^{-2} \left( \frac{T_\mathrm{eff}}{T_{\mathrm{eff},\odot}} \right)^{-0.5}.
  \label{eq:numax}
\end{equation}
The height attained by a single $p$-mode profile is related to the intrinsic amplitude of the underlying stellar oscillation but is also affected by the lifetime of the mode.  Modes with shorter lifetimes have profiles with greater linewidths, spreading power over many frequency bins.  If the power spectrum is converted to power spectral density by dividing by the frequency resolution, the intrinsic RMS amplitude of a radial mode can be obtained as
\begin{equation}
  A_0 = \sqrt{\frac{\pi H_0 \Gamma}{2}},
  \label{eq:ampfromheight}
\end{equation}
where $H_0$ is the height of the underlying limit-spectrum profile, usually modelled by a Lorentzian function, and $\Gamma$ is the corresponding linewidth.  To obtain the intrinsic amplitude of non-radial modes, an extra factor appears in Eq.~\ref{eq:ampfromheight} to account for the averaging of the oscillation eigenfunction over the visible stellar disk and the effects of limb darkening.  The amplitude that would be attained by a radial mode with a frequency of $\nu_\mathrm{max}$ we refer to as $A_\mathrm{max}$ and reflects the excitation mechanism present in the star.  If we assume an adiabatic stellar atmosphere, the value of $A_\mathrm{max}$, once corrected to a bolometric amplitude, is expected to scale with the RMS velocity perturbation ($v_\mathrm{osc}$) and inversely with the square root of effective temperature \citep{kjeldsen1995}:
\begin{equation}
   A_\mathrm{max} \propto \frac{v_\mathrm{osc}}{T_\mathrm{eff}\,^{0.5}}
   \label{eq:amax}
\end{equation}

The surface layers of a star introduce a significant frequency-dependent perturbation into the measured frequencies of $p$ modes \citep[\textit{e.g.}][]{kjeldsen2008a}.  The modes of degree $l=0$ and $l=2$ are closely spaced in frequency and are therefore affected similarly by the surface layers.  The small frequency separation between these modes ($\delta\nu_{02}$) is sensitive to the physical conditions in the deep interior and is therefore an important indicator of stellar structure, age and evolutionary state.  Additionally, the ratio of large to small frequency separations has been shown to discriminate well between stellar evolutionary models and is insensitive to variations in the surface layers \citep{roxburgh2003,roxburgh2005,roxburgh2010}.

\section{Validation and verification of the different methods}

The prepared time series were analysed by the automated methods of nine teams.  Each method returned some or all of the globally averaged seismic parameters outlined above for each star.  Although the exact methods used by each team varied, many of the methods used similar techniques based on autocorrelation analyses \citep[e.g.][]{roxburgh2006,roxburgh2009} to obtain $\Delta\nu$ and $\nu_\mathrm{max}$ or the power spectrum smoothing method of \citet{kjeldsen2008b} to obtain $A_\mathrm{max}$ and $\nu_\mathrm{max}$.  The methods used are briefly outlined in Table~\ref{tab:methods}.

\subsection{Simulations}

\begin{figure}
\dimen0=\hsize
\dimen1=2mm
\advance\dimen0 by -1\dimen1
\dimen1=0.5\dimen0
\centerline{%
  \vl{\includegraphics[width=\dimen1]{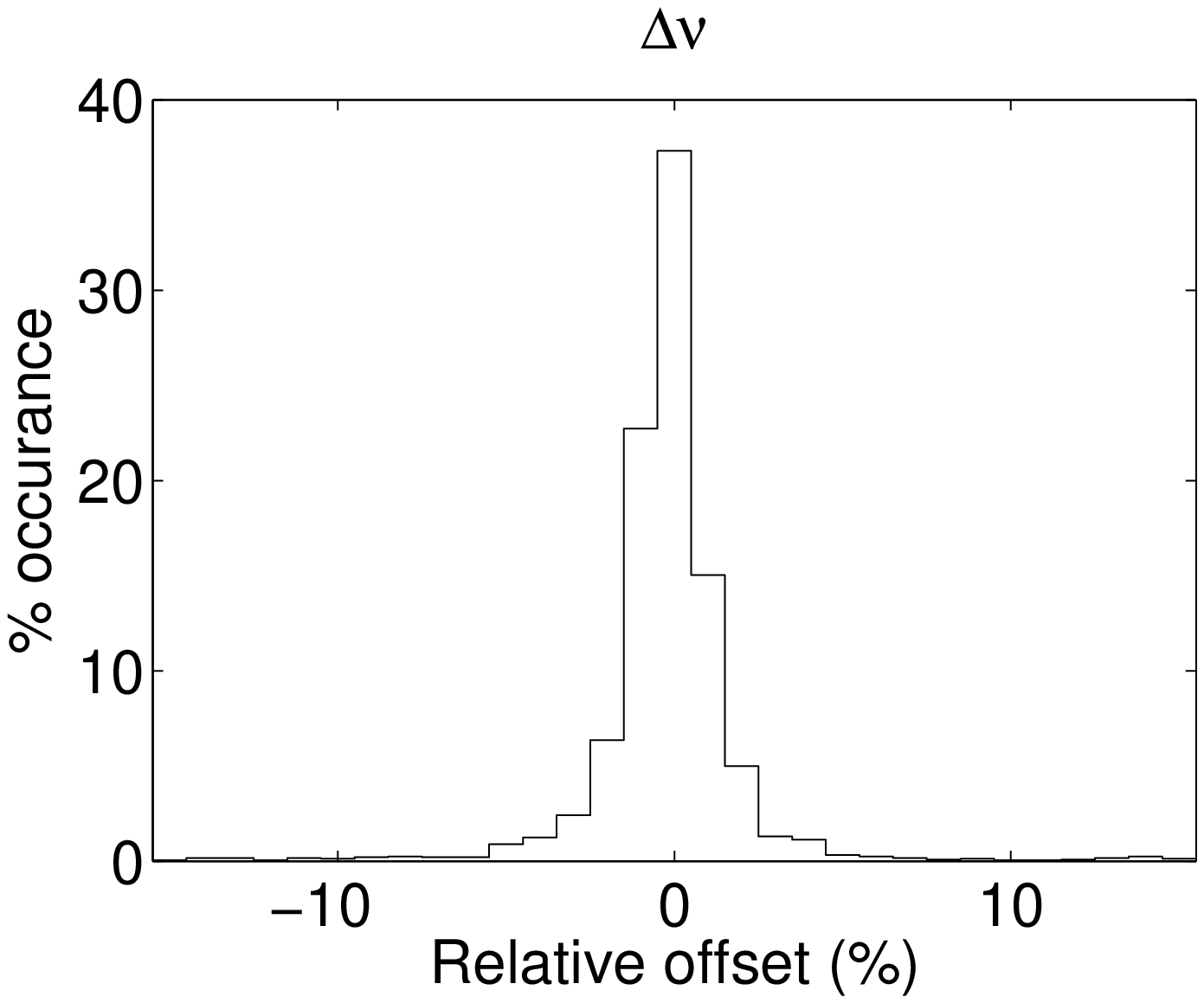}}
  \hfill
  \vl{\includegraphics[width=\dimen1]{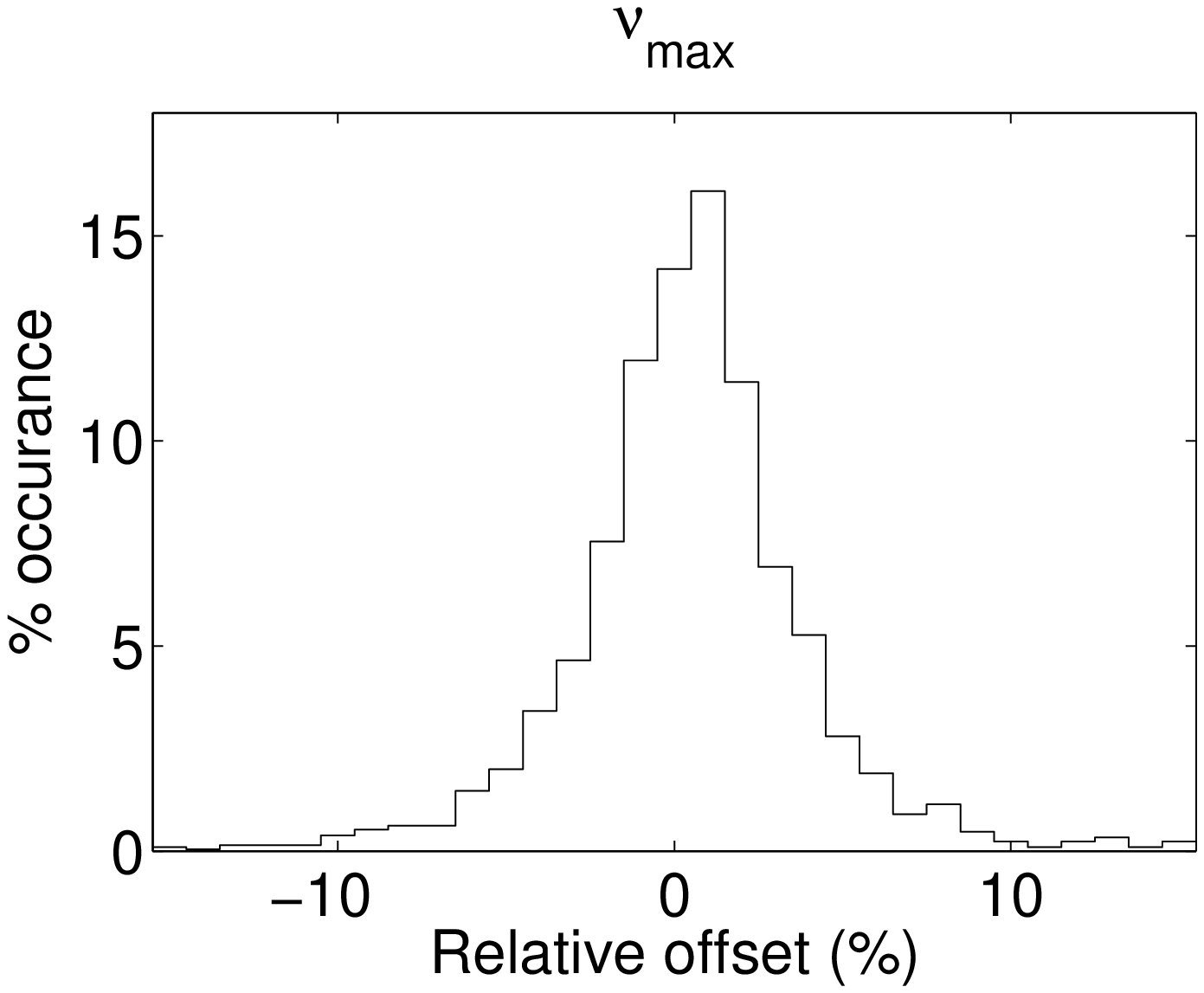}}
}
\medskip
\centerline{%
  \vl{\includegraphics[width=\dimen1]{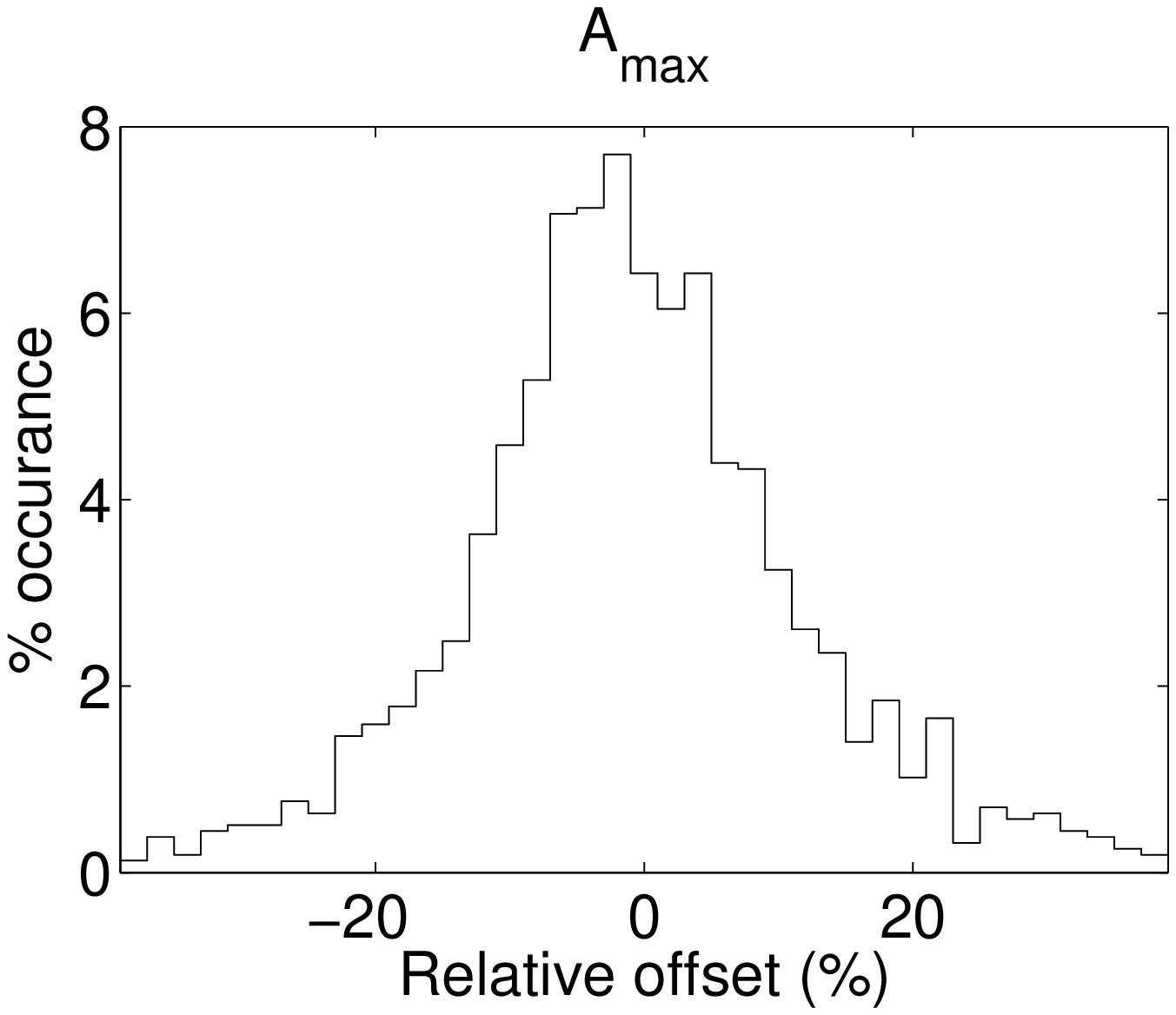}}
  \hfill
  \vl{\includegraphics[width=\dimen1]{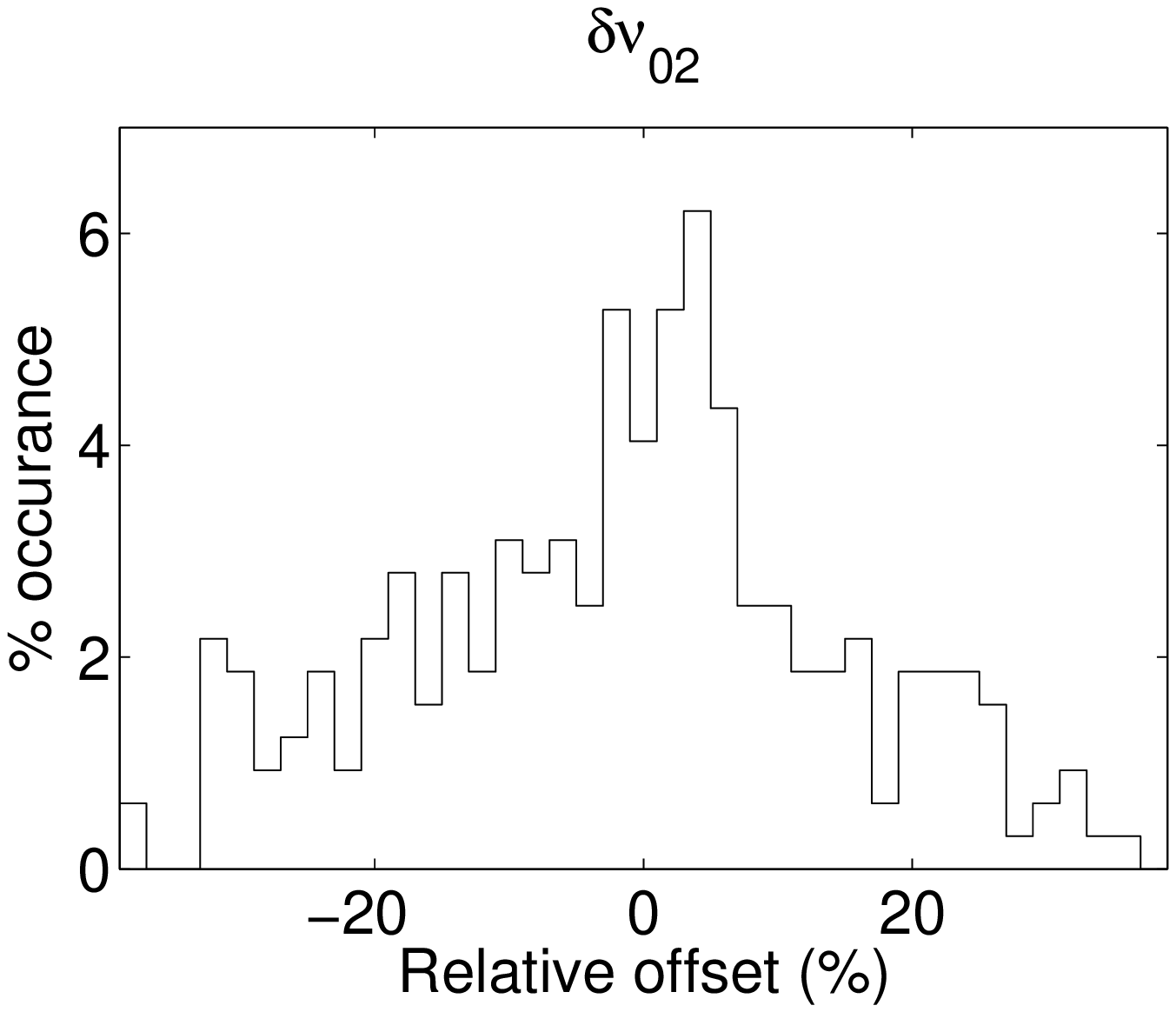}}
}
\caption{Histograms showing the relative deviation of the fitted values of $\Delta\nu$, $\nu_\mathrm{max}$, $A_\mathrm{max}$ and $\delta\nu_{02}$ from their true values as determined from the simulated data results across all methods.}
\label{fig:arthist}
\end{figure}

In order to test the performance of a representative sample of the analysis methods, a set of 309 simulated {\em Kepler}-like time series each of one-month duration were prepared and analysed by six teams (A2Z, COR, KAB, OCT, QML, SYD) in a blind hare-and-hounds exercise.  The stellar parameters were chosen to cover a range in values of $\Delta\nu$, $\nu_\mathrm{max}$, $A_\mathrm{max}$ and $\delta\nu_{02}$ in accordance with the expected parameters of the stars showing solar-like oscillations in the {\em Kepler} field.  The values returned by the teams were compared with the input values of asteroseismic parameters for each simulated star to determine the bias, precision, number of outliers and accuracy of formal uncertainties for each parameter.  Due to the large variation of the parameters over the stars in the cohort, and the assumption that the inverse granulation timescale and width of the $p$-mode power excess both scale with frequency \citep{huber2009}, we have assumed that the formal uncertainties scale with frequency also and therefore used relative uncertainties in our analysis.  The histograms showing the relative deviations of each parameter from the corresponding input values across all methods are shown in Fig.~\ref{fig:arthist}.

The simulated data were generated using realistic stellar-model frequencies as input parameters, thus the values of $\Delta\nu$ and $\delta\nu_{02}$ for each simulated star vary with frequency.  The comparison reference values of $\Delta\nu$ were determined using a cubic spline interpolation to the spacing of the seven radial modes in the vicinity of $\nu_\mathrm{max}$ as a function of frequency, evaluated at $\nu_\mathrm{max}$.  The reference values for $\delta\nu_{02}$ were obtained in a similar way.  However, the $\delta\nu_{02}$ spacing, particularly for evolved solar-like oscillators, may not vary smoothly with frequency due to mode `bumping' caused by coupling with $g$ modes confined within the stellar interiors (White et al., in preparation).  Stars in the simulated set which did not have a smoothly varying $\delta\nu_{02}$ spacing in the region around $\nu_\mathrm{max}$ were not included in the $\delta\nu_{02}$ comparison analysis.

\subsubsection{$\Delta\nu$}

The analysis of simulated data showed that $\Delta\nu$ was the most precisely determined parameter.  For the results of each method, we identified outliers as those points lying outside of a relative range of $5$\,$\sigma$ from the true value.  We used the median absolute deviation, scaled by $1.4826$ assuming a normal distribution, for our estimate of $\sigma$ to reduce the influence of outliers on our measure of the spread.  The average relative number of outliers we found across all methods was $6$\,\%, although this varied from $1$\,\% to $11$\,\%.  With the outliers removed from the results of each method, the average standard deviation of the remaining distributions was $1.2$\,\% and varied between $0.7$\,\% and $2.0$\,\%.  We applied the t-test to the distributions and found that four (out of twelve) had centroids sufficiently far from zero that the null hypothesis (mean is zero) could be rejected at the $95$\,\% level, hence indicating a small bias.  The centroids of these distributions were all between $0.3$\,\% and $0.4$\,\%, indicating that any overestimation bias is at a level that is well within the expected $1$\,$\sigma$ error distribution.  These biases were not apparent in the later analysis of {\em Kepler} data.

\subsubsection{$\nu_\mathrm{max}$}

The simulations showed that the spread in values of $\nu_\mathrm{max}$ is wider than that for $\Delta\nu$.  The values for comparison were determined from the frequency at which the amplitude envelope of radial modes attained its maximum value.  In the simulations, as in real stars, the non-radial $l=1$ modes have a slightly higher visibility than radial modes due to their increased spatial contribution when averaged over the visible stellar disk and the effects of limb darkening.  The amplitude envelope for non-radial modes is assumed to peak at the same frequency as that for radial modes, however the determined value of $\nu_\mathrm{max}$ may be sensitive to which modes are present in the region around $\nu_\mathrm{max}$.  Many of the methods use a wide smoothing filter in the power spectrum to minimise this effect.

We identified outliers in the same way as for $\Delta\nu$ and found that the average proportion of outliers was $5$\,\%, varying from $1$\,\% to $9$\,\% over the methods.  Once these outliers were removed, the average standard deviation of the remaining distributions was $3.0$\,\% and varied between $2.1$\,\% and $4.3$\,\%.  Applying the t-test to the distributions indicated that three (out of eight) had centroids that rejected the zero-mean hypothesis at the $95$\,\% level.  The mean value of these distributions were $-0.9$\,\%, $1.1$\,\% and $1.2$\,\%, the largest of which corresponded to $0.4$\,$\sigma$.

The presence of a bias in the estimation of $\nu_\mathrm{max}$ may be due to the model used by each method to fit and remove the granulation background or the function used to fit the $p$-mode power excess.  The methods that were identified as having a small bias in the analysis of simulated data were subsequently improved upon before the analysis of real data.

\subsubsection{$A_\mathrm{max}$}

\begin{figure*}
\dimen0=\hsize
\dimen1=2mm
\advance\dimen0 by -3\dimen1
\dimen1=0.25\dimen0
\centerline{%
   \vl{\includegraphics[width=\dimen1]{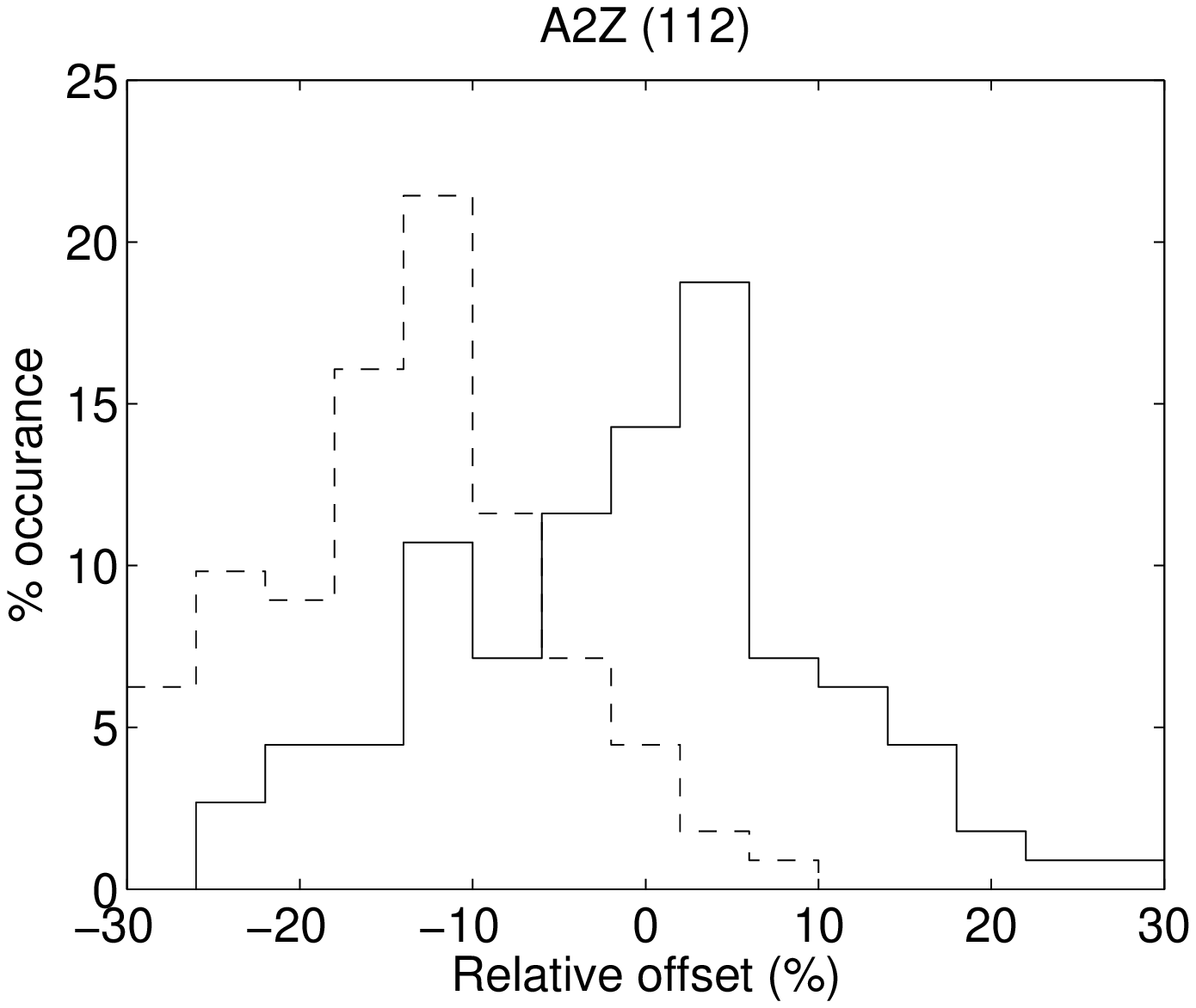}}
   \hfill
   \vl{\includegraphics[width=\dimen1]{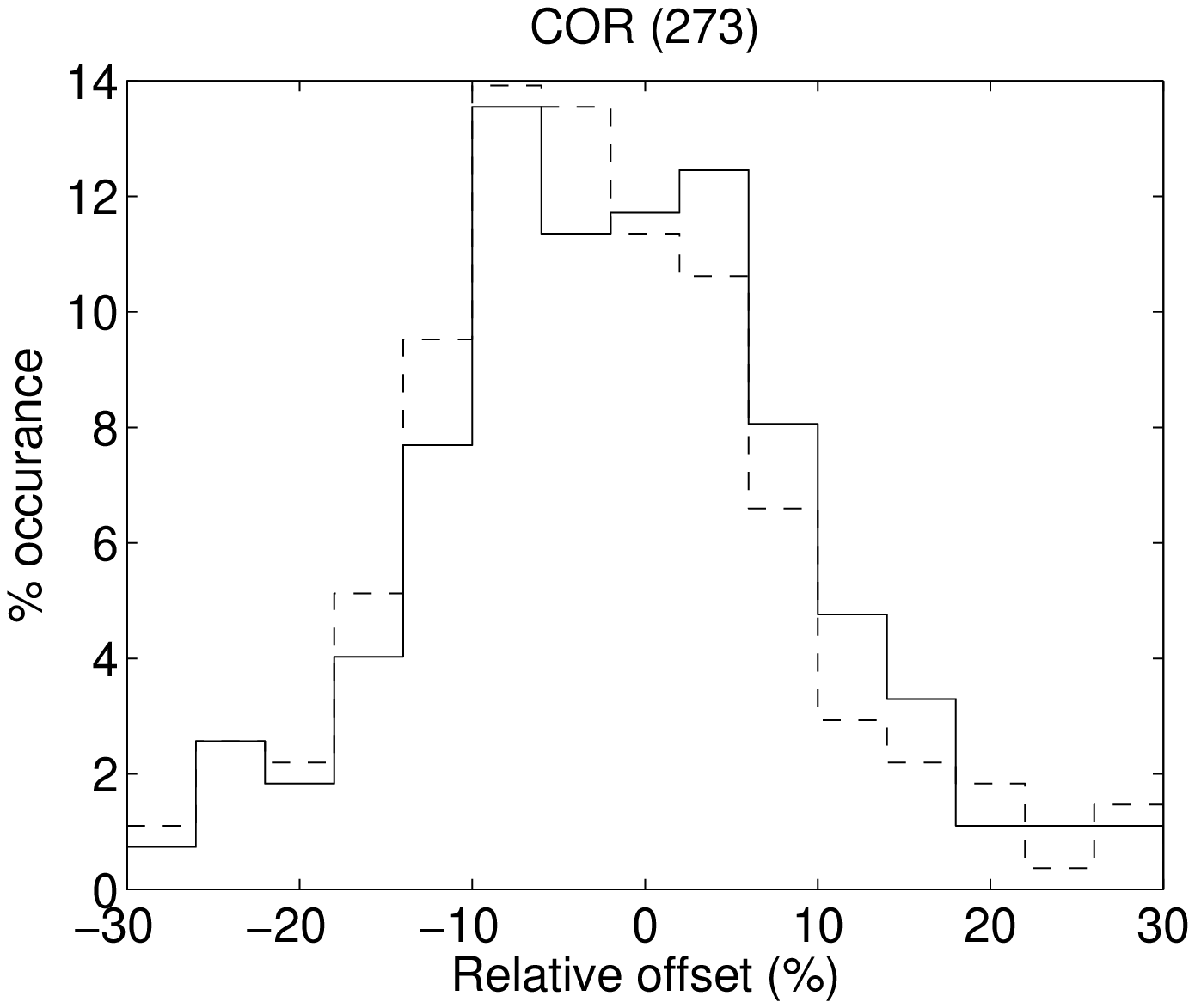}}
   \hfill
   \vl{\includegraphics[width=\dimen1]{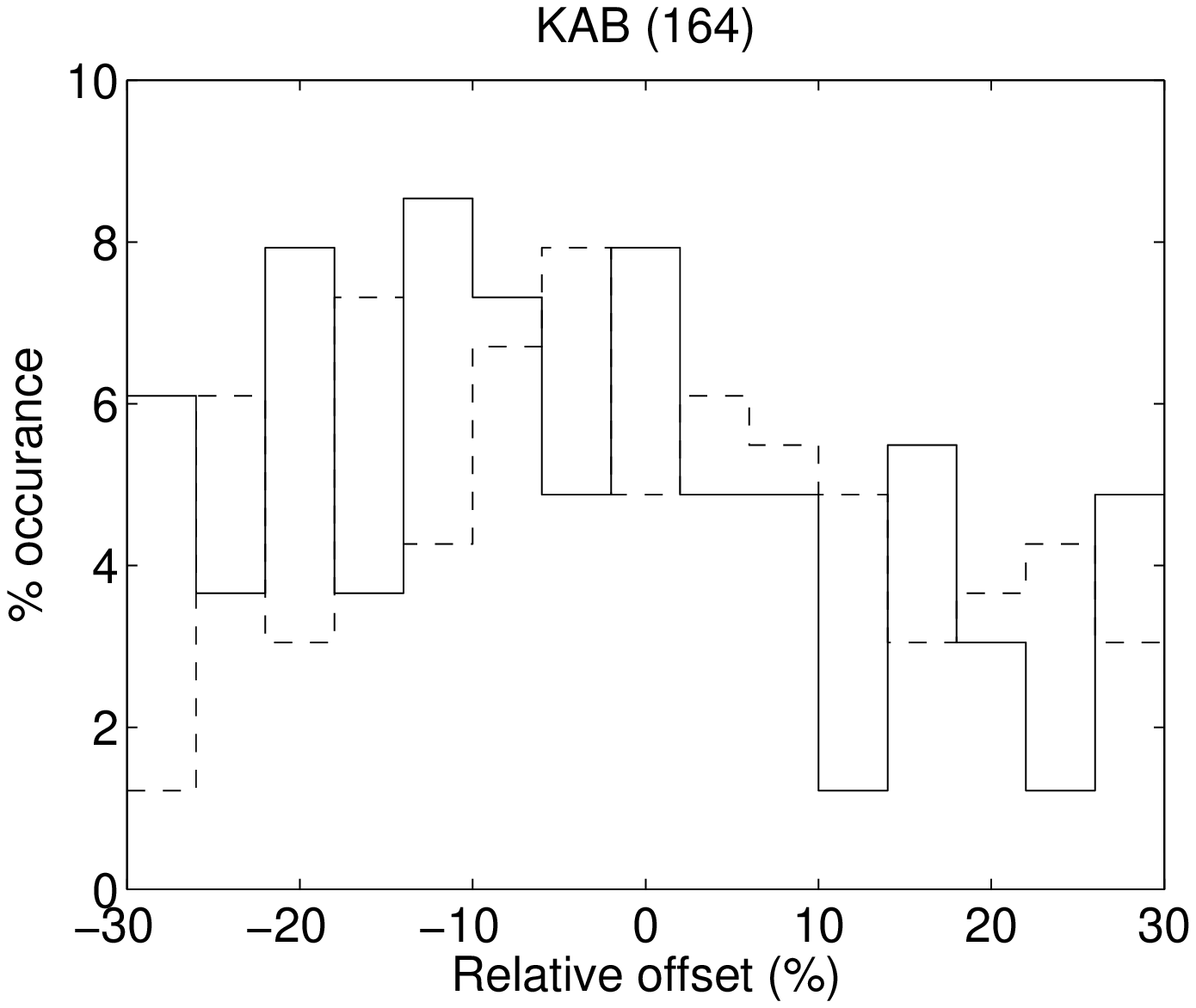}}
   \hfill
   \vl{\includegraphics[width=\dimen1]{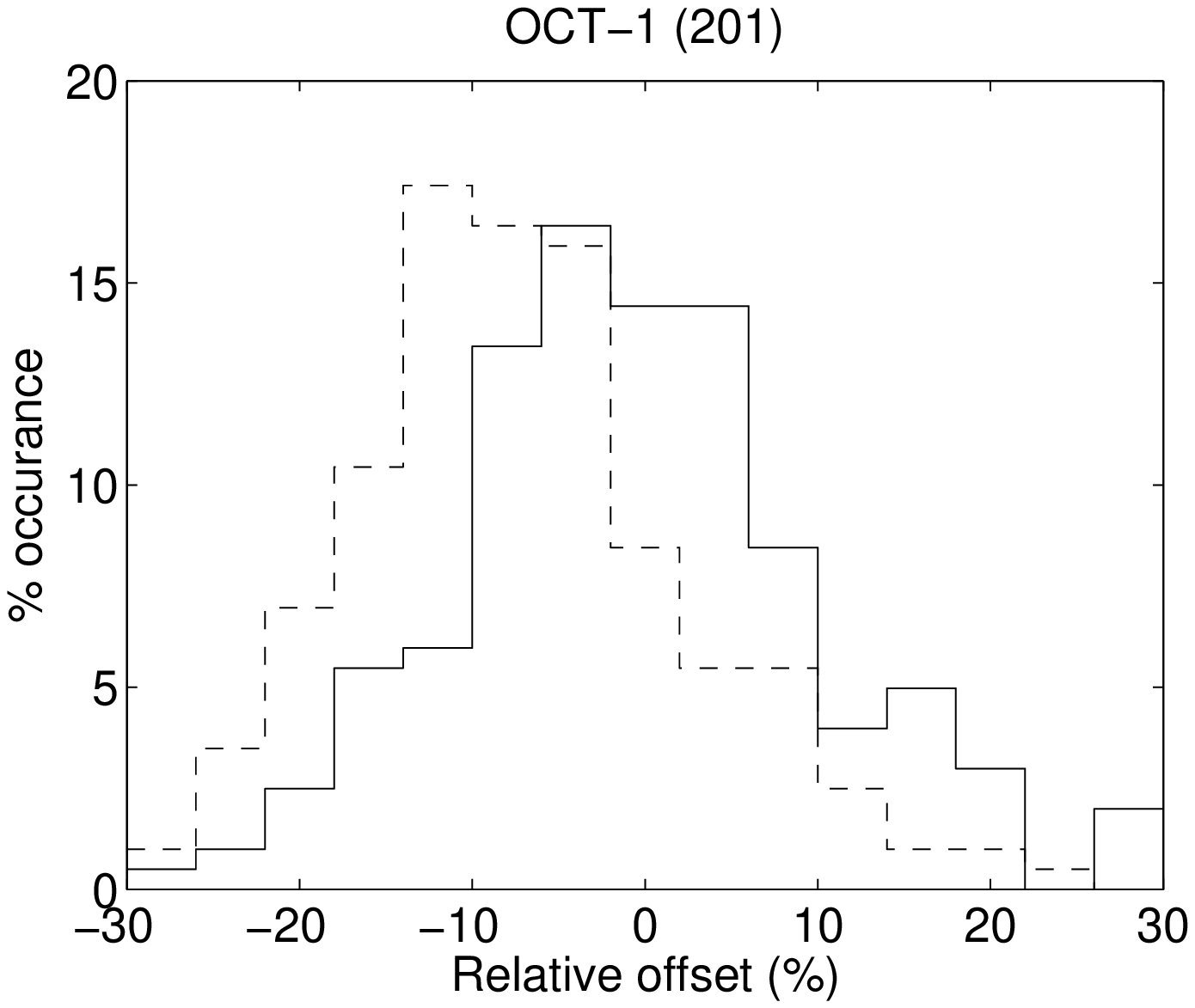}}
}
\medskip
\centerline{%
   \vl{\includegraphics[width=\dimen1]{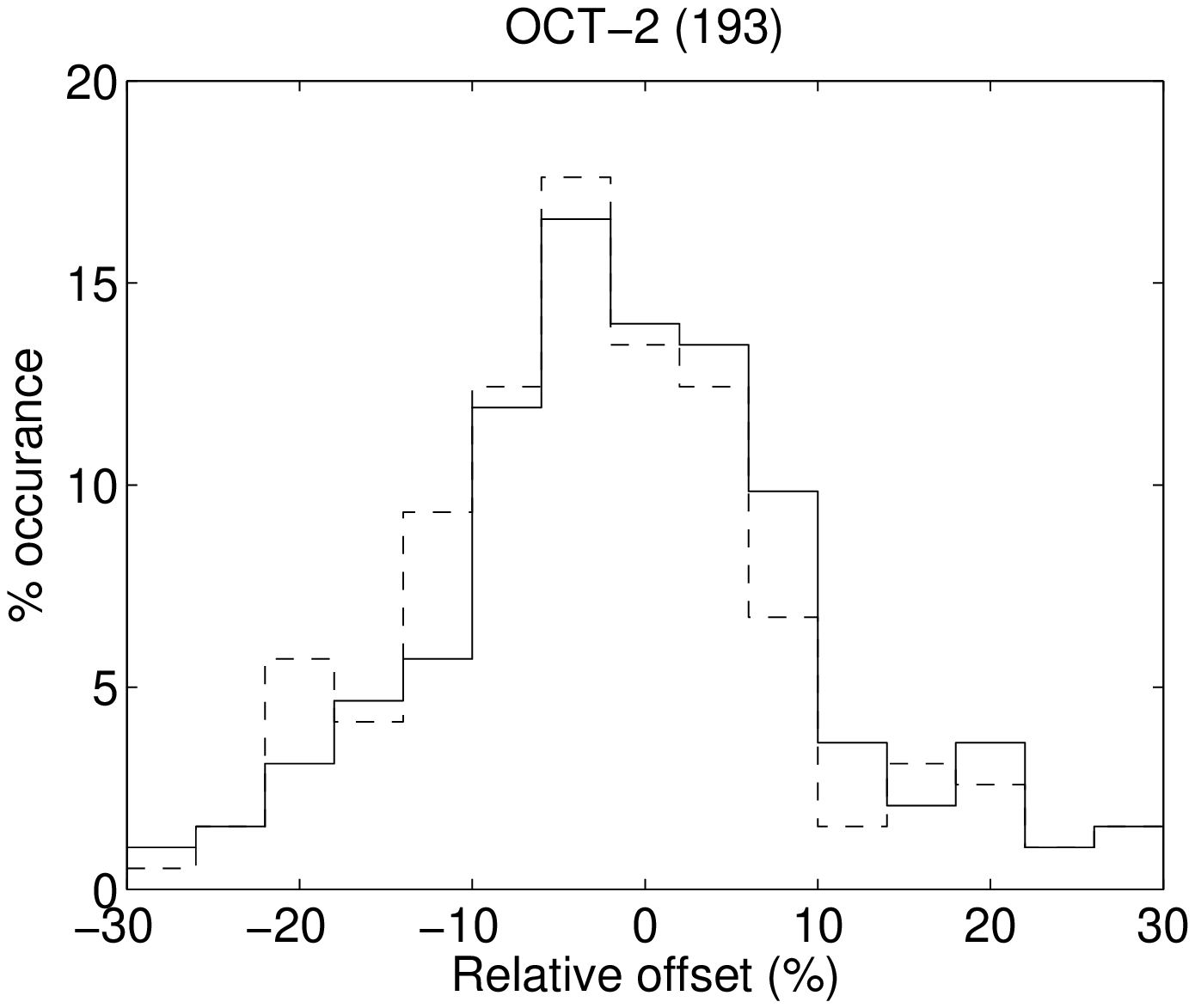}}
   \hfill
   \vl{\includegraphics[width=\dimen1]{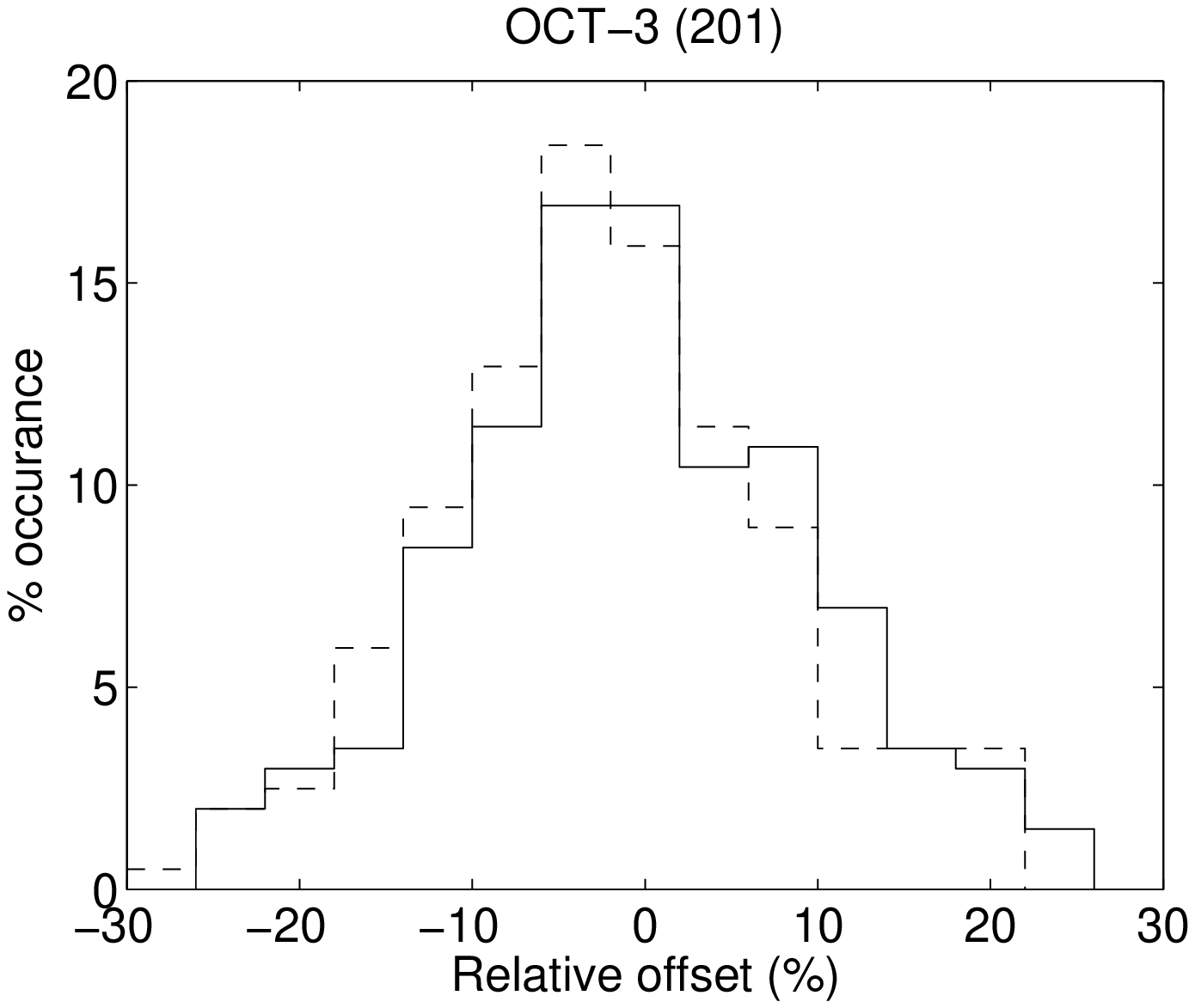}}
   \hfill
   \vl{\includegraphics[width=\dimen1]{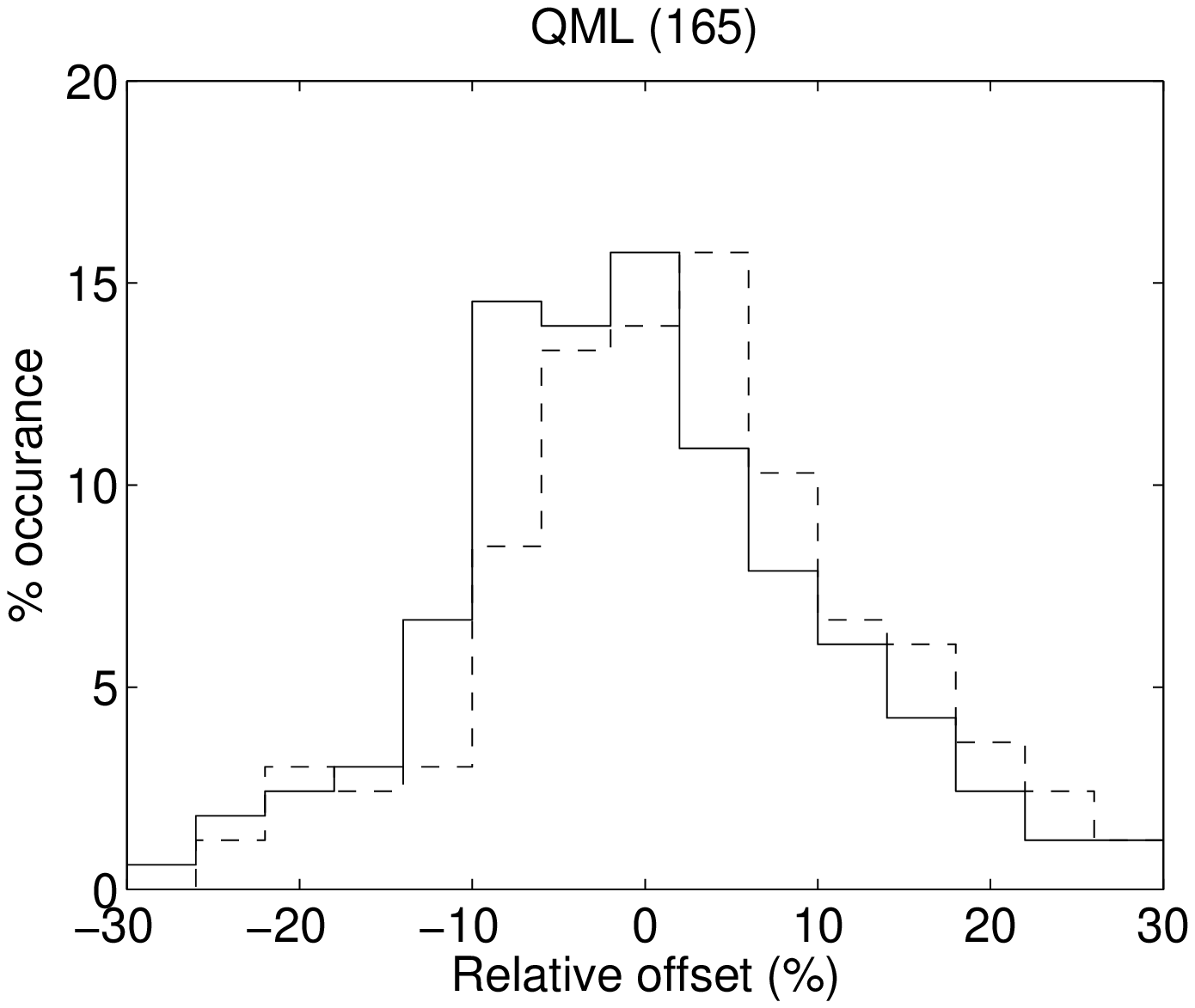}}
   \hfill
   \vl{\includegraphics[width=\dimen1]{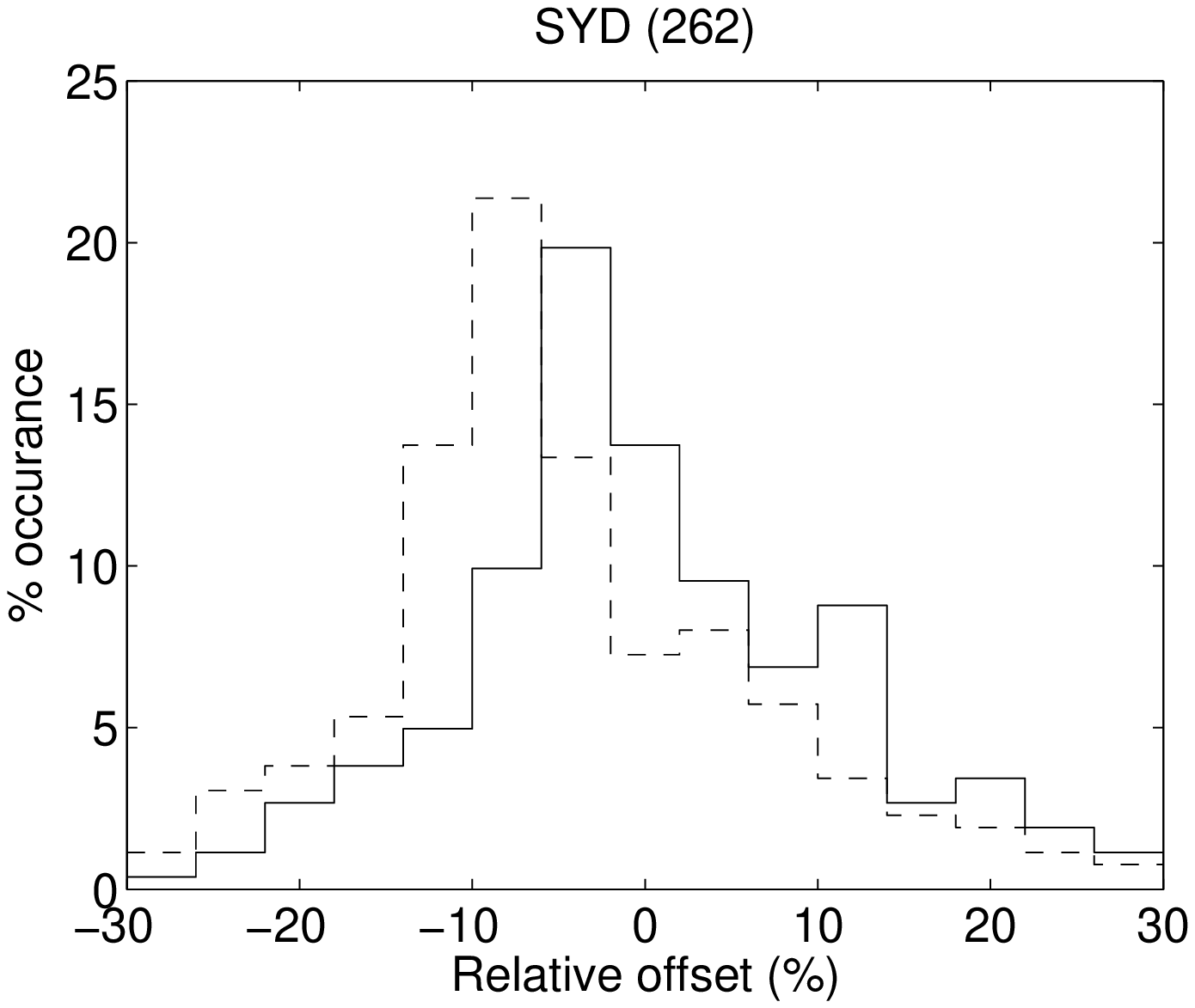}}
}
\caption{Histograms showing the distributions of the relative deviation of $A_\mathrm{max}$ from the correct value for each method determined from the analysis of simulated asteroseismic data.  The method names and number of artificial stars included are given in the titles.  Dotted lines show the distributions based on the $A_\mathrm{max}$ values returned by each method, solid lines show the distributions after a calibration scaling to match the true values.}
\label{fig:amaxcorr}
\end{figure*}

The results obtained for $A_\mathrm{max}$ showed a much wider spread than for $\Delta\nu$ and $\nu_\mathrm{max}$.  The relative number of outliers was slightly lower than for the other parameters due to the increase in the measured $1$\,$\sigma$ spread of values, with an average of $3$\,\% and varying between $1$\,\% and $6$\,\%.  The outlier-removed mean standard deviation over all methods was $14.4$\,\%, and varied between $9.7$\,\% and $15.6$\,\%.

By applying the t-test to each distribution, the centroids of only one distribution (out of eight) was sufficiently close to zero to be consistent with the zero-mean hypothesis at the $95$\,\% level.  The other distributions had mean values between $-15$\,\% and $2$\,\%, which indicated an underestimation bias in all cases except one.  Despite the presence of a bias, all of the methods showed a strong correlation between the values of $A_\mathrm{max}$ determined and the simulation input values, with correlation coefficients between $0.92$ and $0.96$.  Fitting a linear relation to the variation of the determined $A_\mathrm{max}$ as a function of the input values suggested that a simple scaling is sufficient to calibrate the amplitudes determined by each method.

Determining the maximum mode amplitude in an automated way requires assumptions to be made about the relative contribution of each mode degree to the amplitude per radial order (\textit{e.g.} the $c$ parameter in \citet{kjeldsen2008b}), the amount of smoothing applied to the power spectrum (a wider smoothing window can reduce the amplitude of the $p$-mode power excess) and also an appropriate contribution to model the stellar background (e.g. \citealt{harvey1985,karoff2010b}).  As each method addressed these problems in a slightly different way, the amplitudes they obtained were scaled differently from each other.  To calculate a calibration factor for each method, a linear regression was performed on the $A_\mathrm{max}$ values from each team against the true values for the simulated data.  The $A_\mathrm{max}$ values were then scaled by this calibration factor before any further analysis.  The relative deviations of the $A_\mathrm{max}$ results from each method before and after the calibration was applied are shown in Fig.~\ref{fig:amaxcorr}.

\subsubsection{$\delta\nu_{02}$}

The values of the $\delta\nu_{02}$ small separation were only determined by two of the methods (COR, QML).  Repeating the same outlier detection procedure applied to the other parameters indicated that the relative number of outliers was $17$\,\% and $8$\,\% for these two methods.  The average $1$\,$\sigma$ spread in values of $\delta\nu_{02}$ was the largest of all the parameters at $26$\,\%, reflecting the difficulty with which one can precisely determine this small spacing in an automated way from data of just one-month duration.  Despite the wide distributions, applying a t-test found that both methods had distribution centroids that were consistent with the zero-mean hypothesis at the 95\,\% level, indicating that no bias was present.

\subsubsection{Formal uncertainties}

The formal uncertainties of each parameter from each method were tested for consistency against the known deviations from true values using the $\chi^2$ test.  In many cases the formal uncertainties were found to be inconsistent with the accuracy of the results.  To obtain a $\chi^2$ per degree of freedom of unity for the parameters from each method would require scaling the formal uncertainties by a significant amount.  Despite this inconsistency, the relative formal uncertainties were correlated with the relative absolute deviations of the results from their true values, particularly in the case of $\Delta\nu$ (with a correlation coefficient of 0.30) and $\nu_\mathrm{max}$ (0.29).  The formal uncertainties and relative deviations of $A_\mathrm{max}$ were less strongly correlated (0.20), however all of these correlations are sufficient to reject the null hypothesis (\textit{i.e.} random points with zero correlation) at the 0.1\,\% level.  In the case of $\delta\nu_{02}$, the stated uncertainties and deviations were not significantly correlated, however the sample size was relatively small because only two of the methods returned results on $\delta\nu_{02}$ for the simulated stars.

\subsection{Application to {\em Kepler} data}

\begin{figure*}
\dimen0=\hsize
\dimen1=2mm
\advance\dimen0 by -3\dimen1
\dimen1=0.25\dimen0
\centerline{%
   \vl{\includegraphics[width=\dimen1]{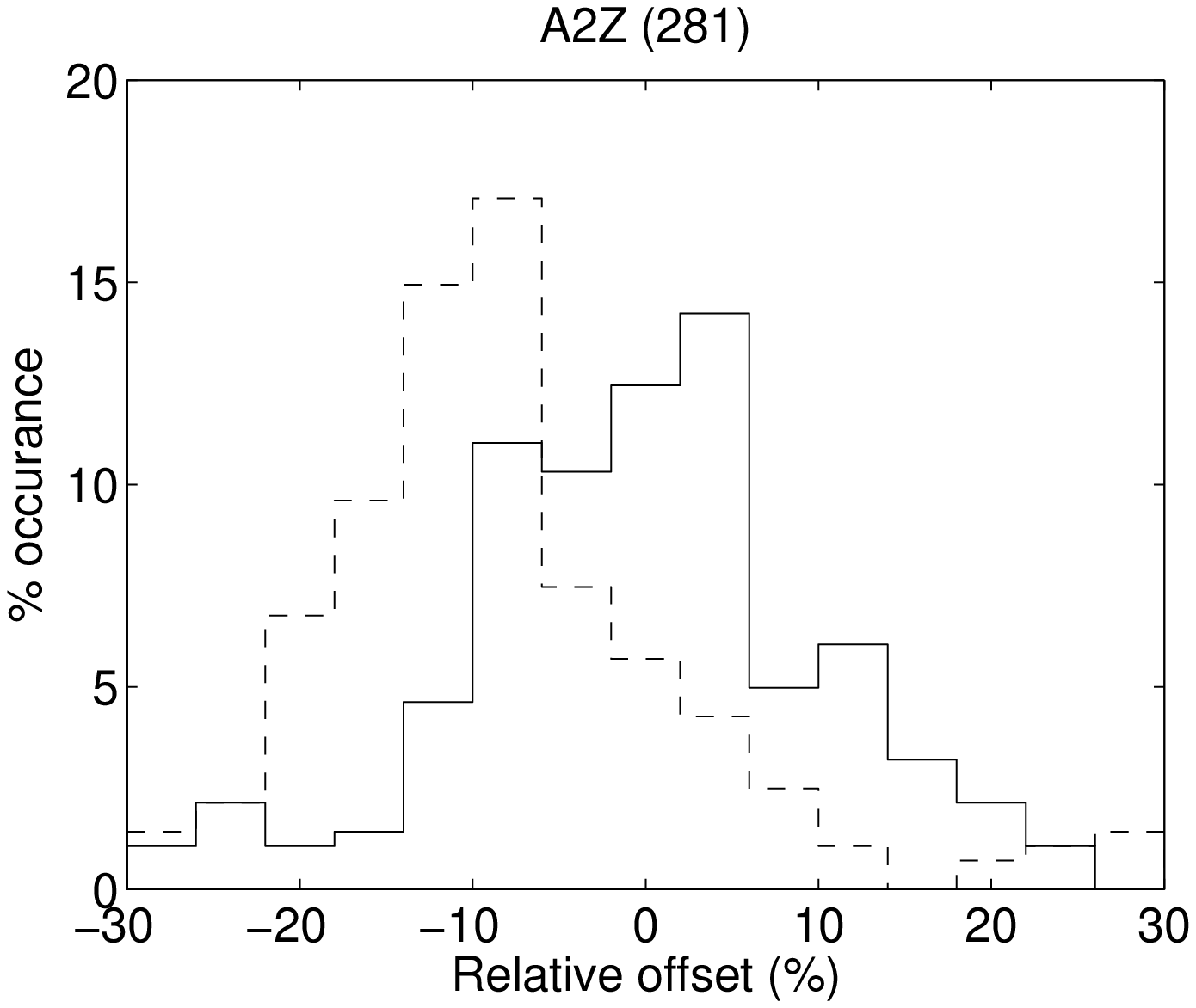}}
   \hfill
   \vl{\includegraphics[width=\dimen1]{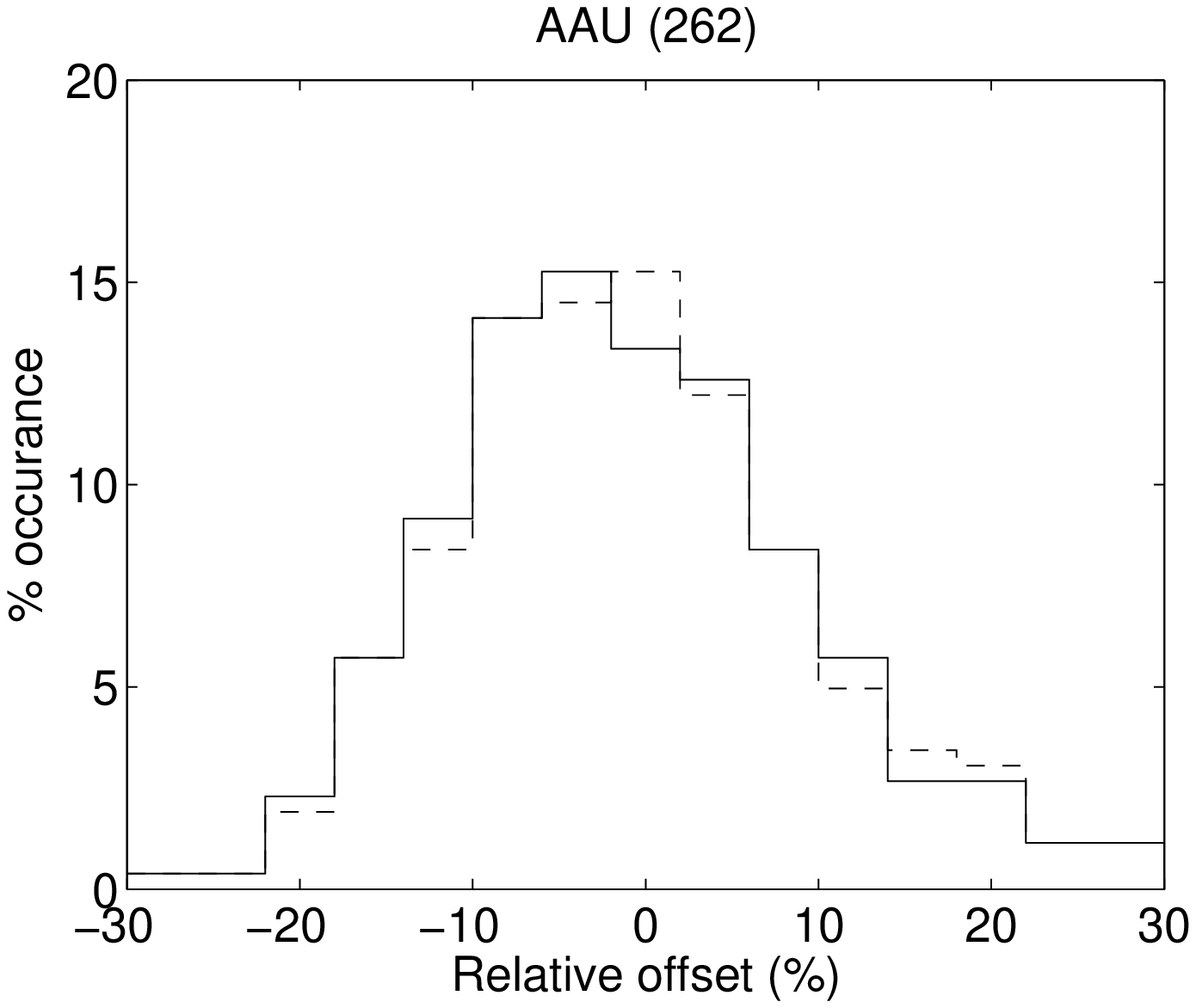}}
   \hfill
   \vl{\includegraphics[width=\dimen1]{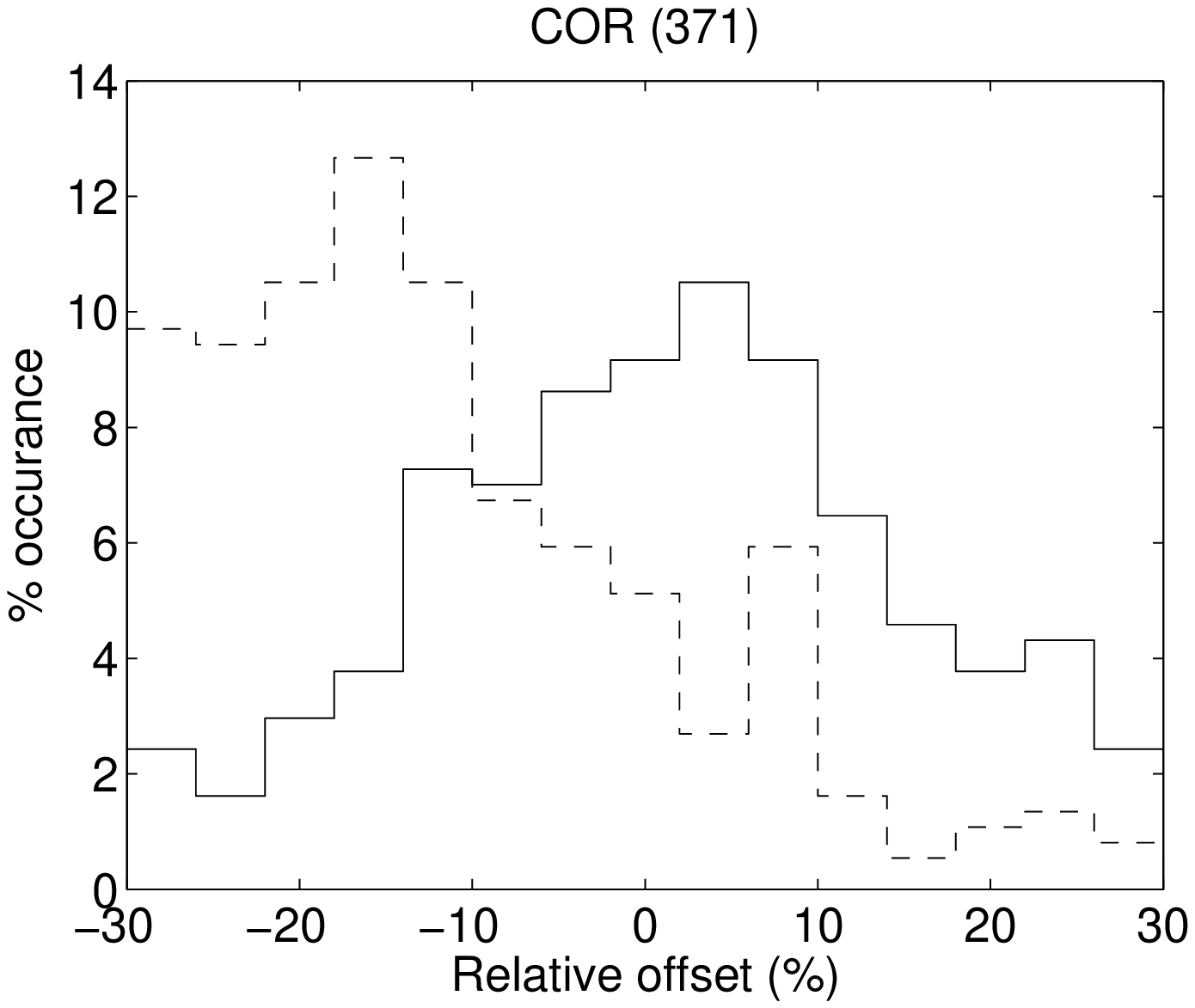}}
   \hfill
   \vl{\includegraphics[width=\dimen1]{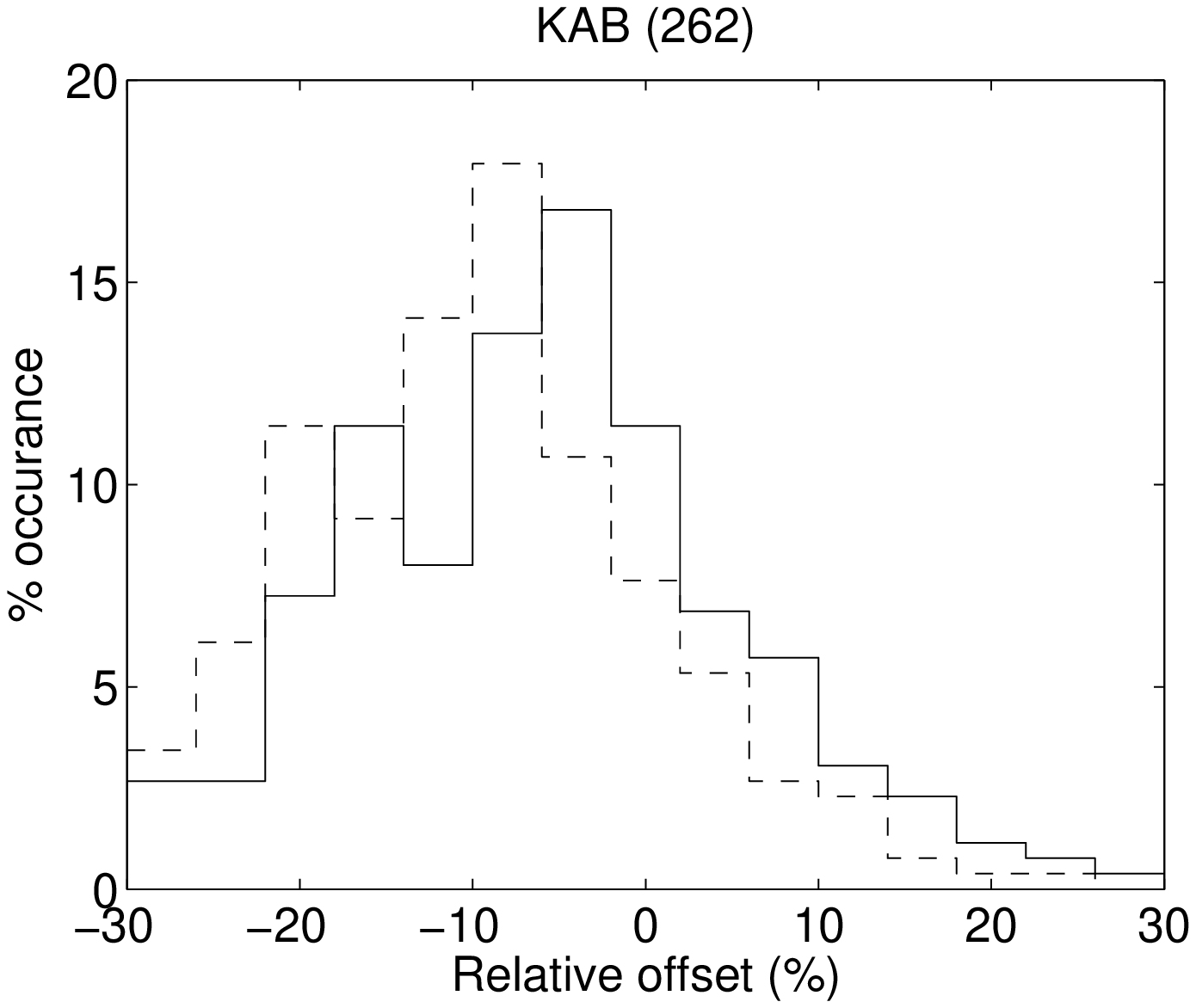}}
}
\medskip
\centerline{%
   \vl{\includegraphics[width=\dimen1]{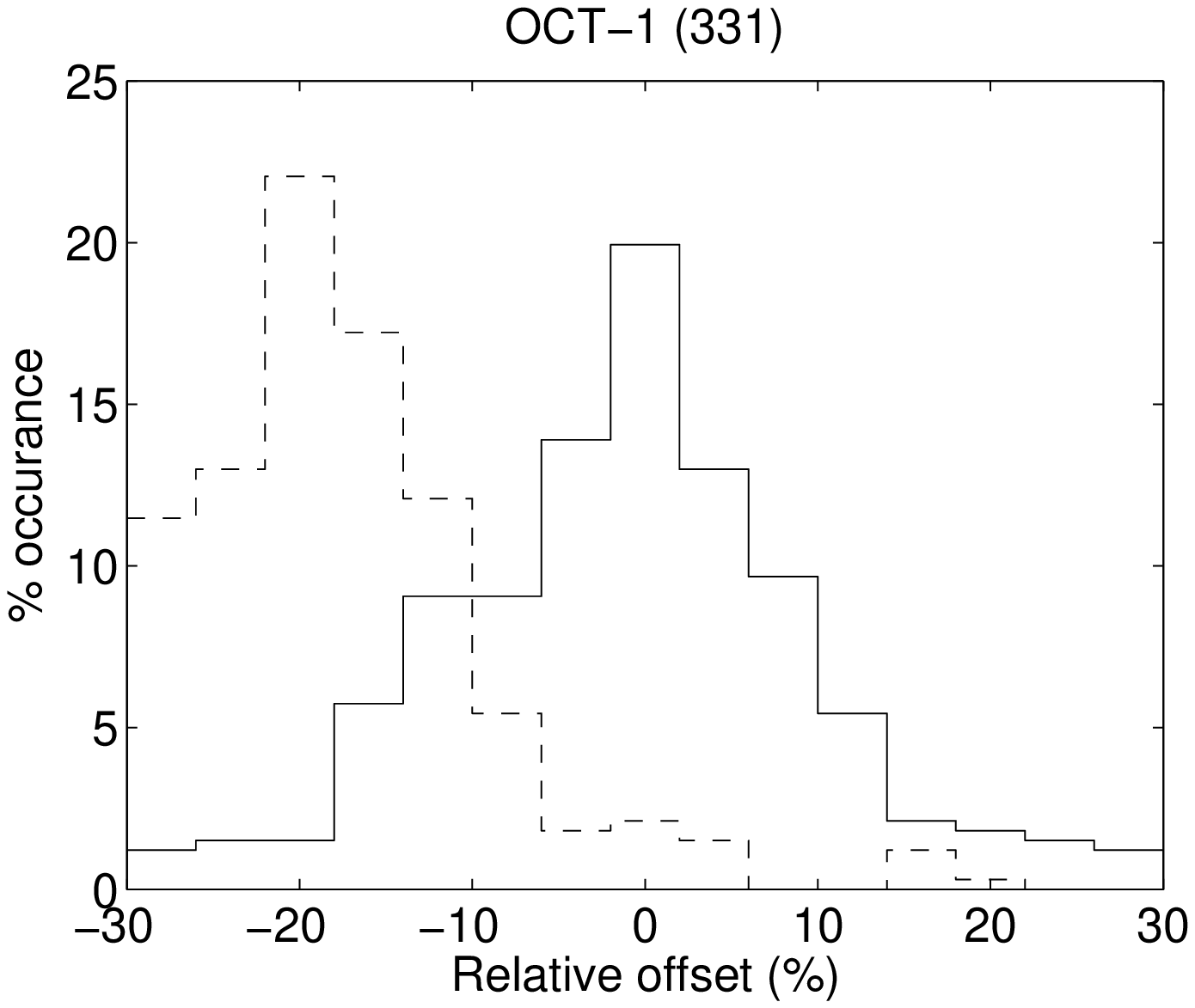}}
   \hfill
   \vl{\includegraphics[width=\dimen1]{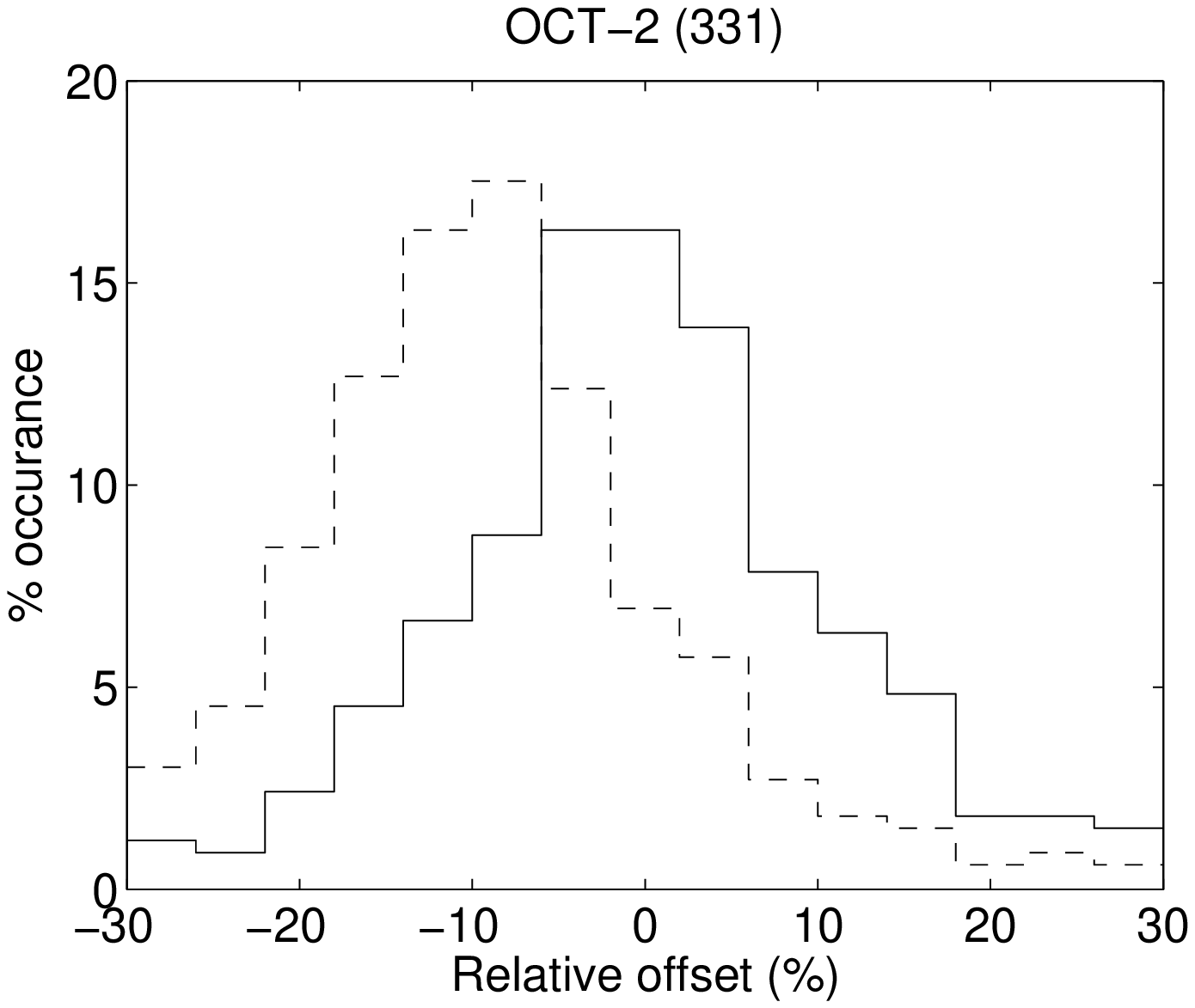}}
   \hfill
   \vl{\includegraphics[width=\dimen1]{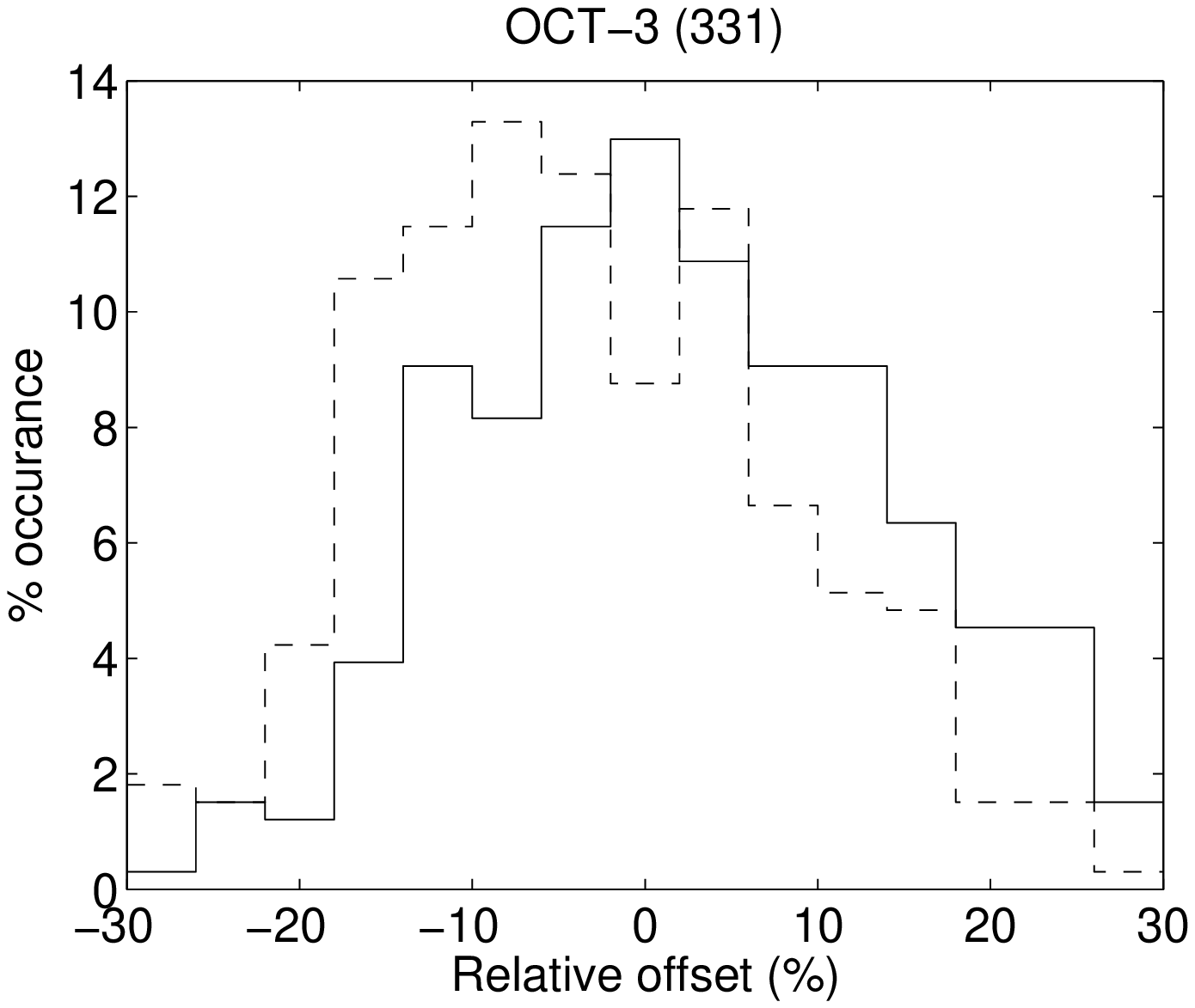}}
   \hfill
   \vl{\includegraphics[width=\dimen1]{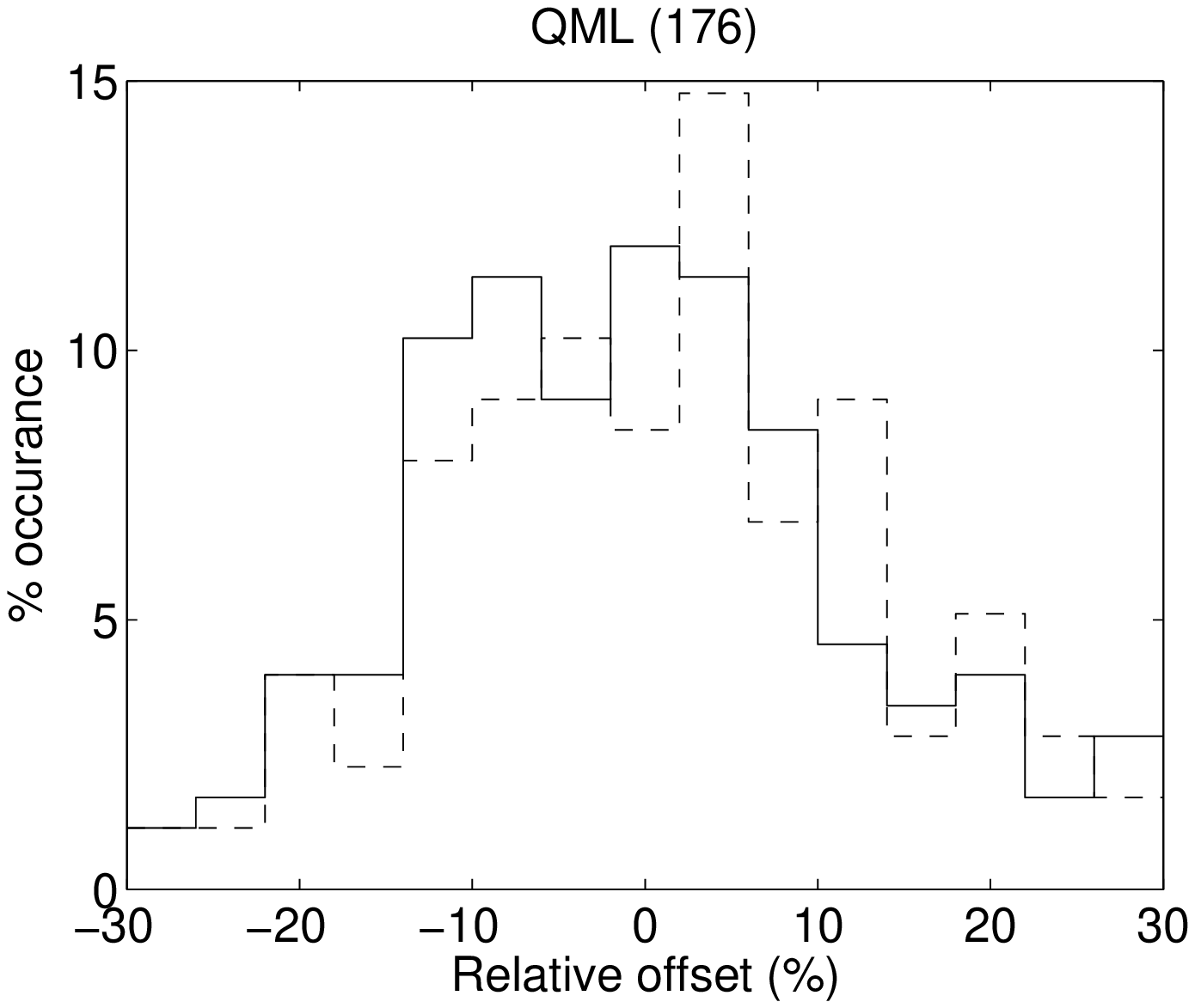}}
}
\caption{Histograms showing the distributions of the relative deviation of $A_\mathrm{max}$ from that given by the SYD-1 method for {\em Kepler} data.  The method names and number of stars included are given in the titles.  Dotted lines show the distributions before calibration by peak-bagging results, solid lines show the distributions using the corrected amplitudes (including those by SYD-1).}
\label{fig:amaxcorr_real}
\end{figure*}

The prepared {\em Kepler} light curves were analysed using the methods of all nine teams (see Table~\ref{tab:methods}).  The methods did not all detect oscillations on the same targets so there was some non-overlap in the sets of stars for which oscillations were detected by each method.  In order to maximise the number of stars for which we obtained verified global asteroseismic parameters, it was desirable to combine the results from all methods rather than ranking them in some way and selecting only the best.  The methods used previously to analyse the results from simulated data required knowing the correct parameters {\em a priori}.  In the case of real data we constructed alternative methods to detect outliers, calibrate the maximum mode amplitudes and test the consistency of formal errors.

\subsubsection{Outlier rejection}

The results from the simulated data showed that for all of the parameters, there were significantly more outliers at $5$\,$\sigma$ than might be expected assuming a normal error distribution.  This is an indication that the error distribution is not solely random and has a dependence on the varying signal-to-noise ratio of the stars present.  It also highlights the limitations of the automated analysis methods in cases where the signal-to-noise ratio of the data is low.

As a verification procedure, we required that for each parameter determined for each star the results from at least two independent teams were contained within a range of fixed relative size centred on the median value.  Results outside of this range were iteratively removed until either all results were in agreement or fewer than two results remained.  The range needed to be small enough to reject outliers but not so small that results were removed due to the inherent scatter in the data.

The range size for each parameter was defined using the results from the simulated data.  We used the smallest range such that the outliers identified at the $5$\,$\sigma$ levels described previously would also be detected using the fixed-range method.  The relative ranges that we calculated were $\pm3.5$\,\% ($\Delta\nu$), $\pm10.5$\,\% ($\nu_\mathrm{max}$) and $\pm21.5$\,\% ($A_\mathrm{max}$ and $\delta\nu_{02}$).

\subsubsection{$A_\mathrm{max}$ calibration}

To identify any significant biases in the asteroseismic parameters derived from {\em Kepler} data, we compared the results of each method to those of a reference method (SYD-1).  As was the case for the simulations, the parameter with problematic biases was $A_\mathrm{max}$.  In order to calibrate the amplitudes of each method, the maximum mode amplitudes of 22 high signal-to-noise stars within the cohort were determined using an established peak-bagging technique and used as a reference set to obtain $A_\mathrm{max}$ correction factors for each method.  The peak-bagging technique was tested using a subset of 48 simulated stars with high signal-to-noise ratios and found to give an unbiased estimate of $A_\mathrm{max}$ to a relative precision of 6.2\,\%.  The relative differences between the $A_\mathrm{max}$ values for {\em Kepler} data determined by each method and that given by the reference method before and after calibration by the peak-bagging results are shown in Fig.~\ref{fig:amaxcorr_real}.  This illustrates the improved agreement in the values determined for $A_\mathrm{max}$ by each method after calibration.

\subsubsection{Formal uncertainty correction}

\begin{figure*}
\dimen0=\hsize
\dimen1=2mm
\advance\dimen0 by -2\dimen1
\dimen1=0.3333333\dimen0
\centerline{%
   \vl{\includegraphics[width=\dimen1]{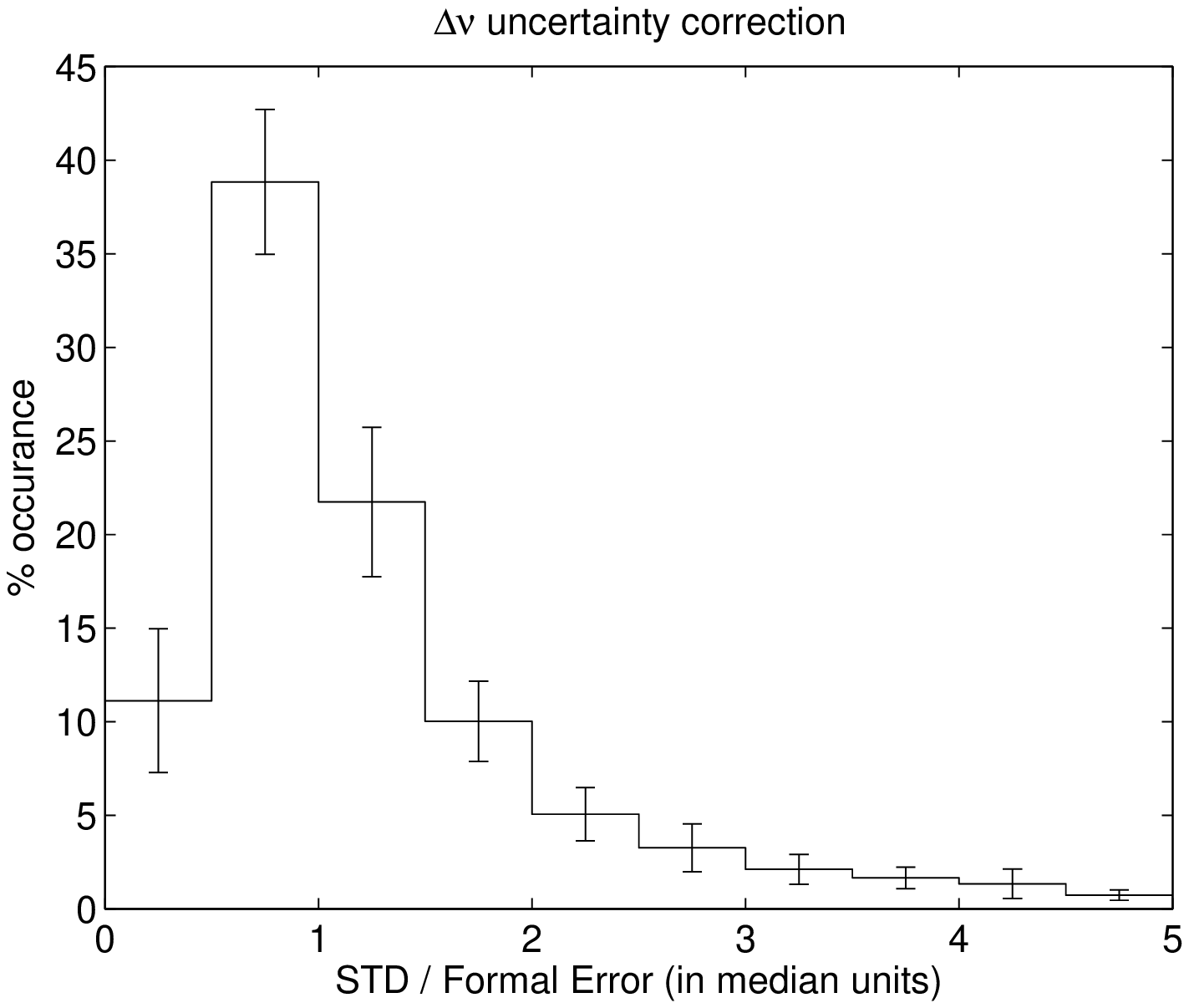}}%
   \hfill
   \vl{\includegraphics[width=\dimen1]{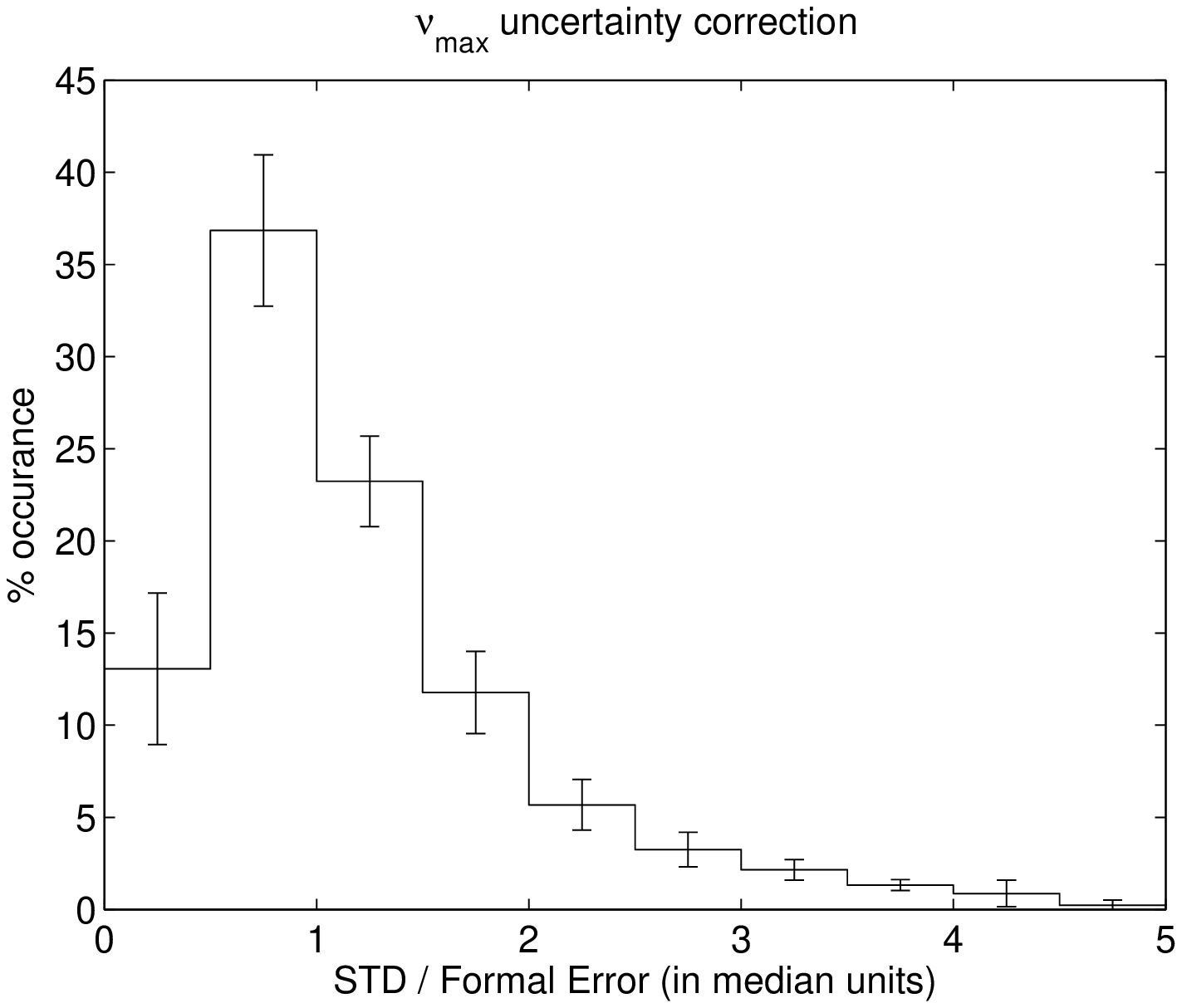}}%
   \hfill
   \vl{\includegraphics[width=\dimen1]{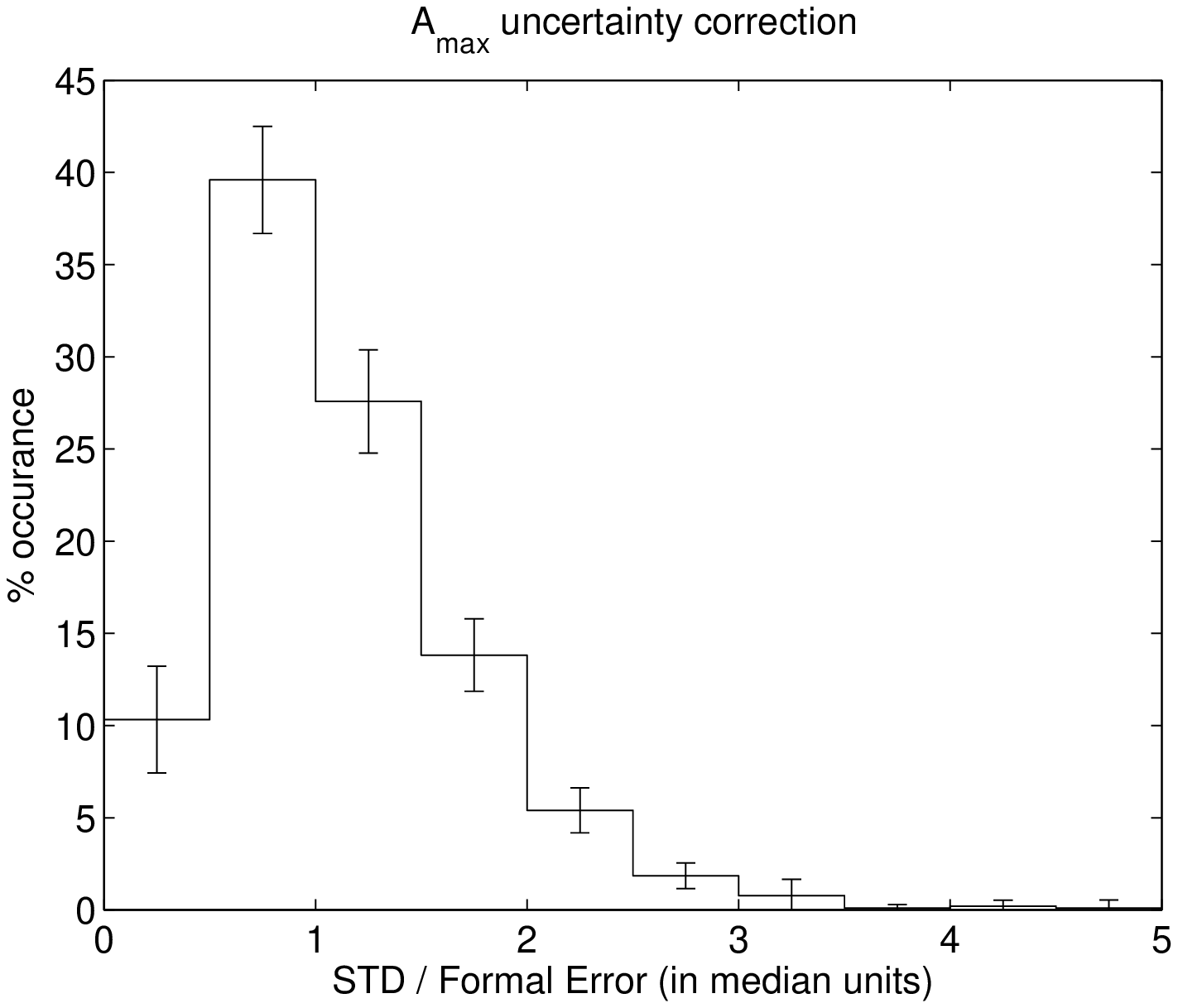}}%
}
\caption{Distributions of uncertainty-correction factors (\textit{i.e.} standard deviation divided by formal uncertainty) for $\Delta\nu$, $\nu_\mathrm{max}$ and $A_\mathrm{max}$.  The correction factors for each method have been scaled by their median values (Table~\ref{tab:medvals}) and combined in the histograms for each parameter.  The error bars show the standard deviations in each bin when determined using individual methods.}
\label{fig:errmult_all}
\end{figure*}

In the case of real data, it was not possible to calculate uncertainty-scaling factors using the $\chi^2$ test without knowing the correct values {\em a priori} and so a method was devised to estimate these using the scatter in the results obtained for the same stars from multiple teams.  For those stars where at least five teams gave results that were in agreement within the thresholds defined above, the standard deviation of these results was used as an estimate of the expected uncertainty.  A scaling factor was then determined from the ratio of this standard deviation to the formal uncertainties stated by each team.  Single average scaling factors for each method were then calculated as the median of the individual scaling factors.  The distributions of the uncertainty-scaling factors are shown in Fig.~\ref{fig:errmult_all} and the median values for each method are given in Table~\ref{tab:medvals}.  The distributions in Fig.~\ref{fig:errmult_all} are encouraging and show that once scaled by a single scaling factor, the uncertainties of each method agree well with the spread observed between results obtained from multiple methods.

The error correction procedure ensured that the data were self-consistent within stated uncertainties, however there is an additional internal uncertainty which must be included.  For a given star we expect the scatter in the results obtained from multiple teams to be less than a true measure of the uncertainty because all of the methods were analysing the same data.  If we use the scatter by itself to calculate the correction factors then they will be underestimated.  By estimating the uncertainty-scaling factors using the scatter method on the simulated data and then applying the $\chi^2$ test to check the consistency of the scaled uncertainties, we determined the additional internal uncertainty-scaling factors required for $\Delta\nu$ (1.20), $\nu_\mathrm{max}$ (1.19) and $A_\mathrm{max}$ (1.11).

For certain sets of results, the scaling that must be applied to the formal errors to ensure they are consistent with multiple alternative methods is considerable (see Table~\ref{tab:medvals}).  For cases where this scaling is large (\textit{i.e.} greater than four), the reliability of the scaled uncertainties is questionable.  In addition to this, some of the methods did not calculate any formal uncertainties on their results.  For these data, a constant relative uncertainty was assumed to be more appropriate than using the rescaled values or a constant absolute uncertainty.  The size of these indicative uncertainties was determined from the median relative standard deviation of the results of each parameter where at least five teams gave consistent results and the additional internal uncertainty contribution described above.  This gave an indicative relative uncertainty of 1.8\,\% for $\Delta\nu$, 3.8\,\% for $\nu_\mathrm{max}$ and 9.8\,\% for $A_\mathrm{max}$.

Where these results were used in subsequent analysis to derive model coefficients, we repeated the calculations with both scaled and indicative errors and found that the coefficients derived were in agreement at the level of precision available in the data.

\begin{table}
  \centering
  \begin{tabular}{|c|c|c|c|}
  \hline
  Method & $\sigma_{\Delta\nu}$ factor & $\sigma_{\nu_\mathrm{max}}$ factor & $\sigma_{A_\mathrm{max}}$ factor \\
  \hline
  A2Z   & 0.32 & 28.5  & 1.12 \\
  AAU   & 0.94 & 0.47  & 0.75 \\
  COR   & 1.14 & 3.18  & *    \\
  IAS   & 7.55 & 305.0 & *    \\
  KAB-1 & 0.30 & 6.92  & 8.85 \\
  KAB-2 & --   & 2.48  & --   \\
  OCT-1 & 2.03 & 1.09  & 0.74 \\
  OCT-2 & --   & --    & 1.07 \\
  OCT-3 & --   & --    & 0.89 \\
  ORK   & *    & *     & --   \\
  QML-1 & 0.64 & 0.89  & 0.22 \\
  QML-2 & 0.45 & --    & --   \\
  SYD-1 & 0.93 & 0.86  & 1.18 \\
  SYD-2 & --   & 0.77  & 0.88 \\
  \hline
\end{tabular}
\caption{Median uncertainty-correction factors for each method.  Values greater than unity indicate where formal uncertainties have been underestimated.  Asterisks denote cases where results were provided without formal uncertainties.  For methods where a scaling factor greater than four is required, we have assumed these uncertainties to be unreliable.}
\label{tab:medvals}
\end{table}

\section{Distribution of parameters and detectability}

\begin{figure}
\dimen0=\hsize
\dimen1=2mm
\advance\dimen0 by -1\dimen1
\dimen1=0.5\dimen0
\centerline{%
  \vl{\includegraphics[width=\dimen1]{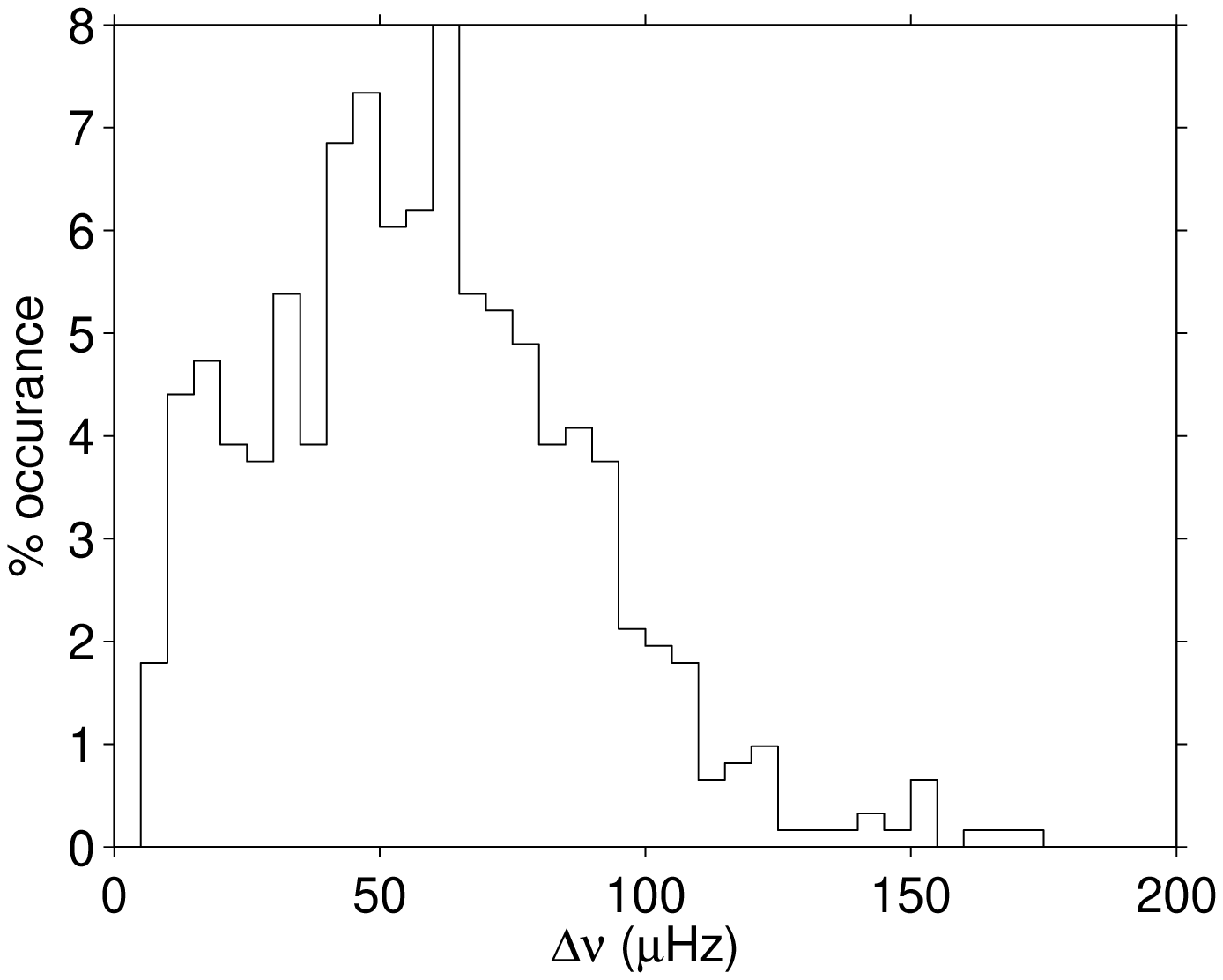}}
  \hfill
  \vl{\includegraphics[width=\dimen1]{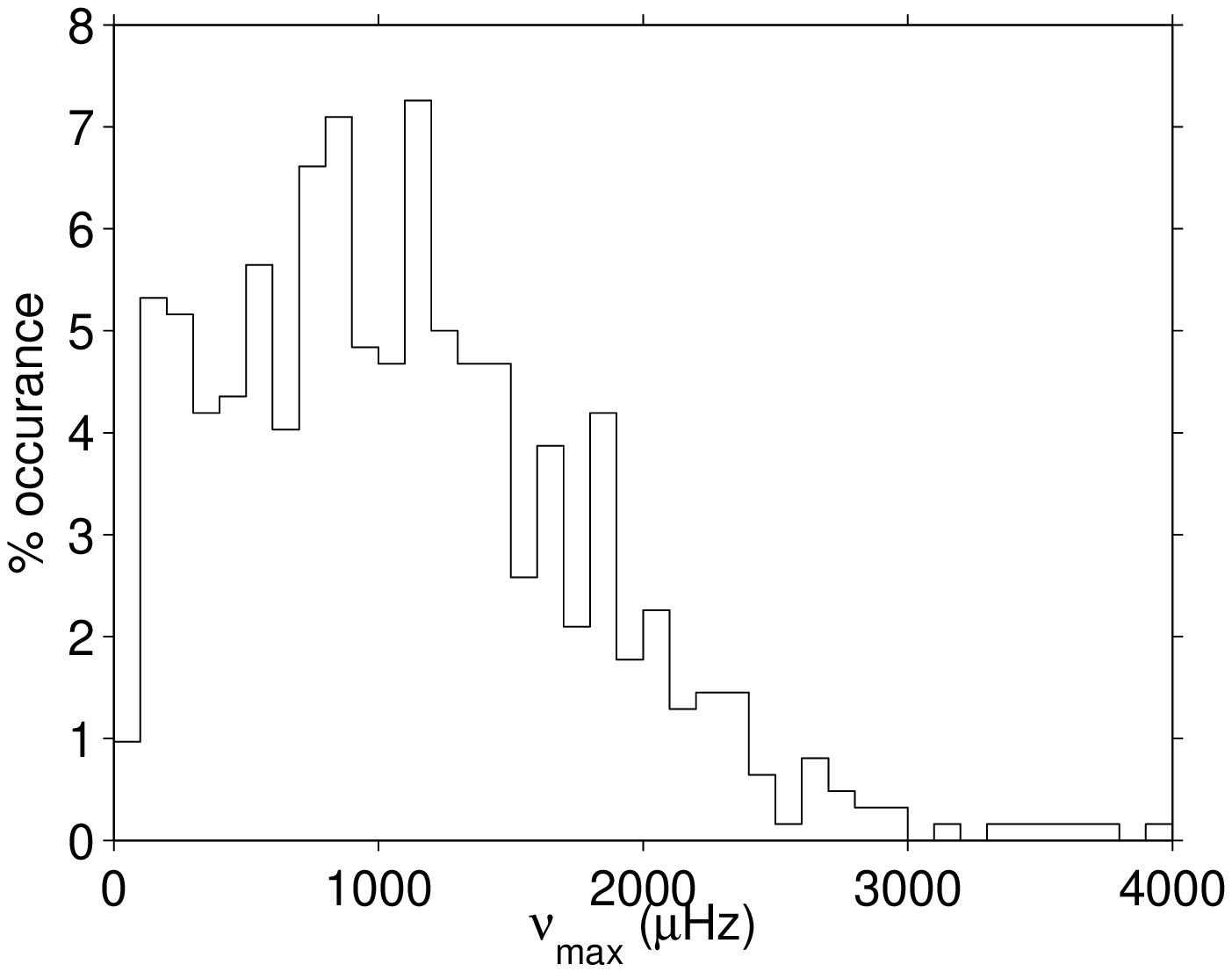}}
}
\medskip
\centerline{%
  \vl{\includegraphics[width=\dimen1]{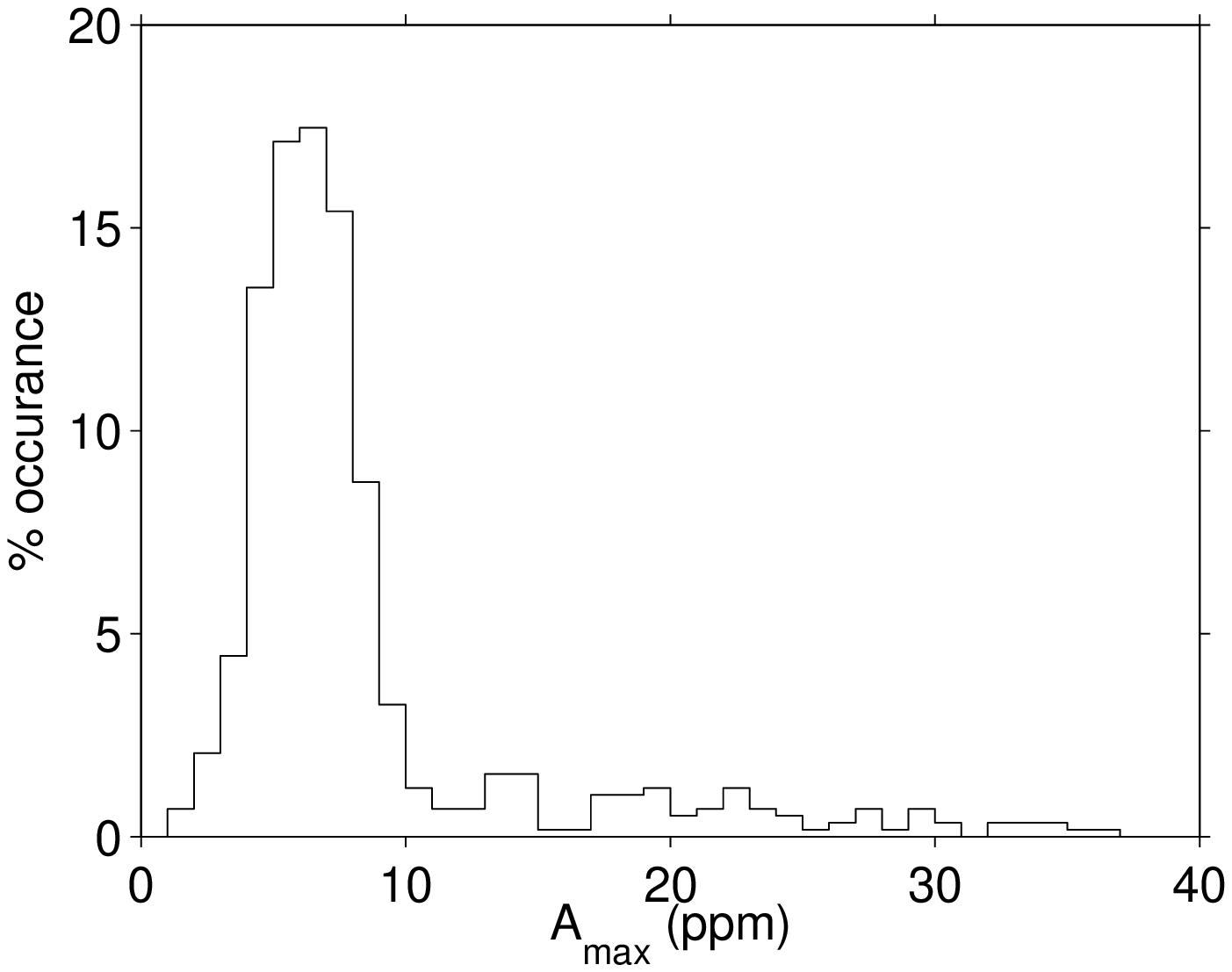}}
  \hfill
  \vl{\includegraphics[width=\dimen1]{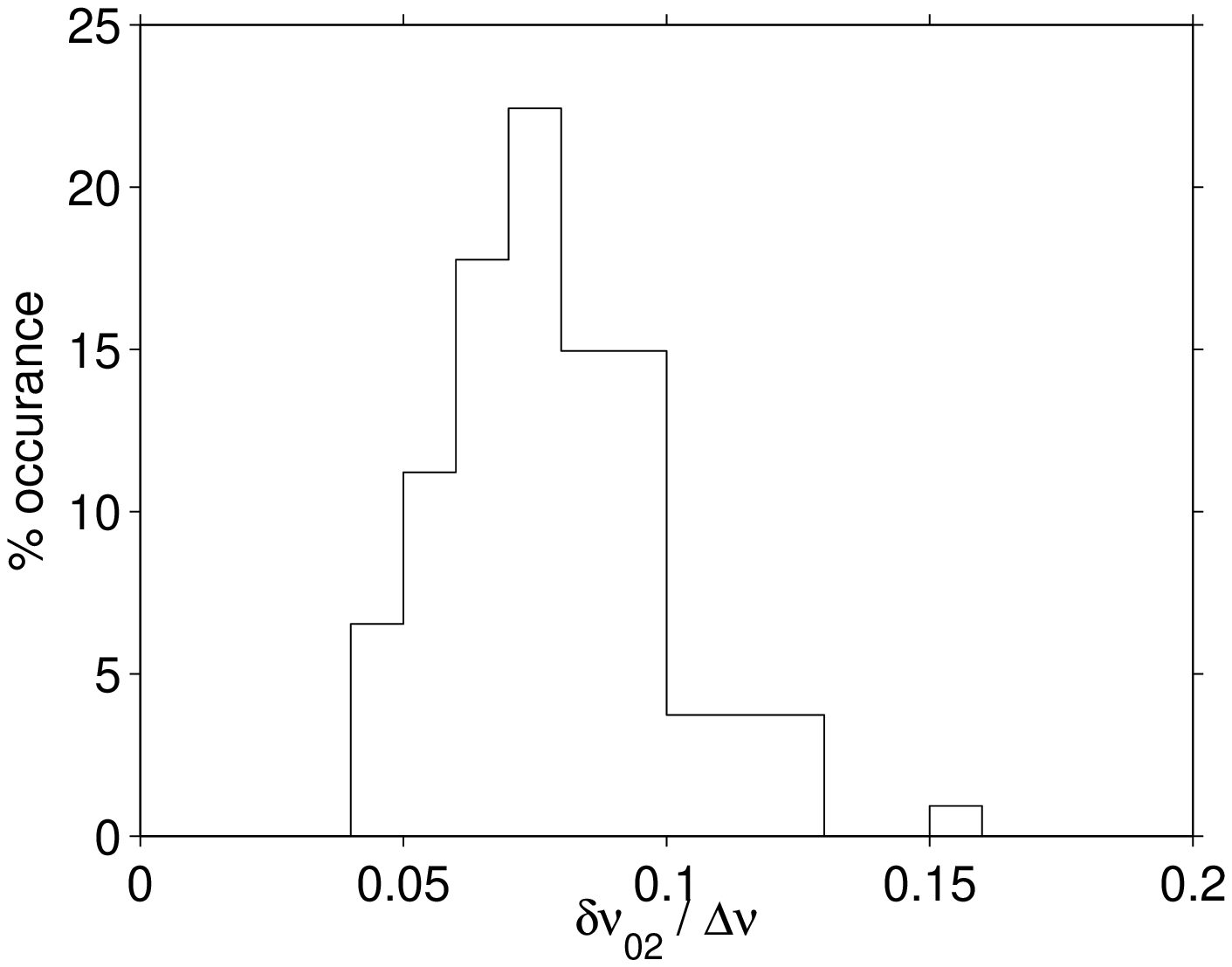}}
}
\caption{Histograms of the average asteroseismic parameters confirmed by at least two methods.  The number of parameters in each range is scaled by the total number to give the relative number in each bin.  The bin widths are 5\,$\mu$Hz ($\Delta\nu$), 100\,$\mu$Hz ($\nu_\mathrm{max}$), 1\,ppm ($A_\mathrm{max}$) and 0.01 ($\delta\nu_{02}/\Delta\nu$).}
\label{fig:realdist}
\end{figure}

The $A_\mathrm{max}$ correction, verification procedure and formal uncertainty scaling was applied to the results of the {\em Kepler} data obtained from nine independent analysis methods.  The results from stars that were observed in multiple one-month runs were averaged to give a single set of results for each star.  Of the 1948 unique stars in the cohort, 642 (33\,\%) were confirmed as solar-like oscillators with their determined asteroseismic parameters consistent with the results of at least two independent teams.  The distributions of the average asteroseismic parameters for each star are shown in Fig.~\ref{fig:realdist}.

\begin{figure*}
\dimen0=\hsize
\dimen1=2mm
\advance\dimen0 by -3\dimen1
\dimen1=0.25\dimen0
\centerline{%
   \vl{\includegraphics[width=\dimen1]{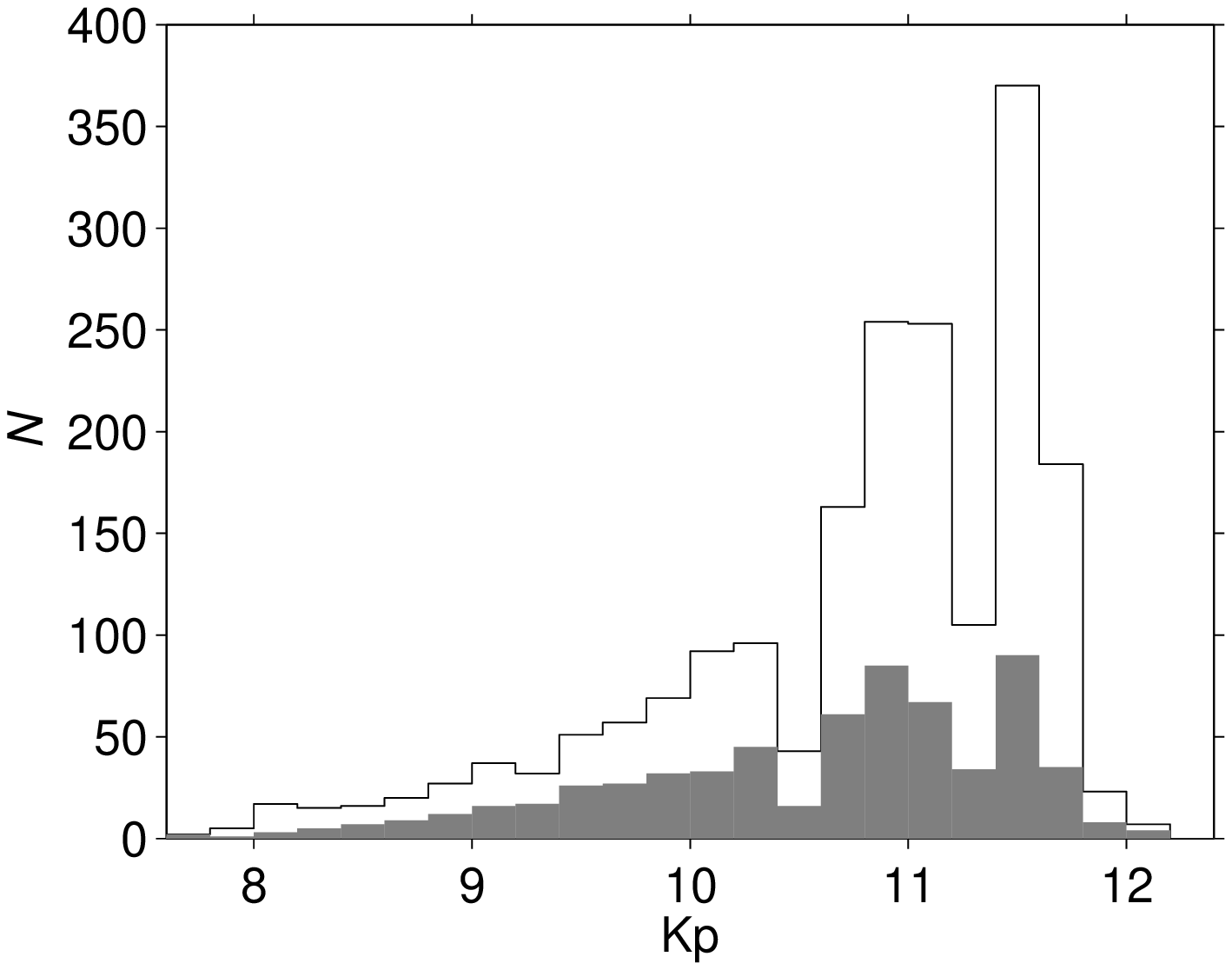}}
   \hfill
   \vl{\includegraphics[width=\dimen1]{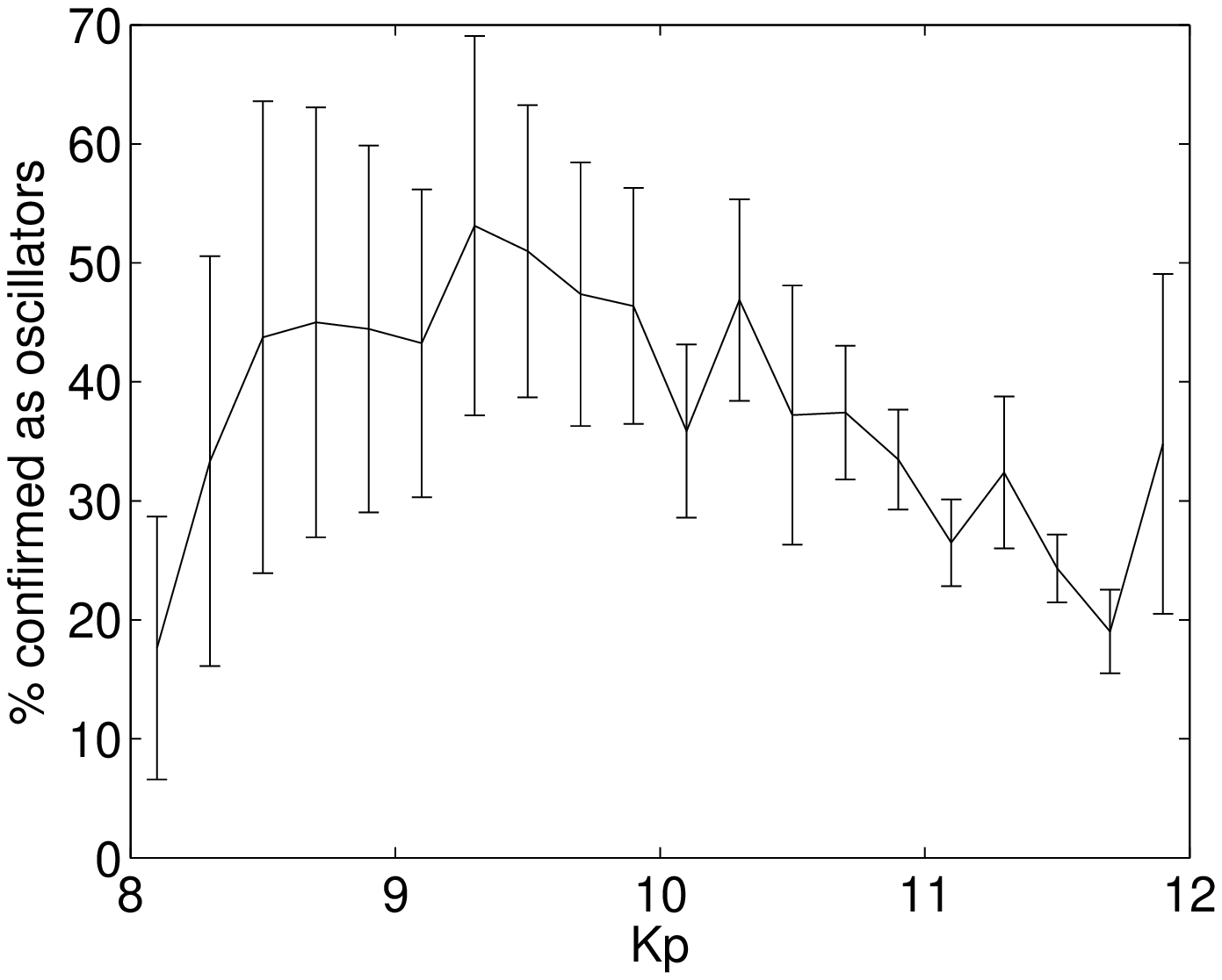}}
   \hfill
   \vl{\includegraphics[width=\dimen1]{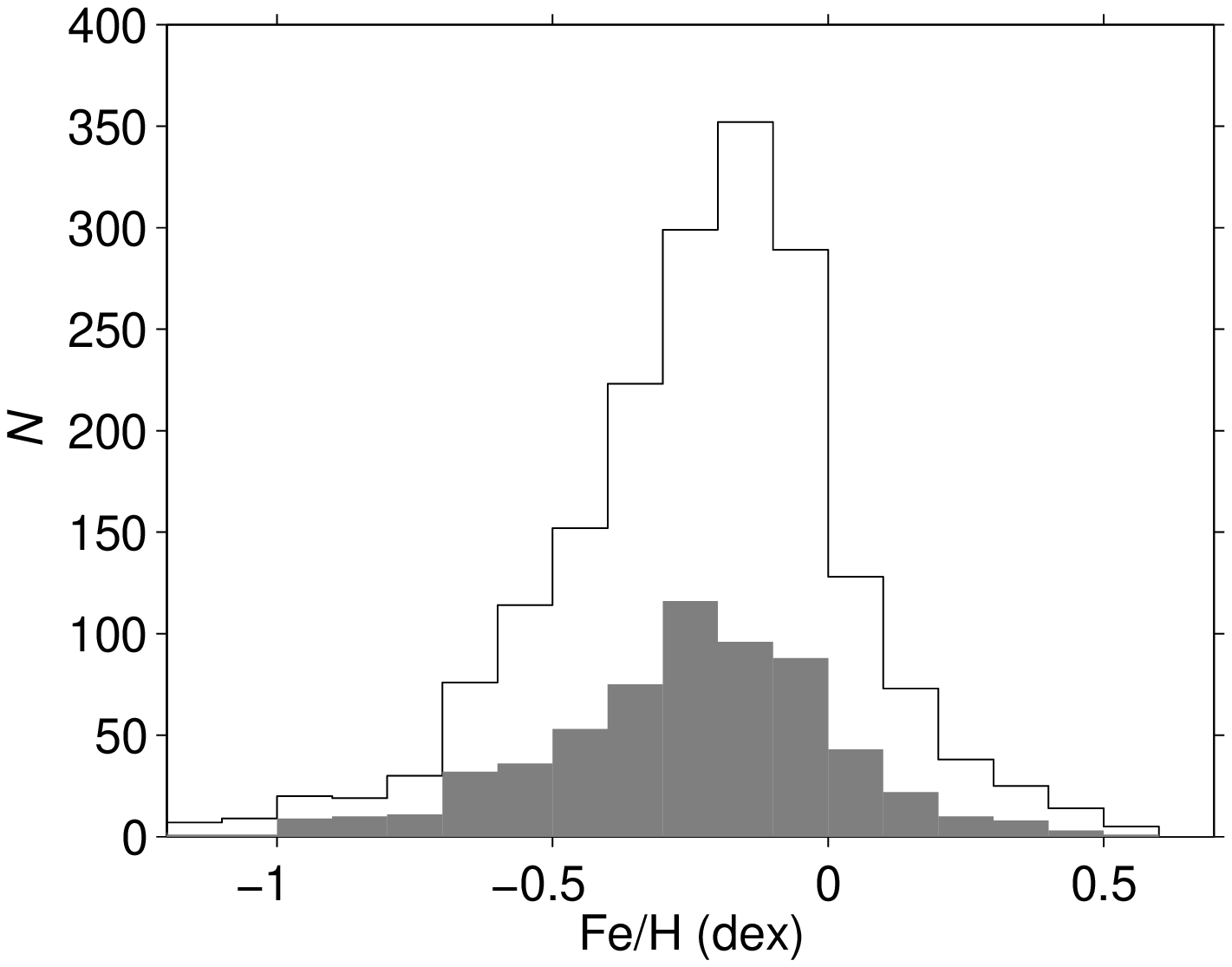}}
   \hfill
   \vl{\includegraphics[width=\dimen1]{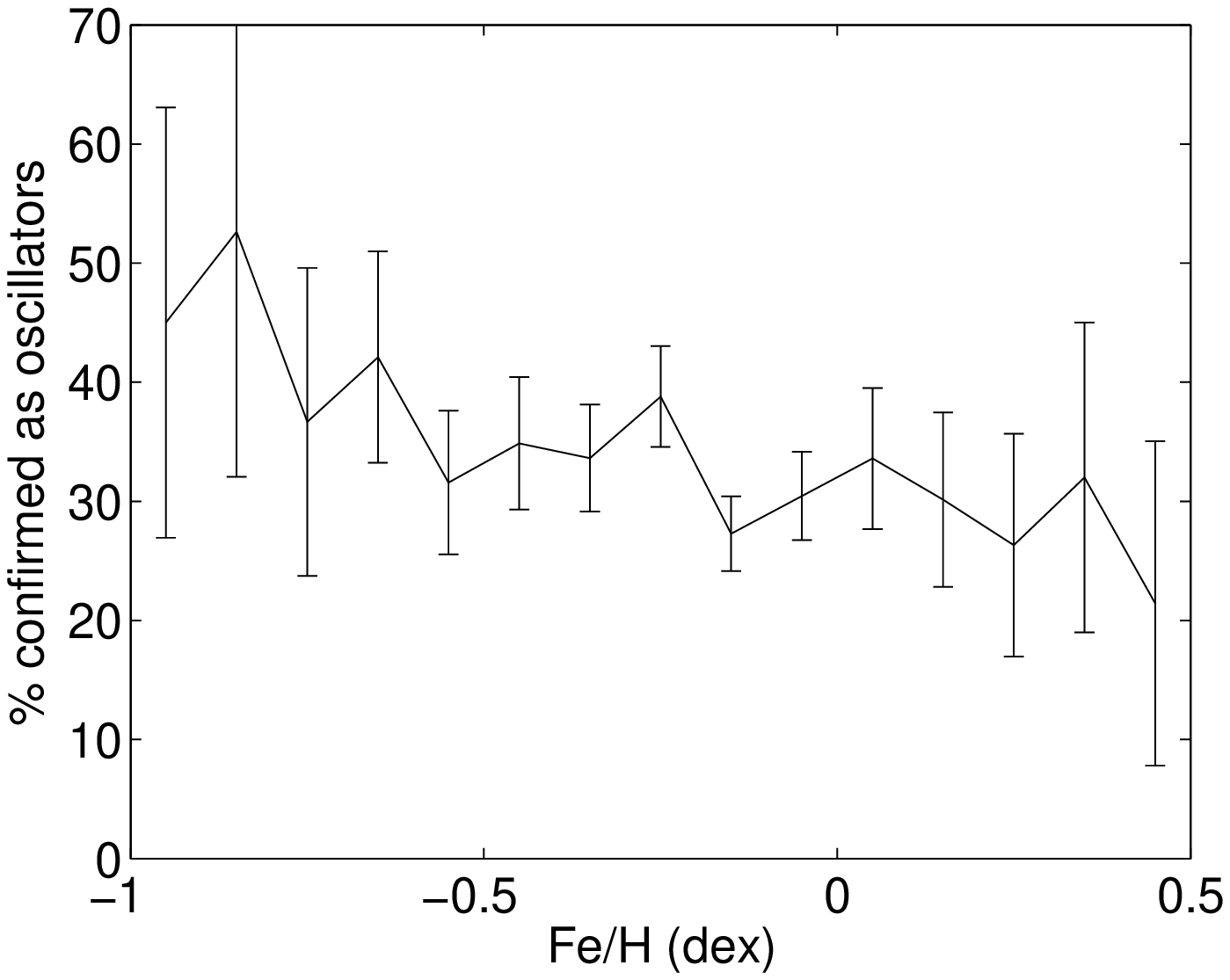}}
}
\medskip
\centerline{%
   \vl{\includegraphics[width=\dimen1]{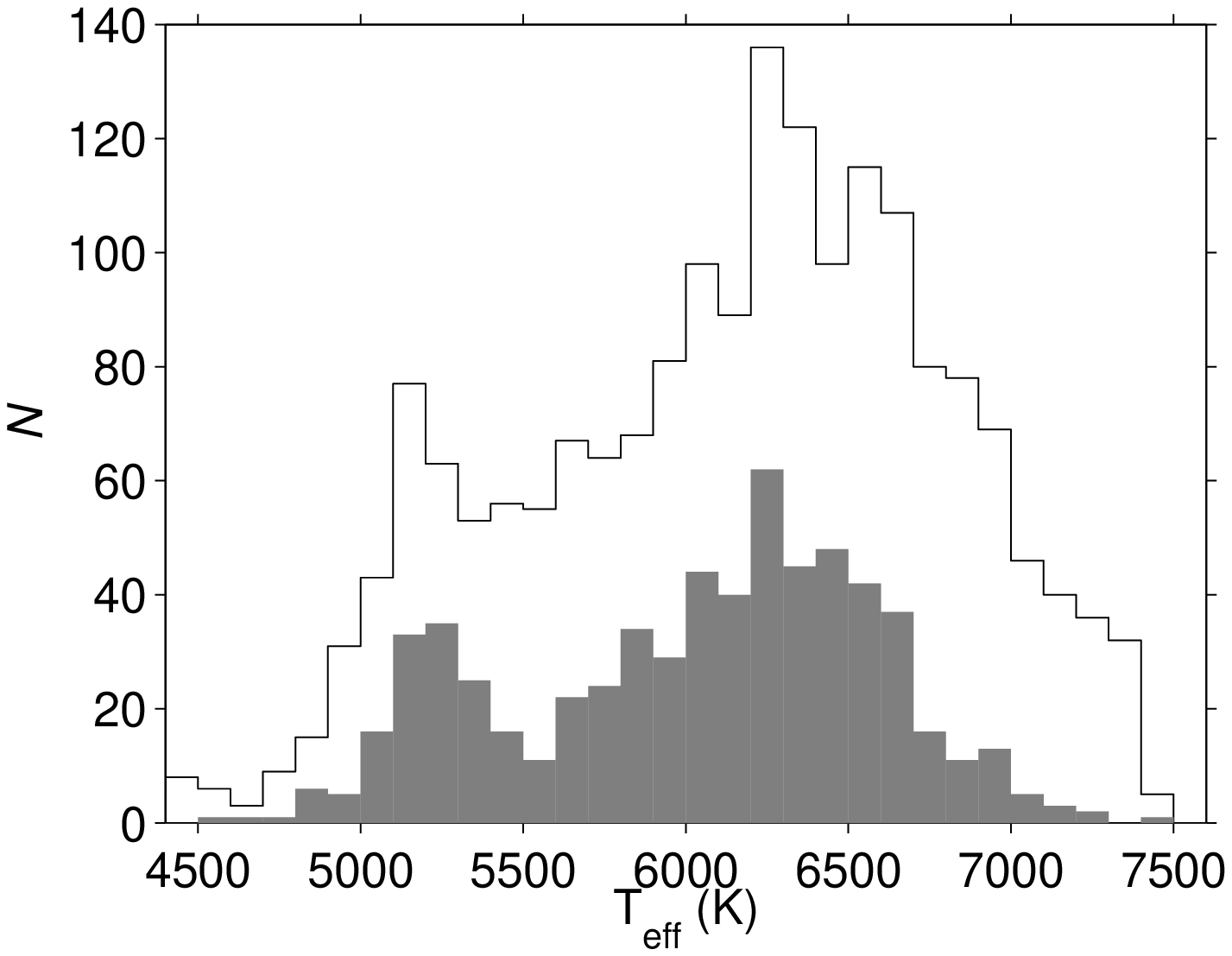}}
   \hfill
   \vl{\includegraphics[width=\dimen1]{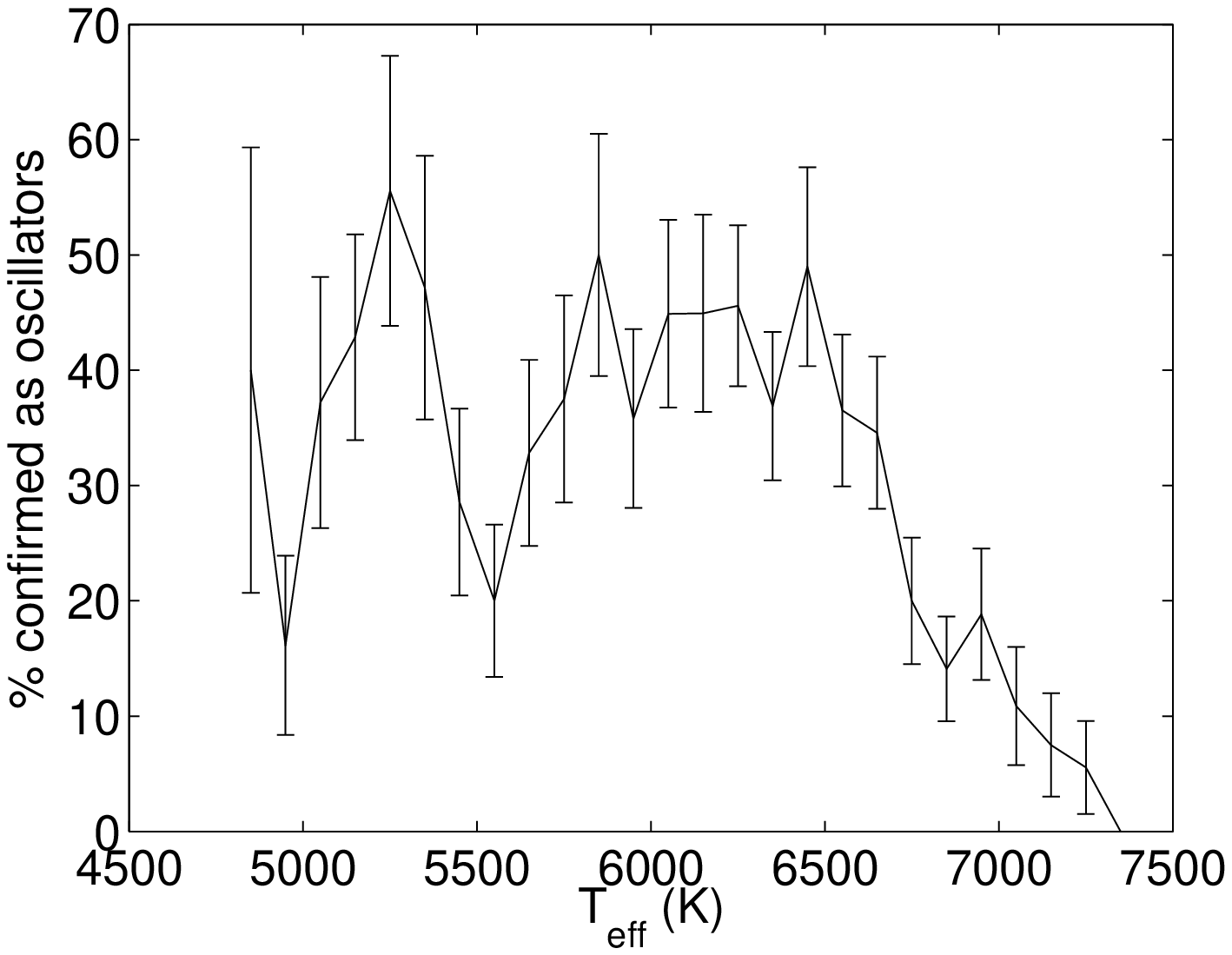}}
   \hfill
   \vl{\includegraphics[width=\dimen1]{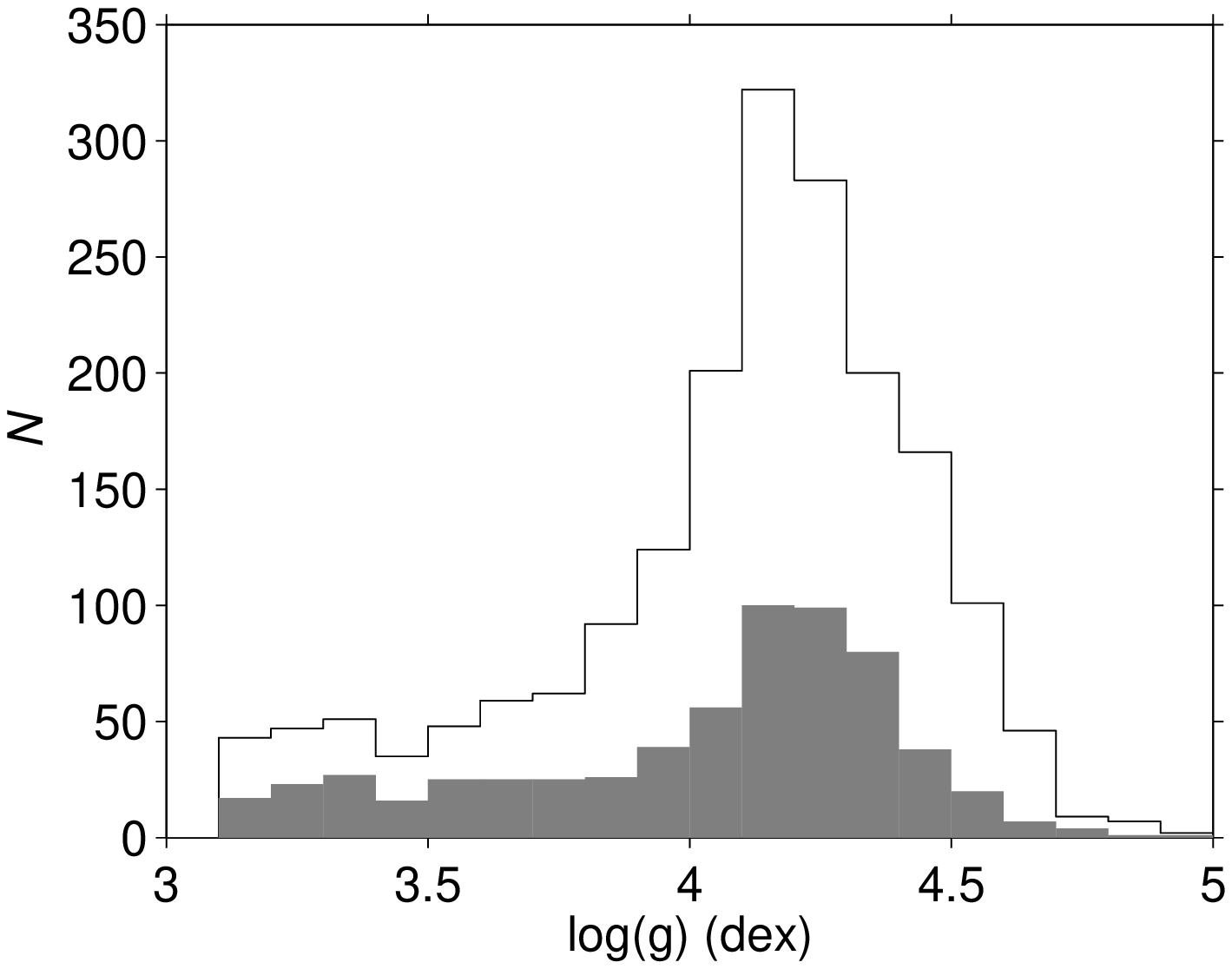}}
   \hfill
   \vl{\includegraphics[width=\dimen1]{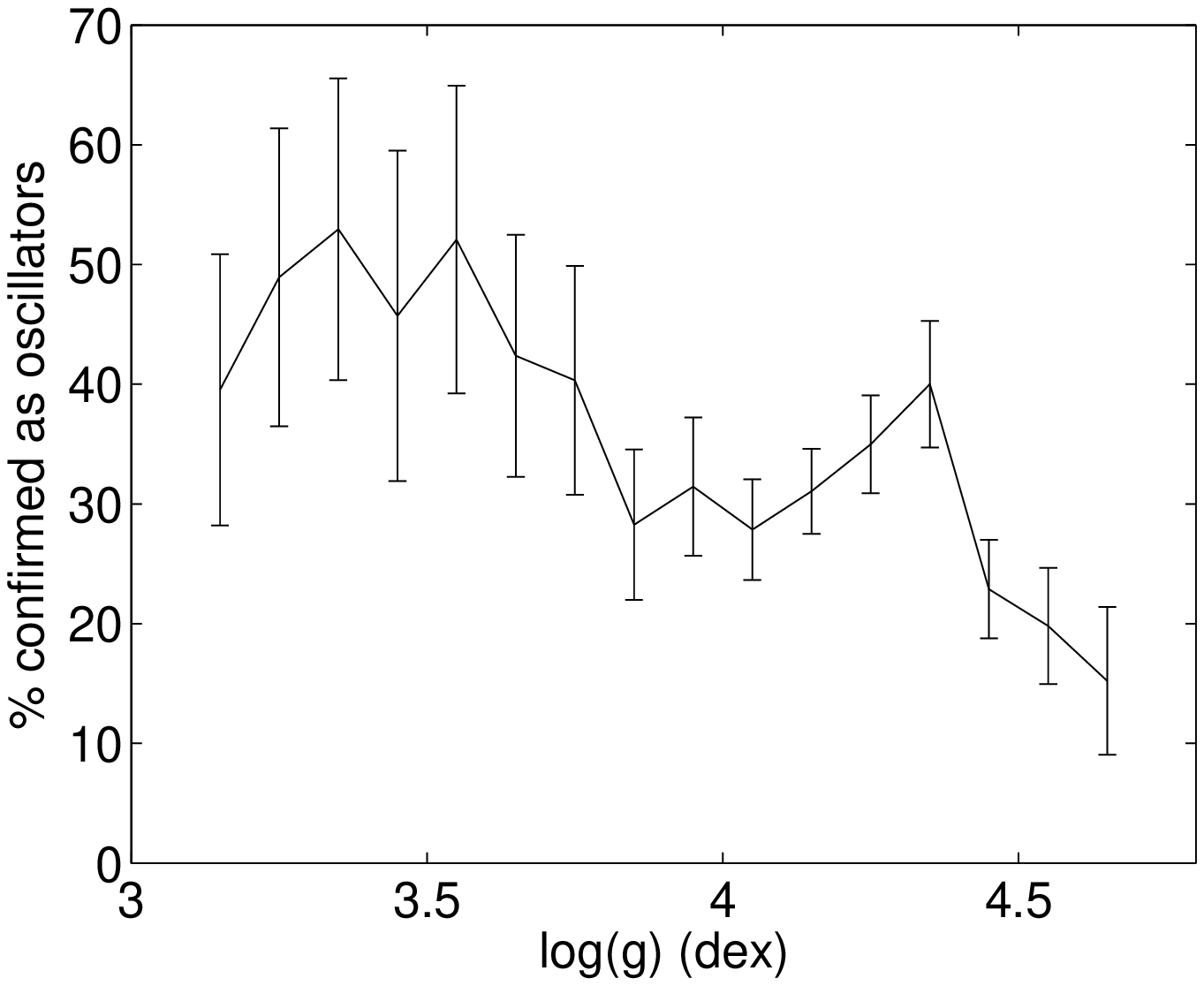}}
}
\caption{Left panels show the distributions of stellar parameters in the full set (open bars) and for confirmed solar-like oscillators (filled bars).  Right panels show the ratio of these distributions with error bars calculated from Poisson statistics.}
\label{fig:propdist}
\end{figure*}

The stellar parameters for the stars observed by {\em Kepler} are provided in the KIC.  This includes the apparent magnitude in the {\em Kepler} band, effective temperature, metallicity and surface gravity derived from Sloan photometry of pre-flight observations.  The KIC effective temperatures are found to differ from those determined from spectroscopic studies which are typically 200-300K higher than those in the KIC.  A recalibration of the KIC Sloan photometry was performed by Pinsonneault et al. (in preparation) who find effective temperatures that are $\sim$280\,K higher than the KIC temperatures for solar-like stars.  We used these recalibrated temperatures with the apparent magnitude, metallicity and surface gravity from the KIC to compare the distribution of stellar parameters in the confirmed solar-like oscillators to those in the complete set (Fig.~\ref{fig:propdist}).

We find an expected correlation of the relative number of confirmed solar-like oscillators with the apparent magnitude of the star.  At $Kp\sim9.5$, we find approximately 50\,\% of the stars to be confirmed solar-like oscillators compared with only 20\,\% at $Kp\sim11.5$.  The effective temperature distributions of the complete cohort and that of the confirmed solar-like oscillators are bimodal, corresponding to a population of subgiants between $\sim$5000 and $\sim$5500\,K and a hotter group of main-sequence stars.  There is a notable reduction in the proportion of confirmed solar-like oscillators in the region between these distributions, for $5300 \lesssim T_\mathrm{eff} \lesssim 5700$\,K.  This has also been noted by \cite{chaplin2011} who suggest that this absence may be due to evolutionary effects that cause an increase in surface magnetic activity and therefore reduces the detectability of oscillations.  The effective temperature dependence also indicates a high-temperature fall-off in the proportion of confirmed solar-like oscillators starting at $\sim$6700\,K with fewer than 10\,\% of the stars with $T_\mathrm{eff}>7000$\,K having confirmed solar-like oscillations.  This high-temperature drop is expected and is in agreement with the location of the red edge of the classical instability strip.  We also observe a slight decreasing trend in the relative number of oscillators with increasing metallicity and surface gravity, which is contrary to what is expected \citep{samadi2007,frandsen2007,stello2009b}.  We find no significant correlation between metallicity and apparent magnitude that would explain this trend for the stars with detected oscillations.  However, it must be stressed that the KIC values for metallicity and surface gravity are assumed to be accurate only to $\sim$0.5\,dex.

\begin{figure*}
\dimen0=\hsize
\dimen1=2mm
\advance\dimen0 by -1\dimen1
\dimen1=0.5\dimen0
\centerline{%
  \vl{\includegraphics[width=\dimen1]{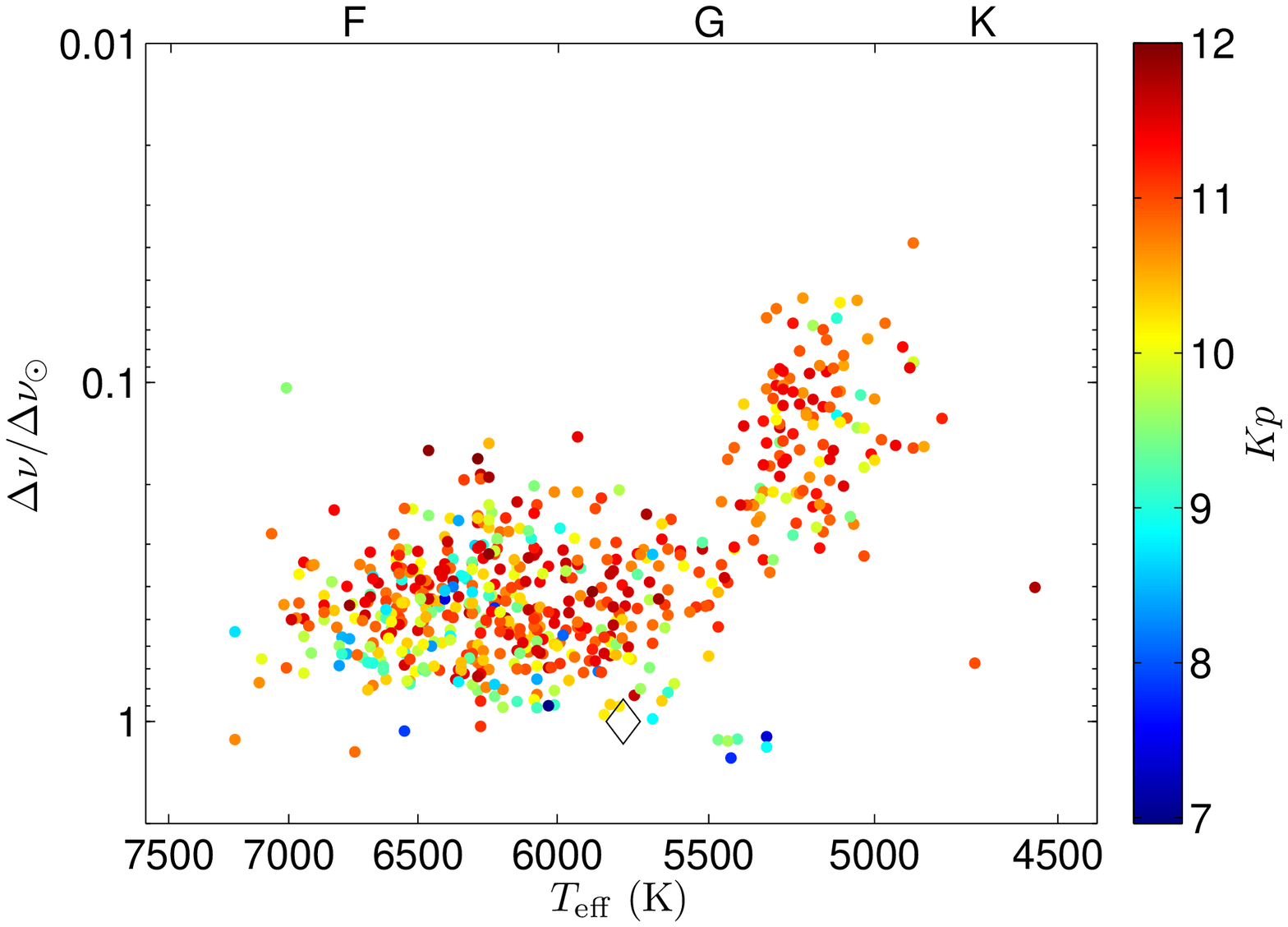}}
  \hfill
  \vl{\includegraphics[width=\dimen1]{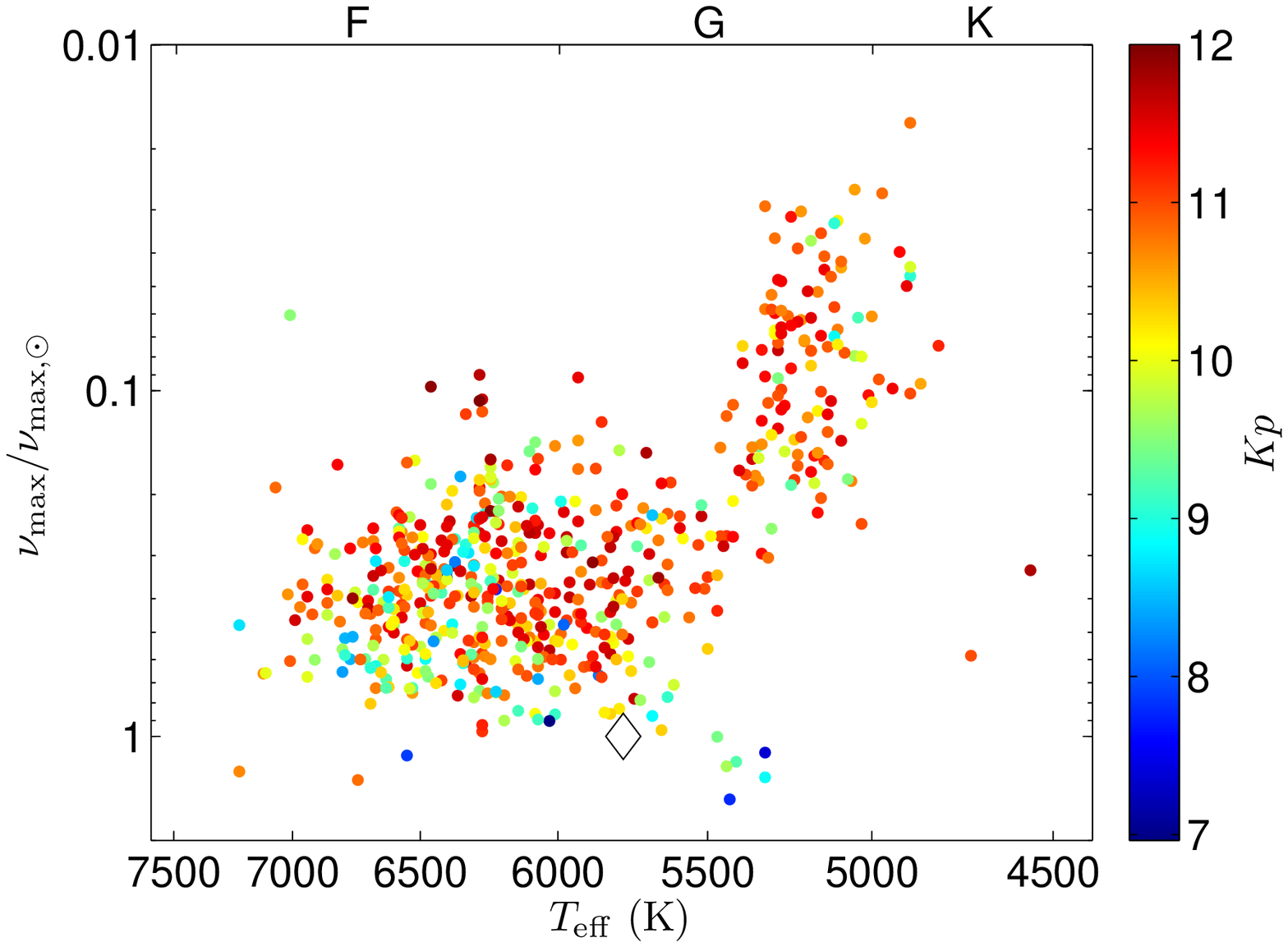}}
}
\caption{Asteroseismic population diagrams produced using the average values of $\Delta\nu$ ({\em left}) and $\nu_\mathrm{max}$ ({\em right}).  The subgiants are grouped in the upper right of the diagrams, with the main sequence stars grouped in the lower left.  The diamonds indicate the position of the Sun.  The points are coloured according to the corresponding apparent magnitude taken from the KIC.}
\label{fig:hr}
\end{figure*}

\begin{figure}
\dimen0=\hsize
\dimen1=2mm
\advance\dimen0 by -1\dimen1
\dimen1=0.5\dimen0
\centerline{%
  \vl{\includegraphics[width=\dimen1]{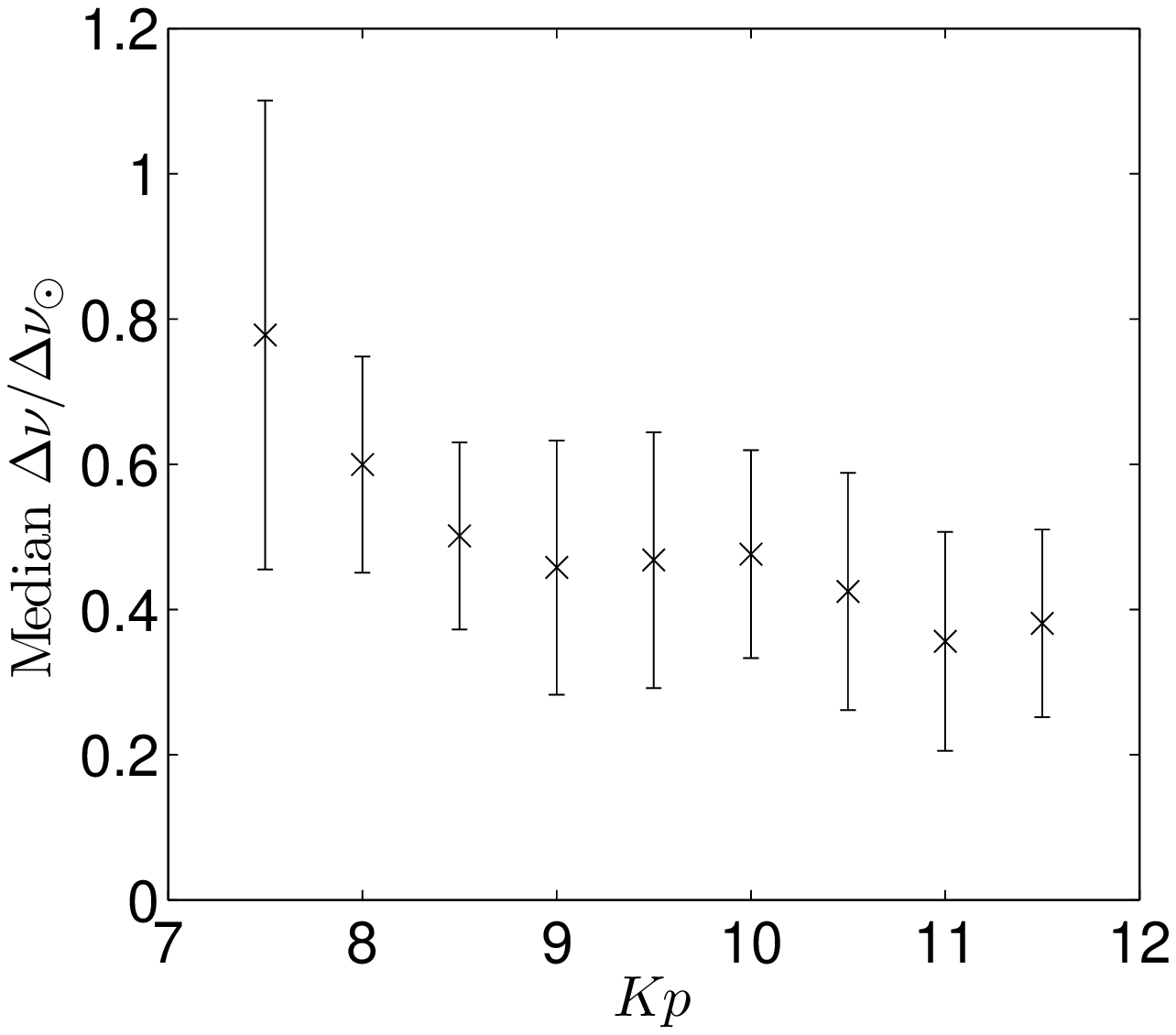}}
  \hfill
  \vl{\includegraphics[width=\dimen1]{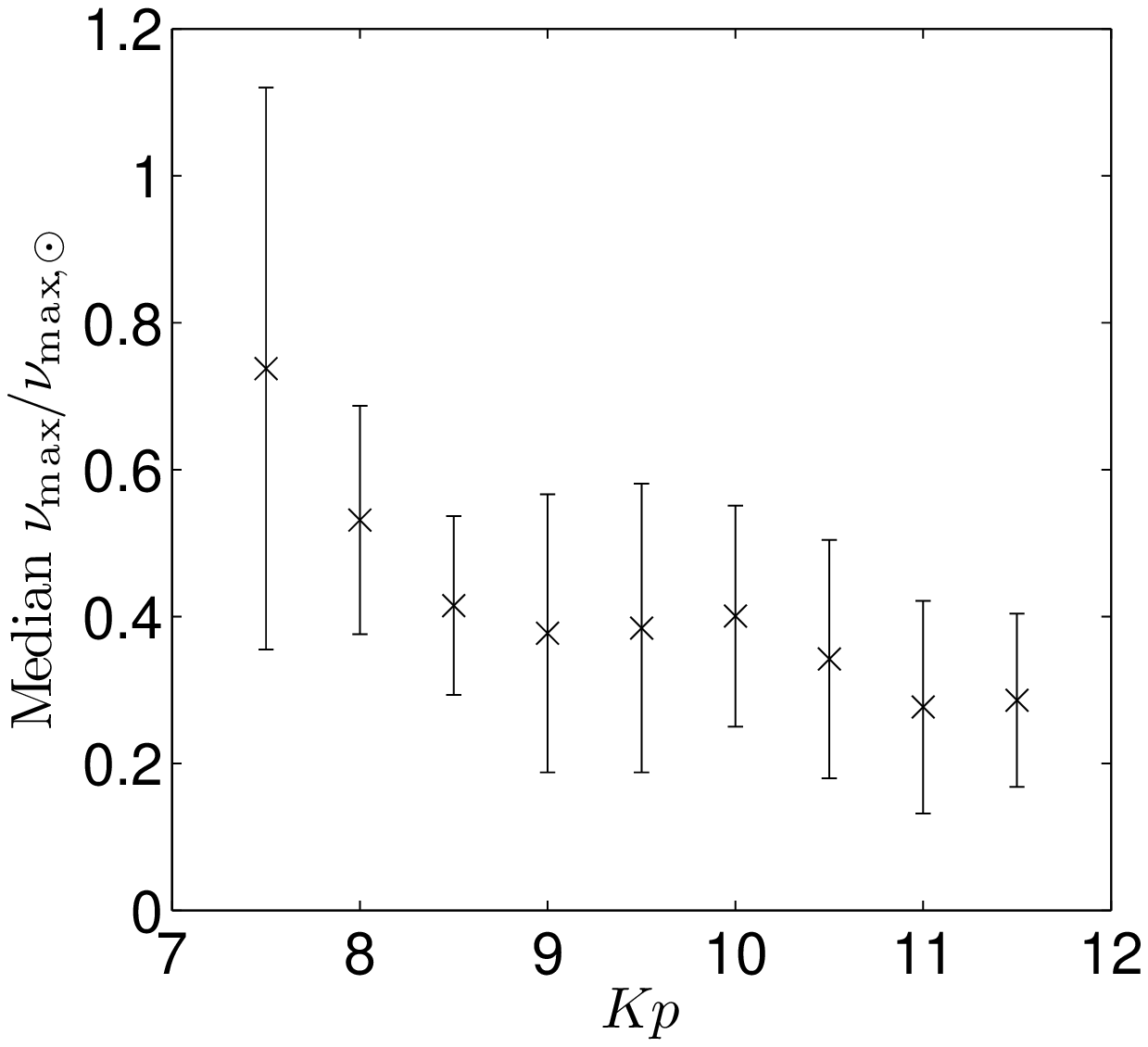}}
}
\medskip
\centering{%
  \includegraphics[width=\dimen1]{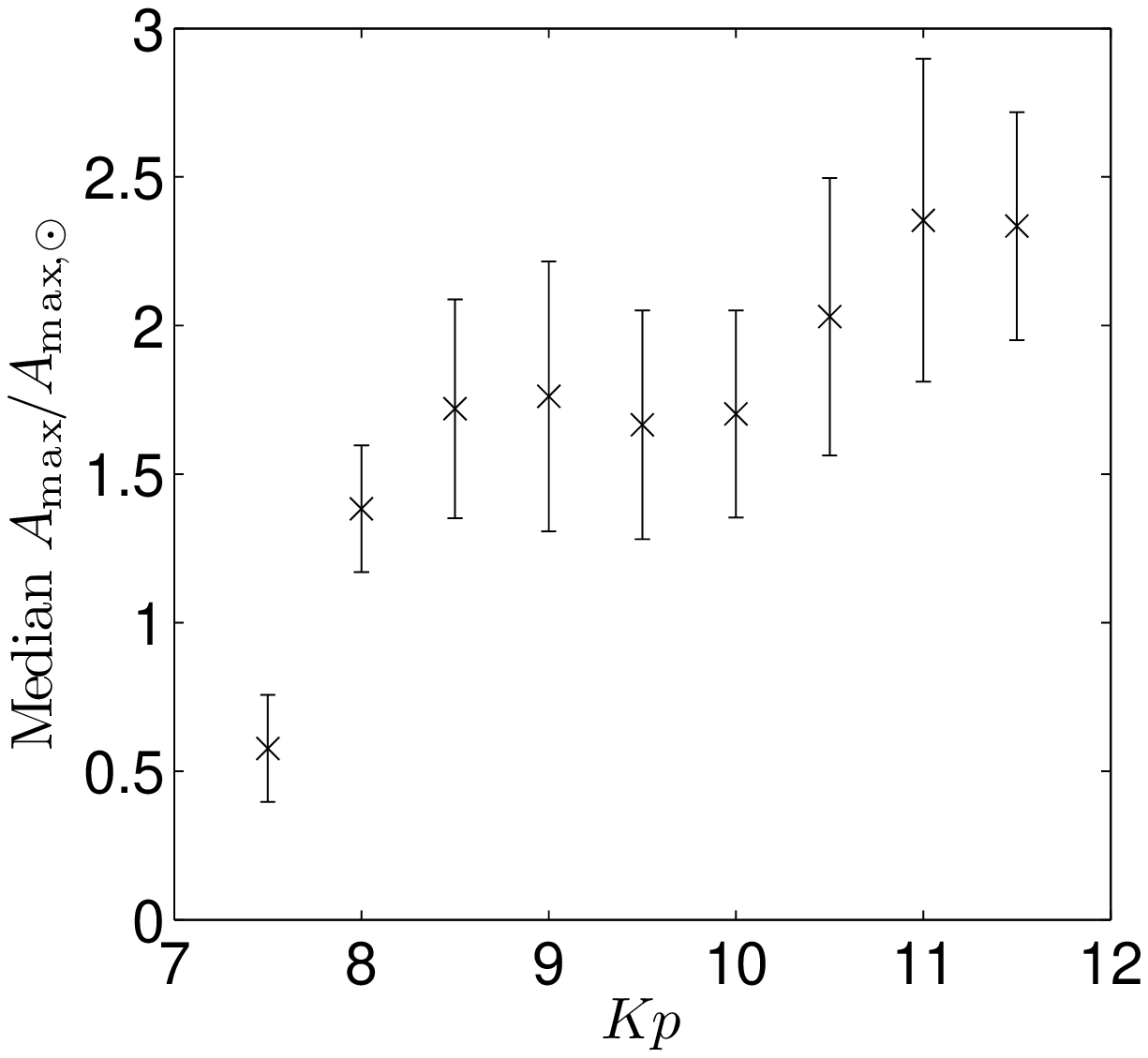}
}
\caption{Median values of $\Delta\nu$, $\nu_\mathrm{max}$ and $A_\mathrm{max}$ as a function of apparent magnitude.  Each cross corresponds to the median in a range covering $\pm0.5$ magnitudes around the point.  Error bars indicate the median absolute deviation in each magnitude range.}
\label{fig:medianparam}
\end{figure}

The average values of $\nu_\mathrm{max}$ and $\Delta\nu$ are plotted against effective temperature in the asteroseismic H-R style diagrams in Fig.~\ref{fig:hr}.  It is clear from these diagrams that the cohort includes a population of subgiants as well as main sequence stars, with the majority of main-sequence stars being hotter than the Sun and with lower values of $\nu_\mathrm{max}$ and $\Delta\nu$.  The stars with asteroseismic parameters closer to those of the Sun also tend to be the brighter stars in the set.  This is illustrated in Fig.~\ref{fig:medianparam}, which shows the median values of $\Delta\nu$, $\nu_\mathrm{max}$ and $A_\mathrm{max}$ as a function of apparent magnitude.

\section{Relationships between global parameters}

\subsection{\boldmath{$\nu_\mathrm{max}$} and \boldmath{$\Delta\nu$}}

It has been observed that $\Delta\nu$ and $\nu_\mathrm{max}$ follow a close power-law relationship both for main sequence stars (e.g. \citealt{stello2009}) and red giants \citep{hekker2009,bedding2010,huber2010,mosser2010,hekker2011}.  By combining the scaling relations Eq.~\ref{eq:numax} and Eq.~\ref{eq:lsep}, \cite{stello2009} showed that the expected relationship between $\Delta\nu$ and $\nu_\mathrm{max}$ is almost independent of luminosity and only weakly dependent on mass and effective temperature.  To a good approximation, we can assume
\begin{equation}
  \Delta\nu \propto \left( \nu_\mathrm{max} \right)^a.
  \label{eq:nmlsprop}
\end{equation}
Using a set of 55 main sequence and giant stars, \cite{stello2009} obtained a power law exponent of $a=0.772\pm0.005$.  This can be compared with the value obtained by \cite{mosser2010}, using data from 930 red giants observed by the {\em COROT} satellite, of $a=0.75\pm0.01$.  \cite{mosser2010} point out that the apparent discrepancy between these values relates to the different temperatures and evolutionary states of the stars used in each study.  \cite{hekker2009} found a slightly higher value for {\em COROT} red-giant stars of $a=0.784\pm0.003$, however this was obtained after scaling by solar values and fitting a power law of the form
\begin{equation}
  \frac{\Delta\nu}{\Delta\nu_\odot} = \left( \frac{\nu_\mathrm{max}}{\nu_{\mathrm{max},\odot}} \right)^a,
  \label{eq:nmlspowlaw}
\end{equation}
which fixes the proportionality constant in Eq.~\ref{eq:nmlsprop} and forces the fit through the solar values of $\Delta\nu$ and $\nu_\mathrm{max}$.  \cite{stello2009} showed that an extra scaling factor appears in Eq.~\ref{eq:nmlspowlaw} that is close to unity but has a weak dependence on mass and $T_\mathrm{eff}$, therefore we expect a spread about this power law, as discussed in \cite{huber2010}.  As the set of stars used in this analysis cover a wide range in $T_\mathrm{eff}$, and each method returns results on different set of stars (\textit{i.e.} covering different ranges in $T_\mathrm{eff}$), we have scaled $\nu_\mathrm{max}$ and $\Delta\nu$ by their solar values and fitted power laws of the form given in Eq.~\ref{eq:nmlspowlaw}.  As this restricts the power law to one degree of freedom, the exponents obtained are not directly comparable to those found by \cite{stello2009} and \cite{mosser2010}, but may be compared with that found for red giants by \cite{hekker2009}.

\begin{figure*}
\dimen0=\hsize
\dimen1=2mm
\advance\dimen0 by -1\dimen1
\dimen1=0.5\dimen0
\centerline{%
  \vl{\includegraphics[width=\dimen1]{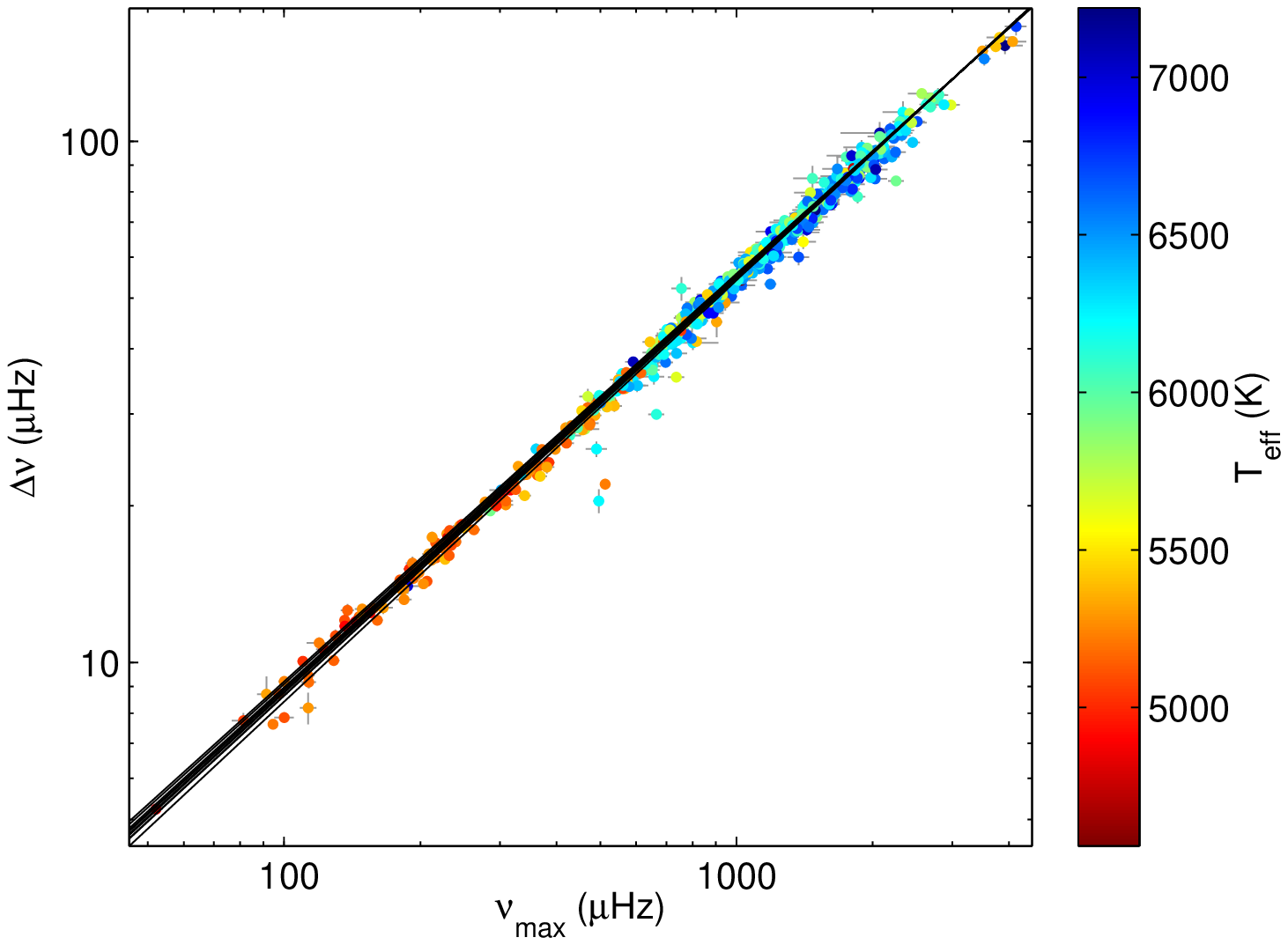}}
  \hfill
  \vl{\includegraphics[width=\dimen1]{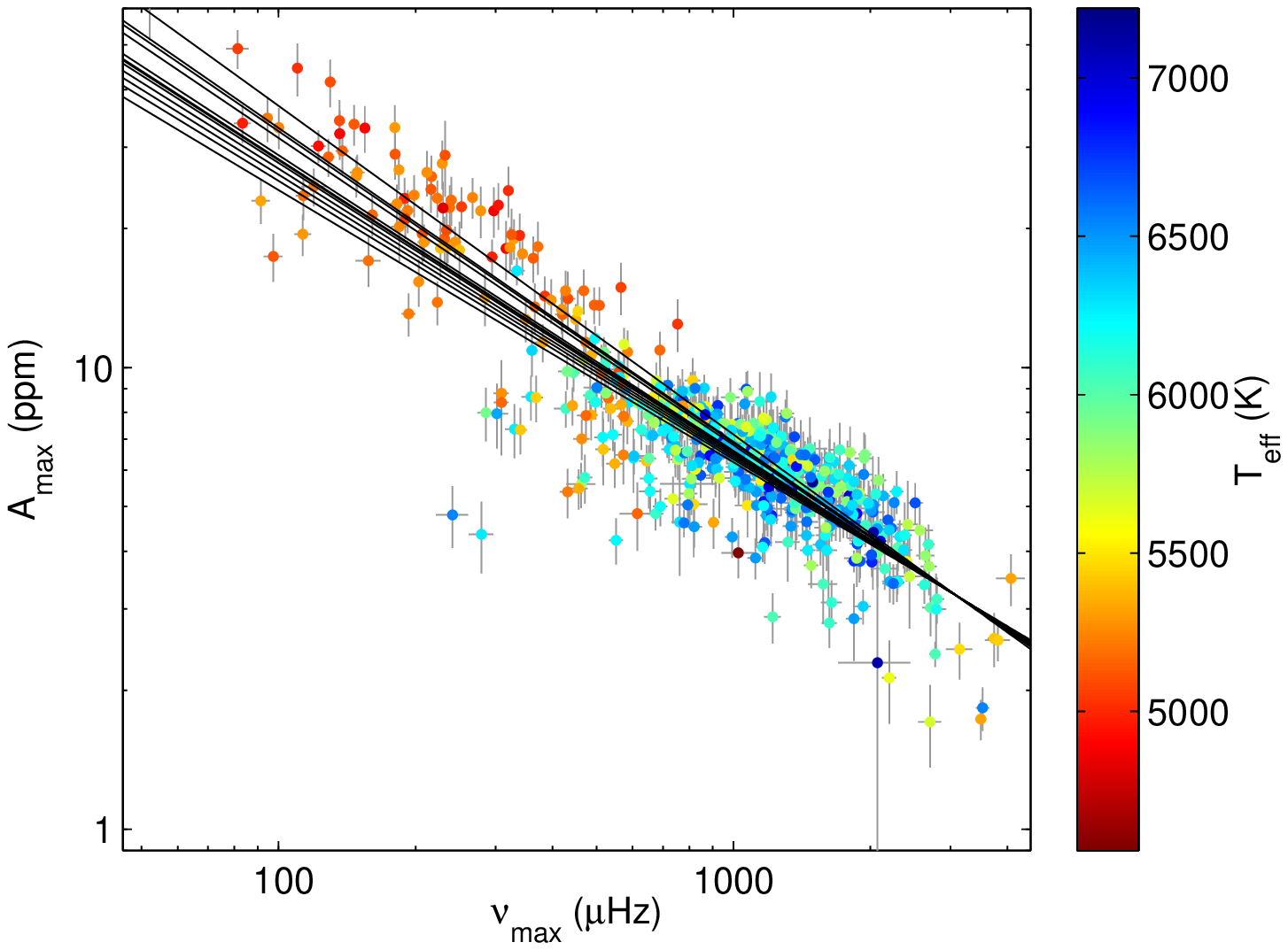}}
}
\medskip
\centerline{%
  \vl{\includegraphics[width=\dimen1]{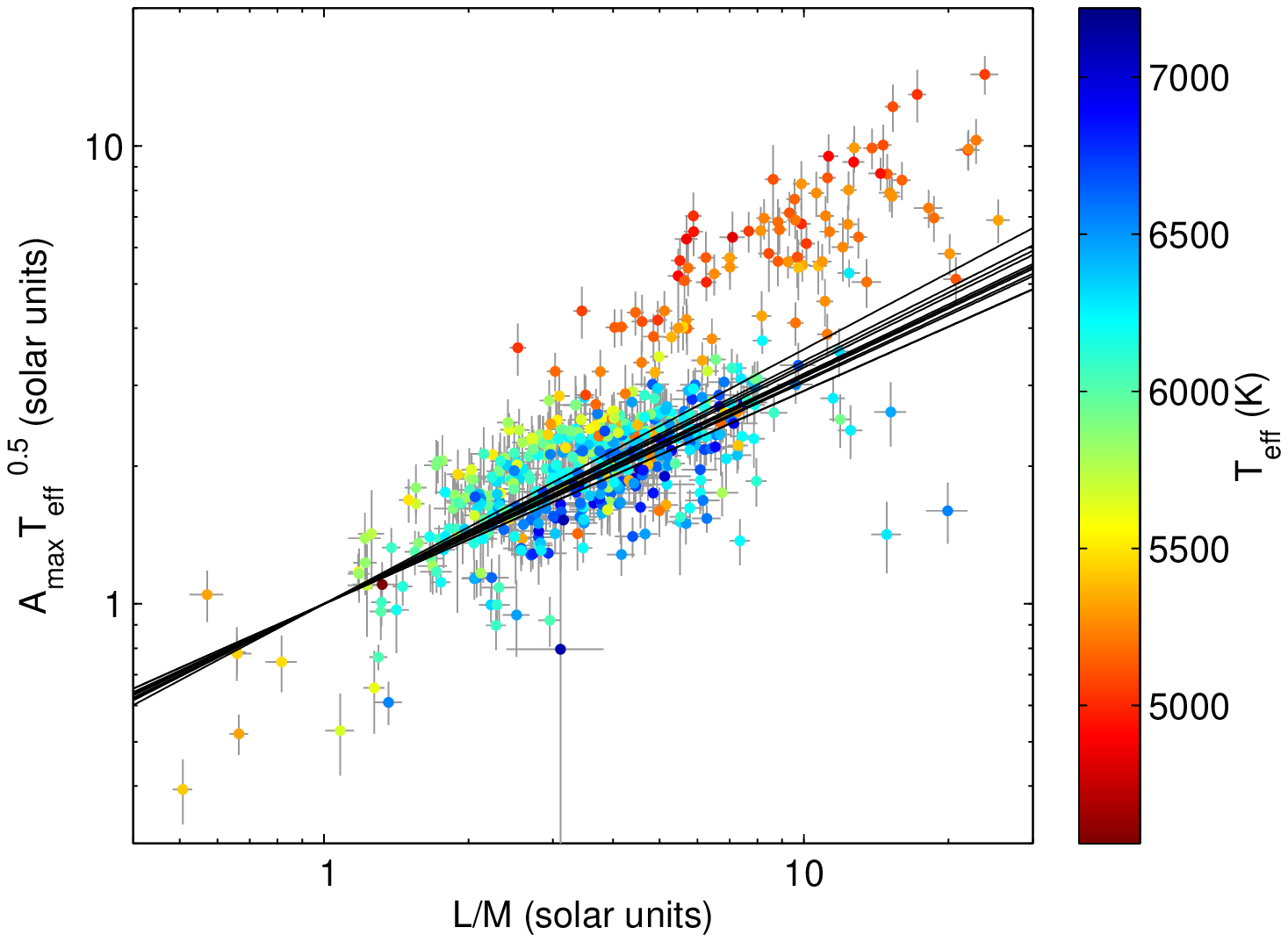}}
  \hfill
  \vl{\includegraphics[width=\dimen1]{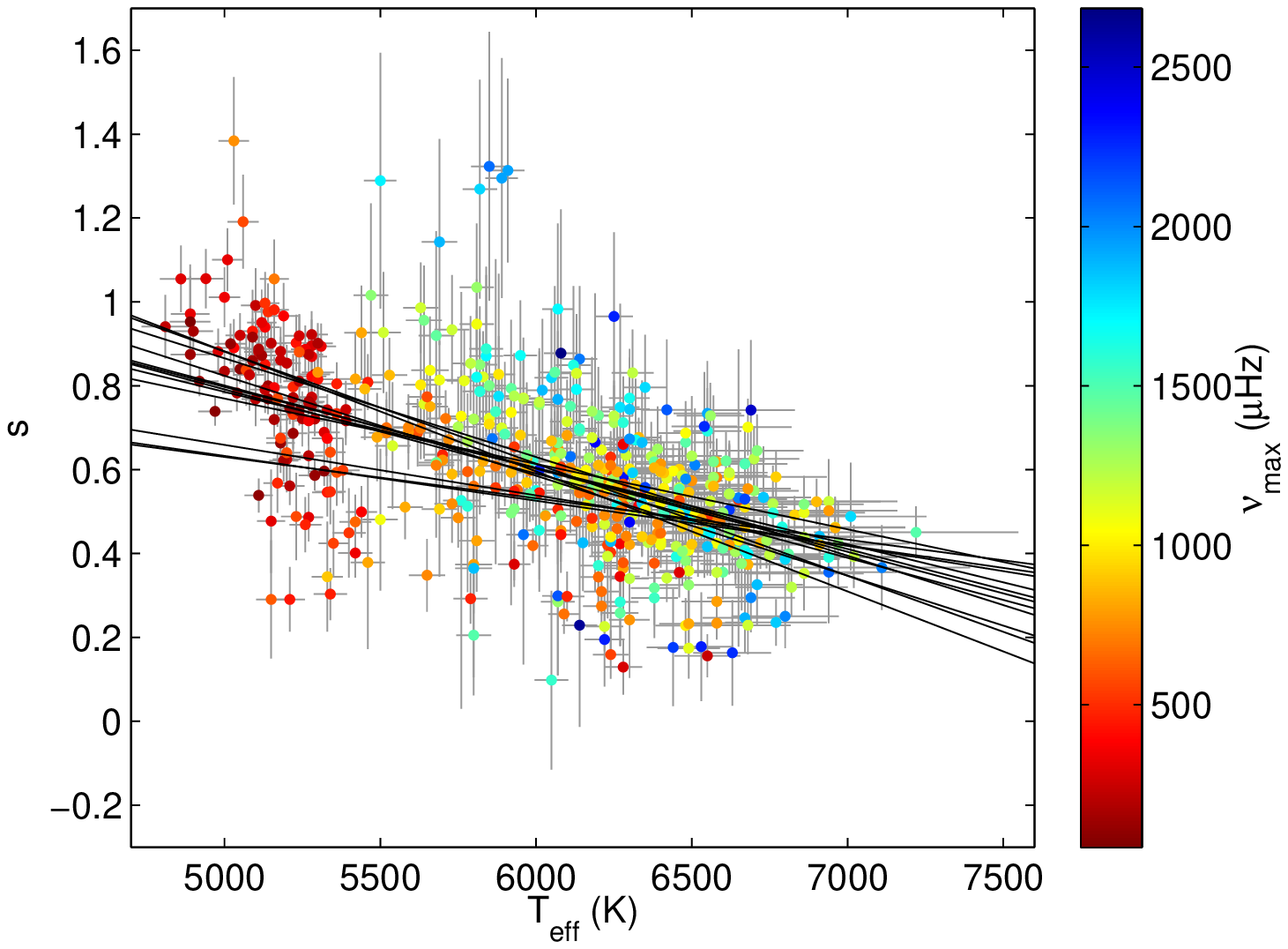}}
}
\caption{Power law fits for stated parameters, solid lines show the fits derived from different methods.  For the upper right and lower left panels, the fits shown are those performed over the range $T_\mathrm{eff}>6250$\,K.  Data points shown are the mean values determined using all methods.}
\label{fig:finfigs}
\end{figure*}

\begin{figure}
\includegraphics[width=\hsize]{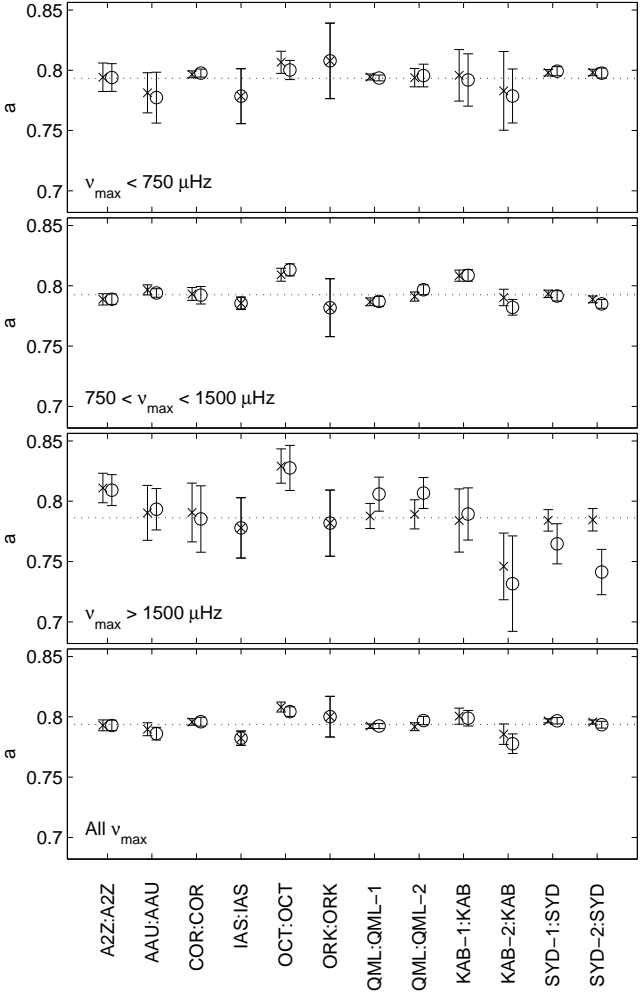}
\caption{$\Delta\nu$ vs. $\nu_\mathrm{max}$ power law exponents (Eq.~\ref{eq:nmlspowlaw}) determined over the $\nu_\mathrm{max}$ ranges given in each plot.  Points with crosses are determined using the scaled formal uncertainties submitted by each group, circles are determined using constant relative uncertainties.  Dotted lines shows the mean values calculated from all methods in each range.  Methods used are given in the format [$\nu_\mathrm{max}$]\,:\,[$\Delta\nu$]}
\label{fig:powlawres}
\end{figure}

We have determined the power law exponent $a$ from the results of each of the methods using a weighted least-squares approach.  This was done using the scaled formal uncertainties returned by the methods and also using constant relative indicative uncertainties estimated from the mean scatter in the results and the internal uncertainties determined from simulated data.  The power law fits were performed over the full range in $\nu_\mathrm{max}$ and also individually in three ranges: $\nu_\mathrm{max}<750$\,$\mu$Hz, $750<\nu_\mathrm{max}<1500$\,$\mu$Hz and $\nu_\mathrm{max}>1500$\,$\mu$Hz.  The power law fits are shown in Fig.~\ref{fig:finfigs} and the derived exponents plotted in Fig.~\ref{fig:powlawres} and given in Table~\ref{tab:powlawexp}.

Taking the weighted average of all methods when we used the corrected uncertainties gave a value of $a=0.795$, with a mean error of $0.005$ and a standard deviation of $0.007$ across the different methods.  When we repeated this using fixed relative uncertainties for all of the methods we obtained a value of $a=0.794$, with mean error $0.005$ and a standard deviation of $0.008$.  It is reassuring that the average value of $a$ is not significantly affected by the way in which we treat the formal uncertainties.  However, in both cases the spread in values obtained for $a$ by the different methods was greater than the average error derived for the parameter.  This apparent disagreement is due to the fact that each method gives results on a different sample of stars, sampling a different range of metallicities, effective temperatures and evolutionary states.  As the degeneracy of Eq.~\ref{eq:nmlspowlaw} is known to be affected by stellar composition, it is not surprising that each method gave slightly different values of $a$.  As a reference value for the uncertainty on $a$ we suggest that it is more appropriate to take this scatter into account rather than to use the error obtained from a single method.  The $\nu_\mathrm{max}$-$\Delta\nu$ relation is discussed further by Huber et al. (in preparation).

\subsection{\boldmath{$\nu_\mathrm{max}$} and \boldmath{$A_\mathrm{max}$}}

The calibrated maximum mode amplitudes from each method ($A_\mathrm{max,obs}$) were scaled to their bolometric equivalent ($A_\mathrm{max}=c_{K-\mathrm{bol}} A_\mathrm{max,obs}$) using the bolometric correction derived from the spectral response of the {\em Kepler} passband \citep{ballot2011},
\begin{equation}
  c_{K-\mathrm{bol}} = 1 + a_1 (T_\mathrm{eff} - T_0) + a_2 (T_\mathrm{eff} - T_0)^2,
  \label{eq:bc}
\end{equation}
with $T_0=5934$\,K, $a_1=1.349\times10^{-4}$\,K$^{-1}$ and $a_2=-3.12\times10^{-9}$\,K$^{-2}$.  The average bolometric maximum mode amplitude as a function of $\nu_\mathrm{max}$ is shown in Fig.~\ref{fig:finfigs}.  The relationship between $A_\mathrm{max}$ and $\nu_\mathrm{max}$ is expected to have a colour dependence (Eq.~\ref{eq:amax}), therefore power law fits were performed over three ranges in $T_\mathrm{eff}$: $T_\mathrm{eff}<5560$\,K, $5560<T_\mathrm{eff}<6250$\,K and $T_\mathrm{eff}>6250$\,K.  The fits were performed according to
\begin{equation}
  \frac{ A_\mathrm{max} }{ A_{\mathrm{max},\odot} } = \left( \frac{ \nu_\mathrm{max} }{ \nu_{\mathrm{max},\odot} } \right)^b.
  \label{eq:nmampowlaw}
\end{equation}
The power laws were determined using the corrected formal uncertainties and also the constant relative uncertainties indicative of the spread in the results from each method.  The power law indices determined for each method are given in Table~\ref{tab:powlawexp}.

Following the scaling relations given in \cite{kjeldsen1995} and \cite{samadi2007}, for an adiabatic stellar atmosphere the maximum mode amplitude is expected to scale as
\begin{equation}
  \frac{A_\mathrm{max}}{A_{\mathrm{max},\odot}} = \left( \frac{L / L_\odot}{M / M_\odot} \right)^s \left( \frac{T_\mathrm{eff}}{T_{\mathrm{eff},\odot}} \right)^{-0.5},
  \label{eq:ampscale}
\end{equation}
where bolometric intensity variations are observed.  For observations obtained in a narrow wavelength band, the wavelength dependence of the measured mode amplitude results in a modification of the effective temperature exponent in Eq.~\ref{eq:ampscale} from $-0.5$ to $-1.5$.  The assumption of an adiabatic stellar atmosphere is a simple approximation and, using narrow-band observations, \cite{kjeldsen1995} showed that an effective temperature exponent of $-2.0$ gave a better fit to the data in their sample.  For main-sequence solar-like stars, \cite{samadi2007} calculated a value of $s=0.7$ based on numerical models.  However, using the bolometric amplitudes of red giants observed by {\em COROT} and an effective temperature exponent of $-0.5$, \cite{mosser2010} obtained a value of $s=0.89\pm0.02$.  It is apparent from these results that there is an additional temperature dependence in Eq.~\ref{eq:ampscale}.  Here we incorporate any additional temperature dependence into the exponent $s$ and fix the effective temperature exponent to $-0.5$.

Rewriting luminosity in terms of radius and $T_\mathrm{eff}$ and using Eq.~\ref{eq:numax} leads to:
\begin{equation}
  \frac{L / L_\odot}{M / M_\odot} = \left( \frac{T_\mathrm{eff}}{T_{\mathrm{eff},\odot}} \right)^{3.5} \left( \frac{\nu_\mathrm{max}}{\nu_{\mathrm{max},\odot}} \right)^{-1}.
  \label{eq:lmnm}
\end{equation}
The value of $s$ can then be found by combining and rearranging Eqs.~\ref{eq:ampscale} and \ref{eq:lmnm} or by fitting a power law to $A_\mathrm{max}$\,$(T_\mathrm{eff})^{0.5}$ vs. $L/M$ \citep{huber2010,mosser2010}.  The power law fits and the variation in $s$ with $T_\mathrm{eff}$ are shown in Fig.~\ref{fig:finfigs}.

We find a clear dependence of $s$ on $T_\mathrm{eff}$ for all of the methods.  The value of $s$ for cooler stars is in agreement with \cite{mosser2010} who find a value of $s=0.89\pm0.02$ in a set of red giants with an average $T_\mathrm{eff}$ of 4500\,K and explains the different values of $s$ obtained for main-sequence and giant stars.  The value of $s$ for $T_{\mathrm{eff},\odot}=5777$\,K and the gradients determined from Fig.~\ref{fig:finfigs} are given in Table~\ref{tab:svals}.  Taking the weighted mean values across all methods gives
\begin{equation}
  s = 0.64 - 1.18 \left( \frac{T_\mathrm{eff} - T_{\mathrm{eff},\odot}}{T_{\mathrm{eff},\odot}} \right).
  \label{eq:sval}
\end{equation}
It is not surprising that there is a remaining temperature dependence in Eq.~\ref{eq:ampscale} as the scaling relation relies on a very simple adiabatic description of the stellar atmosphere.  For further discussion, see \cite{kjeldsen2011}.

\section{Summary and conclusions}

We have detected solar-like oscillations in 642 F-, G- and K-type main-sequence and subgiant stars in the {\em Kepler} field-of-view and characterised them by their large frequency separation, frequency of peak power and maximum mode amplitude.  By combining results from the analysis methods of nine independent teams, we have verified the detections, rejected outliers and devised a method to ensure the results are consistent within an accurate uncertainty range.  It is apparent that the formal uncertainties returned from automated analysis methods are often inconsistent with the actual precision of the results.  Obtaining an accurate uncertainty on the global asteroseismic parameters is essential when using such results to model stellar structure.

Using the standard deviation of the results obtained by multiple methods and the intrinsic uncertainty estimated from the analysis of simulated data, we have determined the average expected relative precision of $\Delta\nu$ (1.8\,\%), $\nu_\mathrm{max}$ (3.8\,\%) and $A_\mathrm{max}$ (9.8\,\%).  In the case of red giants, \cite{bedding2010} and \cite{huber2010} have shown that, when radial modes can be identified, the $\delta\nu_{02}$ separation and the $\epsilon$ factor of the Tassoul asymptotic law scale with $\Delta\nu$.  Furthermore, \cite{mosser2011} have demonstrated that the red-giant oscillation frequencies obey a universal pattern that helps to measure $\Delta\nu$ very precisely and to identify unambiguously the mode angular degree and radial order.  For solar-like stars, the amplitudes of oscillations are much smaller than for red giants and the linewidths of the modes are often broader, reflecting the contribution of increased temperature on mode damping \citep{chaplin2009}.  We have also observed mixed modes and {\em avoided crossings} in evolved solar-type stars, which introduce discontinuities into the small frequency separation.  These effects combine to make the process of accurately estimating the average small frequency separation a more difficult task for solar-like stars than for red giants.  Using the analysis of simulated data, we have shown that the average expected relative precision of $\delta\nu_{02}$, as obtained using automated methods, is only $\sim$26\,\%.  However, for stars with a sufficient signal-to-noise ratio, this can be much improved upon by using a full peak-bagging approach and obtaining individual $p$-mode frequencies.

By characterising the stars by $\Delta\nu$ or $\nu_\mathrm{max}$, we have clearly identified the separate main-sequence and subgiant populations (Fig.~\ref{fig:hr}).  The distributions of these asteroseismic parameters in the observed main-sequence stars show that we typically find smaller values of $\Delta\nu$ and $\nu_\mathrm{max}$ than we find for the Sun (\textit{i.e.} $\Delta\nu_\odot \sim 135$\,$\mu$Hz, $\nu_{\mathrm{max},\odot} \sim 3100$\,$\mu$Hz).  This reflects the increased amplitude of oscillations in stars with lower $\nu_\mathrm{max}$ and the higher intrinsic luminosity of such stars (Fig.~\ref{fig:finfigs}).

We have correlated the fraction of stars for which we detected oscillations with the stellar parameters from the KIC and have found a significant reduction in the proportion of solar-like oscillators with effective temperatures close to $T_\mathrm{eff} \sim 5500$\,K, the temperature range that separates the distributions of main-sequence stars from subgiants.  This is in agreement with the findings of \cite{chaplin2011} who suggest that stars in this region may show interesting evolutionary effects in their stellar dynamos \citep{gilliland1985,dall2010}, which would manifest in the surface magnetic activity and, therefore, decrease the detectability of oscillations.

We have determined the exponents of power-law relationships between verified $\Delta\nu$ and $\nu_\mathrm{max}$ values and have found that the scatter in the results is at a level greater than the formal uncertainties obtained for the individual methods (Fig.~\ref{fig:powlawres} and Table~\ref{tab:powlawexp}).  This illustrates the importance of using results from multiple methods when testing such relationships between parameters.  Taking a weighted average of the exponents, we have determined a power law exponent of $a=0.795\pm0.007$, which is significantly higher than that found using data from a similar number of red giants using the same method \citep{hekker2009}, $a=0.784\pm0.003$.  This discrepancy can be explained by the different temperatures and evolutionary states of the stars in each study and is also seen in the different results obtained for main-sequence stars \citep{stello2009} and red giants \citep{huber2010,mosser2010}.

The maximum mode amplitudes obtained by each method, once calibrated by those found for stars with a high signal-to-noise ratio, have been used to determine the exponent $s$ relating the dependence of mode amplitude on $L/M$ (Eq.~\ref{eq:lmnm}).  Various authors have stated values of $s$ between 0.7 and 1.0 using the amplitudes calculated from main-sequence and giant stars.  We have used bolometric-corrected amplitudes with a temperature dependence exponent of $-0.5$ and found that a strong temperature dependence remains in $s$.  We have constrained this dependence by a linear relationship (Eq.~\ref{eq:sval}) which gives a value of $s$ for stars in the red-giant regime which agrees with that determined in \cite{mosser2010} who used the same expected temperature dependence in their calculations.  This is not an unexpected result and a temperature different to that estimated from a simple adiabatic atmosphere was found by \cite{kjeldsen1995}.  The scaling relations between global asteroseismic parameters obtained by {\em Kepler} have been investigated in more detail and are presented in another paper (Huber at al., in preparation).

The verified global asteroseismic parameters determined using the methods described in this paper will be used in subsequent work to obtain estimates of stellar structural parameters \citep[\textit{e.g.} using the methods described in][]{stello2009c} and constrain models of stellar structure and evolution.  The asteroseismic results obtained in this and other studies of the observations from {\em Kepler} have increased the number of stars on which solar-like oscillations have been detected and characterised by more than an order of magnitude and illustrates the great achievements made by the {\em Kepler Mission} and the {\em Kepler Asteroseismic Science Consortium}.

\section*{Acknowledgements}

GAV, YE, WJC, SH and IWR acknowledge the support of the UK Science and Technology Facilities Council (STFC).  The authors are grateful to IRFU/SAp at CEA-Saclay for providing support for useful meetings in the development of this work.  Funding for the {\em Kepler Mission} is provided by NASA's Science Mission Directorate.  The authors wish to thank the entire {\em Kepler} team, without whom these results would not be possible.  We also thank all funding councils and agencies that have supported the activities of KASC Working Group 1.

\appendix
\section{Tables of fitted power law parameters}

\begin{table*}
  \centering
  \begin{tabular}{|c|c|c|c|c|c|c|c|}
  \hline
  Method & $a_1$ & $a_2$ & $a_3$ & $a_4$ & $b_1$ & $b_2$ & $b_3$ \\
  \hline
  \multicolumn{8}{|c|}{Using scaled formal uncertainties} \\
  \hline
A2Z   & $0.794\pm0.012$ & $0.789\pm0.005$ & $0.811\pm0.012$ & $0.793\pm0.004$ & $-0.74\pm0.12$ & $-0.62\pm0.04$ & $-0.64\pm0.03$ \\
AAU   & $0.781\pm0.017$ & $0.797\pm0.004$ & $0.790\pm0.023$ & $0.790\pm0.005$ & $-0.63\pm0.10$ & $-0.66\pm0.03$ & $-0.63\pm0.03$ \\
COR   & $0.797\pm0.003$ & $0.793\pm0.005$ & $0.791\pm0.024$ & $0.796\pm0.003$ & $-0.74\pm0.04$ & $-0.61\pm0.05$ & $-0.59\pm0.04$ \\
IAS   & $0.778\pm0.023$ & $0.786\pm0.005$ & $0.778\pm0.025$ & $0.782\pm0.006$ & $-0.50\pm0.10$ & $-0.59\pm0.03$ & $-0.60\pm0.03$ \\
OCT-1 & $0.807\pm0.009$ & $0.809\pm0.005$ & $0.829\pm0.014$ & $0.808\pm0.004$ & $-0.67\pm0.02$ & $-0.64\pm0.03$ & $-0.68\pm0.02$ \\
OCT-2 & --              & --              & --              & --              & $-0.67\pm0.03$ & $-0.66\pm0.03$ & $-0.67\pm0.02$ \\
OCT-3 & --              & --              & --              & --              & $-0.73\pm0.03$ & $-0.66\pm0.03$ & $-0.68\pm0.02$ \\
ORK   & $0.808\pm0.031$ & $0.782\pm0.024$ & $0.782\pm0.027$ & $0.800\pm0.017$ & --             & --             & --             \\
QML-1 & $0.794\pm0.002$ & $0.787\pm0.003$ & $0.788\pm0.010$ & $0.792\pm0.002$ & $-0.67\pm0.22$ & $-0.69\pm0.07$ & $-0.71\pm0.04$ \\
QML-2 & $0.794\pm0.008$ & $0.791\pm0.004$ & $0.789\pm0.012$ & $0.792\pm0.003$ & --             & --             & --             \\
KAB-1 & $0.796\pm0.021$ & $0.808\pm0.005$ & $0.784\pm0.026$ & $0.801\pm0.007$ & $-0.56\pm0.09$ & $-0.56\pm0.03$ & $-0.61\pm0.03$ \\
KAB-2 & $0.783\pm0.033$ & $0.790\pm0.007$ & $0.746\pm0.028$ & $0.786\pm0.008$ & $-0.54\pm0.08$ & $-0.55\pm0.03$ & $-0.63\pm0.02$ \\
SYD-1 & $0.798\pm0.003$ & $0.793\pm0.003$ & $0.784\pm0.009$ & $0.797\pm0.002$ & $-0.66\pm0.02$ & $-0.63\pm0.02$ & $-0.63\pm0.02$ \\
SYD-2 & $0.798\pm0.003$ & $0.789\pm0.003$ & $0.785\pm0.009$ & $0.796\pm0.002$ & $-0.69\pm0.02$ & $-0.61\pm0.02$ & $-0.62\pm0.02$ \\
  \hline
  \multicolumn{8}{|c|}{Using constant indicative relative uncertainties} \\
  \hline
A2Z   & $0.794\pm0.012$ & $0.789\pm0.004$ & $0.809\pm0.013$ & $0.793\pm0.004$ & $-0.69\pm0.18$ & $-0.60\pm0.05$ & $-0.63\pm0.03$ \\
AAU   & $0.777\pm0.021$ & $0.794\pm0.004$ & $0.793\pm0.017$ & $0.786\pm0.005$ & $-0.64\pm0.10$ & $-0.67\pm0.03$ & $-0.59\pm0.04$ \\
COR   & $0.798\pm0.003$ & $0.792\pm0.007$ & $0.785\pm0.028$ & $0.796\pm0.004$ & $-0.74\pm0.04$ & $-0.61\pm0.05$ & $-0.59\pm0.04$ \\
IAS   & $0.778\pm0.023$ & $0.786\pm0.005$ & $0.778\pm0.025$ & $0.782\pm0.006$ & $-0.50\pm0.10$ & $-0.59\pm0.03$ & $-0.60\pm0.03$ \\
OCT-1 & $0.800\pm0.008$ & $0.813\pm0.005$ & $0.828\pm0.019$ & $0.804\pm0.004$ & $-0.67\pm0.03$ & $-0.62\pm0.03$ & $-0.64\pm0.02$ \\
OCT-2 & --              & --              & --              & --              & $-0.68\pm0.03$ & $-0.65\pm0.03$ & $-0.65\pm0.02$ \\
OCT-3 & --              & --              & --              & --              & $-0.73\pm0.03$ & $-0.64\pm0.03$ & $-0.65\pm0.02$ \\
ORK   & $0.808\pm0.031$ & $0.782\pm0.024$ & $0.782\pm0.027$ & $0.800\pm0.017$ & --             & --             & --             \\
QML-1 & $0.794\pm0.002$ & $0.787\pm0.004$ & $0.806\pm0.014$ & $0.792\pm0.002$ & $-0.56\pm0.12$ & $-0.69\pm0.05$ & $-0.66\pm0.05$ \\
QML-2 & $0.796\pm0.009$ & $0.797\pm0.004$ & $0.807\pm0.013$ & $0.797\pm0.003$ & --             & --             & --             \\
KAB-1 & $0.792\pm0.022$ & $0.809\pm0.005$ & $0.789\pm0.022$ & $0.799\pm0.006$ & $-0.56\pm0.09$ & $-0.56\pm0.03$ & $-0.61\pm0.03$ \\
KAB-2 & $0.779\pm0.022$ & $0.782\pm0.007$ & $0.732\pm0.040$ & $0.778\pm0.008$ & $-0.54\pm0.08$ & $-0.54\pm0.03$ & $-0.63\pm0.02$ \\
SYD-1 & $0.799\pm0.004$ & $0.792\pm0.004$ & $0.765\pm0.017$ & $0.797\pm0.003$ & $-0.65\pm0.03$ & $-0.64\pm0.03$ & $-0.63\pm0.02$ \\
SYD-2 & $0.798\pm0.004$ & $0.785\pm0.004$ & $0.741\pm0.019$ & $0.794\pm0.003$ & $-0.67\pm0.04$ & $-0.61\pm0.02$ & $-0.59\pm0.02$ \\
  \hline
\end{tabular}
\caption{Values of the $\nu_\mathrm{max}$ - $\Delta\nu$ ($a$) and $\nu_\mathrm{max}$ - $A_\mathrm{max}$ ($b$) power law exponents calculated over the ranges in $\nu_\mathrm{max}$ and $T_\mathrm{eff}$ indicated below. $a_1$ : $\nu_\mathrm{max}<750$\,$\mu$Hz; $a_2$ : $750<\nu_\mathrm{max}<1500$\,$\mu$Hz; $a_3$ : $\nu_\mathrm{max}>1500$\,$\mu$Hz; $a_4$ : All $\nu_\mathrm{max}$; $b_1$ : $T_\mathrm{eff}<5560$\,K; $b_2$ : $5560<T_\mathrm{eff}<6250$\,K; $b_3$ : $T_\mathrm{eff}>6250$\,K.}
\label{tab:powlawexp}
\end{table*}

\begin{table*}
  \centering
  \begin{tabular}{|c|c|c|c|c|c|}
  \hline
  Method & $T_\mathrm{eff}<5560$\,K & $5560<T_\mathrm{eff}<6250$\,K & $T_\mathrm{eff}>6250$\,K & $s$ at $T_\mathrm{eff}=T_{\mathrm{eff},\odot}$ & $ds/d(T_\mathrm{eff}/T_{\mathrm{eff},\odot})$ \\
  \hline
A2Z   & $0.85\pm0.16$ & $0.57\pm0.04$ & $0.50\pm0.02$ & $0.67\pm0.01$ & $-1.54\pm0.13$ \\
AAU   & $0.72\pm0.13$ & $0.60\pm0.03$ & $0.50\pm0.02$ & $0.65\pm0.01$ & $-1.15\pm0.12$ \\
COR   & $0.83\pm0.05$ & $0.56\pm0.05$ & $0.47\pm0.03$ & $0.66\pm0.01$ & $-1.65\pm0.08$ \\
IAS   & $0.56\pm0.12$ & $0.54\pm0.03$ & $0.46\pm0.02$ & $0.55\pm0.01$ & $-0.62\pm0.12$ \\
OCT-1 & $0.75\pm0.03$ & $0.58\pm0.03$ & $0.52\pm0.01$ & $0.64\pm0.01$ & $-1.05\pm0.06$ \\
OCT-2 & $0.76\pm0.03$ & $0.60\pm0.03$ & $0.52\pm0.02$ & $0.65\pm0.01$ & $-1.12\pm0.07$ \\
OCT-3 & $0.83\pm0.03$ & $0.60\pm0.03$ & $0.53\pm0.02$ & $0.68\pm0.01$ & $-1.36\pm0.07$ \\
QML   & $0.73\pm0.27$ & $0.61\pm0.06$ & $0.56\pm0.03$ & $0.65\pm0.01$ & $-0.90\pm0.14$ \\
SYD-1 & $0.64\pm0.10$ & $0.52\pm0.03$ & $0.49\pm0.02$ & $0.57\pm0.01$ & $-0.70\pm0.12$ \\
SYD-2 & $0.62\pm0.10$ & $0.51\pm0.03$ & $0.50\pm0.01$ & $0.55\pm0.01$ & $-0.57\pm0.10$ \\
KAB-1 & $0.75\pm0.02$ & $0.58\pm0.02$ & $0.50\pm0.01$ & $0.64\pm0.01$ & $-1.16\pm0.06$ \\
KAB-2 & $0.78\pm0.03$ & $0.56\pm0.02$ & $0.48\pm0.02$ & $0.64\pm0.01$ & $-1.38\pm0.06$ \\
  \hline
  \end{tabular}
  \caption{Table of $s$ values derived from power law fits and the parameters obtained from a linear fit to $s$ as a function of $T_\mathrm{eff}$.}
  \label{tab:svals}
\end{table*}

\label{lastpage}

\end{document}